\documentclass[11pt]{article}
\pdfoutput=1

\usepackage{jheppub}
\usepackage{url}
\usepackage{times}
\usepackage{latexsym}
\usepackage{graphicx,graphics,hyperref,amsmath,amssymb,xcolor,amsthm,array,bm}
\usepackage{subcaption}
\usepackage{listings}
\usepackage{float}
\usepackage{appendix}
\def\twinHe{$\hat{\text{H}}\text e$}
\def\twinH{$\hat{\text{H}}$}
\def\Ntwin{$\Delta \hat N$}
\def\vevratio{$\hat{v}/v$}
\newcommand{\CLASS}{\texttt{CLASS}}
\def\nur{$\Delta N_{\rm eff}$}
\def\lcdm{$\Lambda$CDM}
\def\lnur{$\Lambda$CDM~+~$\Delta N_{\rm eff}$}
\def\rhat{$\hat r$}
\newcommand\ie{{\it i.e.}}
\graphicspath{{plots/}}

\title{Mirror Twin Higgs Cosmology: Constraints and a Possible Resolution to the $H_0$ and $S_8$ Tensions}
\preprint{FERMILAB-PUB-21-371-T}

\author[a,b]{Saurabh Bansal,}
\author[c,d,e]{Jeong Han Kim,}
\author[a]{Christopher Kolda,}
\author[f]{Matthew Low,}
\author[a]{Yuhsin Tsai}

\affiliation[a]{Department of Physics, University of Notre Dame, IN 46556, USA}
\affiliation[b]{Department of Physics, University of Cincinnati, Cincinnati, Ohio 45221, USA}
\affiliation[c]{Department of Physics, Chungbuk National University, Cheongju, Chungbuk 28644, Republic of Korea}
\affiliation[d]{Center for Theoretical Physics of the Universe,
Institute for Basic Science, Daejeon 34126 Korea}
\affiliation[e]{School of Physics, KIAS, Seoul 02455, Korea}
\affiliation[f]{Theoretical Physics Department, Fermilab, P.O. Box 500, Batavia, IL 60510, USA}

\emailAdd{bansalsh@ucmail.uc.edu}
\emailAdd{jeonghan.kim@cbu.ac.kr}
\emailAdd{ckolda@nd.edu}
\emailAdd{mattlow@fnal.gov}
\emailAdd{ytsai3@nd.edu}

\abstract{The mirror twin Higgs model (MTH) is a solution to the Higgs hierarchy problem that provides well-predicted cosmological signatures with only three extra parameters: the temperature of the twin sector, the abundance of twin baryons, and the vacuum expectation value (VEV) of twin electroweak symmetry breaking.  These parameters specify the behavior of twin radiation and the acoustic oscillations of twin baryons, which lead to testable effects on the cosmic microwave background (CMB) and large-scale structure (LSS).  While collider searches can only probe the twin VEV, through a fit to cosmological data we show that the existing CMB (Planck18 TTTEEE+lowE+lowT+lensing) and LSS (KV450) data already provide useful constraints on the remaining MTH parameters.  Additionally, we show that the presence of twin radiation in this model can raise the Hubble constant $H_0$ while the scattering twin baryons can reduce the matter fluctuations $S_8$, which helps to relax the observed $H_0$ and $S_8$ tensions simultaneously. This scenario is different from the typical $\Lambda$CDM~+~$\Delta N_{\rm eff}$ model, in which extra radiation helps with the Hubble tension but worsens the $S_8$ tension.  For instance, when including the SH0ES and 2013~Planck~SZ data in the fit, we find that a universe with $\gtrsim 20\%$ of the dark matter comprised of twin baryons is preferred over $\Lambda$CDM by $\sim4\sigma$.  If the twin sector is indeed responsible for resolving the $H_0$ and $S_8$ tensions, future measurements from the Euclid satellite and CMB Stage 4 experiment will further measure the twin parameters to $O(1-10\%)$-level precision.  Our study demonstrates how models with hidden naturalness can potentially be probed using precision cosmological data.}
\begin{document}
\begin{flushright}
\small{.}
\end{flushright}
\maketitle

%%%%%%%%%%%%%%%%%%%%%%%
\section{Introduction}
\label{sec.intro}
%%%%%%%%%%%%%%%%%%%%%%%

The electroweak hierarchy problem remains one of the focal questions in particle physics.  While the big hierarchy is usually addressed through supersymmetry or compositeness, there are a number of approaches to solving the little hierarchy problem.  Traditional solutions utilize symmetries that predict new particles near the electroweak scale.  Consequently, these particles are either present only deep in the primordial dark ages or they persist until the present day as cold dark matter (CDM).  Therefore, typical solutions to the little hierarchy problem are expected to produce new signals at high energy colliders, but not in cosmological observations.

More recently, motivated by the increasingly strong limits on new particles from the Large Hadron Collider (LHC), solutions to the little hierarchy problem have been proposed that subvert this expectation.  These models permit the Higgs to be the only observable resonance within reach of the LHC, but, often unavoidably, predict deviations from the standard cosmological scenario ($\Lambda$CDM).  The twin Higgs model~\cite{Chacko:2005pe} is one of the early examples, but this class of ``hidden naturalness" models has expanded significantly~\cite{Chacko:2005un,Barbieri:2005ri,Chang:2006ra,Falkowski:2006qq,Cai:2008au,Craig:2014aea,Craig:2014roa,Chacko:2005vw,Burdman:2006tz,Craig:2013fga,Geller:2014kta,Low:2015nqa,Barbieri:2015lqa,Craig:2015pha,Batell:2015aha,Katz:2016wtw,Craig:2016kue,Badziak:2017wxn,Badziak:2017syq,Badziak:2017kjk,Serra:2019omd}.  Other variations include  $N$-naturalness~\cite{Arkani-Hamed:2016rle} and relaxion mechanisms~\cite{Graham:2015cka,Dvali:2004tma,Dvali:2003br} that also produce minimal collider signatures but have a  cosmological history that is outside the standard paradigm.

In fact, even if hidden naturalness were first discovered at the LHC, the predicted signatures are usually indirect in nature, such as missing transverse energy or deviations in Higgs couplings.  As a result, it would be challenging to attribute such observations uniquely to a hidden sector.

Constraints from cosmological observations, on the other hand, provide a complementary handle on new physics.  For example, additional radiation can be probed through cosmological measurements that are sensitive to the expansion rate of the universe.  Also, the matter power spectrum is sensitive to the presence of new particles that interact with the added radiation. As the precision of such measurements improve, there is more and more that can be learned about a possible hidden sector.

In this work, we specifically study the mirror twin Higgs model (MTH).  In this model, the entire particle content and symmetries of the Standard Model (SM) are replicated, yielding a ``twin sector.''  Each twin particle is heavier than its SM counterpart by a small factor.  While variations of twin Higgs models that only replicate part of the SM particle spectrum or symmetries have been proposed,\footnote{One example is the fraternal twin Higgs model~\cite{Craig:2015pha} where only the third generation of fermions, the color force, and the weak force are replicated.  In the hadrosymmetric twin Higgs model~\cite{Freytsis:2016dgf} the twin sector contains twin copies for all three generations of quarks, but the twin lepton sector is absent.} we consider the case with the minimal assumption of the  mirror-symmetry breaking, in which all particles and symmetries of the SM are also present in the twin sector.  Therefore, the particle content of the twin sector is expected to be similar to that of the SM. 

The presence of additional states in the twin sector, especially so many potentially stable particles, creates immediate problems for the MTH model.  At late times, there are stable twin electrons and baryons, which could together represent some or all of the universe's dark matter; and there are stable, ultra-relativistic states in the form of twin photons and neutrinos, the latter of which will come in three species.  It is these relativistic degrees of freedom that present the first challenge to the model, because such light degrees of freedom will alter the expansion rate of the universe and modify the well-known cosmological processes such as nucleosynthesis and the generation of CMB.\footnote{See also~\cite{Chacko:2021vin} for a study of the DM halo and direct detection signals of the MTH model.}

In order to preserve the successes of traditional Big Bang cosmology, one must constrain the number of additional light degrees of freedom present in any model of new physics. Following the standard practice of expressing this bound in terms of the number of additional effective species of neutrinos allowed, $\Delta N_{\rm eff}$, one finds that $\Delta N_{\rm eff}\lesssim 0.30$~\cite{Aghanim:2018eyx}.  On the other hand, the MTH model predicts $\Delta N_{\rm eff}\approx 5-6$, assuming that the temperature of the twin sector matches that of the SM~\cite{Chacko:2016hvu,Craig:2016lyx}.  That prediction, however, can be reduced if the temperatures in the twin sector are lower than those of the SM sector.  In fact, this is at least somewhat expected.  If the twin and SM sectors decouple as expected near a few GeV~\cite{Barbieri:2005ri}, then the twin sector will be naturally colder because at decoupling more twin species will have already left the thermal bath due to their heavier masses.  One could posit additional sources of cooling in the twin sector, for example, an asymmetric post-inflationary reheating that preferentially reheats the SM more effectively than the twin sector~\cite{Chacko:2016hvu,Craig:2016lyx,Beauchesne:2021opx}.  Alternative ways to cool the twin sector, and thereby lower $\Delta N_{\rm eff}$, have also been studied~\cite{Farina:2015uea,Barbieri:2016zxn,Csaki:2017spo,Barbieri:2017opf,Harigaya:2019shz}.  Regardless of the source, one requirement for a realistic MTH cosmology is that the twin sector must be colder than the SM sector.

It is not only the extra radiation, however, that leads to cosmological constraints on the MTH model.  Since the twin sector is a copy of the SM, twin baryons and electrons will undergo a sequence of events similar to the SM, the most important of which are twin Big Bang nucleosynthesis (twin BBN) and, later, twin recombination.  However, since the twin sector is heavier and colder than the SM, these processes will occur earlier than the analogous processes in $\Lambda$CDM.  As we shall see, these two processes will leave indelible marks on the universe, so we pause to give a brief overview of each process and its signatures.

Twin BBN is the process by which twin nuclei composed of twin protons and twin neutrons are created.  As in the SM, the vast majority of twin baryonic matter ends up in twin hydrogen and twin helium, although with different relative abundances.\footnote{For additional studies of BBN occurring in a dark sector, see~\cite{Krnjaic:2014xza,Redi:2018muu}.}  At later times, twin baryon density perturbations begin to oscillate as they enter the horizon due to an interplay between the gravitational potential and the twin photon pressure.  These twin baryon acoustic oscillations (twin BAO) are visible in the matter power spectrum via both a decrease in power at small scales and oscillations with a period in momentum space that is longer than that of BAO from the SM sector~\cite{Chacko:2018vss}. These oscillations are, in fact, a generic feature of hidden sector models with both dark radiation and matter to which it couples; but unlike the generic case, we will find that these oscillations turn off at twin recombination, which leaves the matter power spectrum unmodified at the largest scales.

Once the temperature of the universe cools sufficiently, the twin electrons and the twin hydrogen and helium nuclei combine into neutral twin atoms, again mimicking recombination in the SM sector.  At this point, twin BAO ends and the twin atoms begin to fall into the gravitational wells already seeded by any non-interacting cold dark matter.  Prior studies have made semi-analytic estimates showing that the size of the suppression in the matter power spectrum at small scales depends on the abundance of twin baryonic matter, while the scale at which the oscillations stop depends on the time of twin recombination~\cite{Chacko:2018vss}.  Such estimates provide a crucial first test of the MTH cosmology, but are not precise enough to make true comparisons to the high precision data on the CMB from Planck and on the matter power spectrum from, {\it e.g.}, the KV450 catalogue~\cite{Hildebrandt:2018yau}.  Nor do they allow us to predict with any confidence the ability of future surveys, such as Euclid~\cite{Amendola:2012ys} or CMB Stage 4 (CMB-S4)~\cite{Abazajian:2016yjj}, to constrain the parameter space of the MTH model.  Given the increasing accuracy of the datasets available, such comparisons are only possible when calculating the full evolution of the universe within the MTH paradigm. 

To that end, we have implemented the MTH model inside the Cosmic Linear Anisotropy Solving System, \CLASS~(v2.9)~\cite{Blas:2011rf}, which is one of the leading programs used to calculate the evolution of the universe.  For comparison with data and to extract parameter bounds, we use the cosmological inference package \texttt{MontePython} (v3.4)~\cite{Brinckmann:2018cvx}, interfaced with our version of \CLASS{}.\footnote{Our modifications to \CLASS{} are available at \url{https://github.com/srbPhy/class_twin}.}

Interestingly, as we will show, addressing the little hierarchy problem using a twin sector may in fact help to alleviate some puzzling tensions in the current cosmological data.  The first of these is the Hubble tension which is an inconsistency between the value of $H_0$ inferred from the CMB~\cite{Aghanim:2018eyx} and the locally measured value~\cite{Riess:2020fzl,Breuval:2020trd,Riess:2019cxk} obtained by the SH0ES collaboration.\footnote{There are other local $H_0$ measurements, for instance, based on the tip of the red-giant branch distance ladder measurement~\cite{Soltis:2020gpl,Freedman:2020dne,Freedman:2021ahq} or measurements using strong gravitational lensing systems~\cite{Suyu:2016qxx,Wong:2019kwg,Birrer:2020tax}. While the obtained $H_0$ varies between these different measurements, we will focus on the SH0ES result, which has the largest deviation from the Planck measurement.}  The inferred $H_0$ value from the $\Lambda$CDM fit to the Planck data is $4-5\sigma$ lower than the locally measured values~\cite{DiValentino:2021izs}.

The second is the $S_8$ tension where $S_8\equiv\sigma_8(\Omega_m/0.3)^{0.5}$~\cite{Hildebrandt:2018yau}, $\sigma_8$ is the root mean square of matter fluctuations around the $8h^{-1}~{\rm Mpc}$ scale, and $\Omega_m$ is the total matter abundance.  The value of $S_8$ inferred from the CMB is $2-3\sigma$ larger than measurements of large-scale structure from weak lensing and galaxy cluster surveys~\cite{Heymans:2013fya,MacCrann:2014wfa,Hildebrandt:2018yau,Joudaki:2019pmv,Heymans:2020gsg}.  While many models have been proposed to ease one or the other tension,\footnote{For recent reviews and other proposed solutions to the $H_0$ tension see, {\it e.g.}~\cite{Knox:2019rjx,DiValentino:2021izs,Schoneberg:2021qvd}. A mirror sector has been proposed to relax the $H_0$ tension~\cite{ Cyr-Racine:2021alc,Blinov:2021mdk} and both the $H_0$ and $S_8$ tensions~\cite{Prilepina:2016rlq,Chacko:2018vss}.} we will find that the MTH model naturally includes parameter space where both tensions are simultaneously and significantly ameliorated.

One necessity of using a detailed numerical approach is that a specific model needs to be chosen.  Though we describe the modifications made to the \CLASS{} software for working within the MTH model, the same roadmap can be followed to study other, similar models, including some of the variations of twin Higgs scheme mentioned above, or, for example, $N$-naturalness models.

The organization of the paper is as follows.  In Sec.~\ref{sec.twin} we discuss the relevant details of the MTH model, in particular the three parameters that determine the cosmological evolution of the twin sector.  Next, in Sec.~\ref{sec.MTHcosmology} the cosmological history of the twin sector is described, including twin BBN, twin recombination, and the evolution of perturbations.  In Sec.~\ref{sec.signals} the CMB and matter power spectrum are computed and compared to present data to extract limits on model parameters. In so doing, we demonstrate that the MTH model can significantly relax both the Hubble and $S_8$ tensions simultaneously. We also discuss the effect that non-linear corrections to the matter power spectrum will have on our results. Finally, we estimate the sensitivity of future observations and find that the MTH parameter space will be probed at the $O(1-10\%)$ level in next generation surveys.  Our conclusions are summarized in Sec.~\ref{sec.conclusions}.

%%%%%%%%%%%%%%%%%%%%%%%%%%%%%%%%%%%%%%%%%%%%%%%%%%%%%%%%%%%%%%%%%%
\section{Parametrizing the Twin Sector for Cosmological Studies}
\label{sec.twin}
%%%%%%%%%%%%%%%%%%%%%%%%%%%%%%%%%%%%%%%%%%%%%%%%%%%%%%%%%%%%%%%%%%

In the MTH model, the SM is accompanied by a twin sector which is an exact replica of the particle content and gauge symmetries of the SM.  Moreover, the gauge and Yukawa couplings of the two sectors are identical, enforced by a discrete $Z_2$ symmetry which is only softly broken.  This breaking of the $Z_2$ symmetry causes the vacuum expectation value (VEV) in the twin sector $\hat v$\footnote{Our convention is to write all twin particles and parameters using the corresponding SM notations but with a ``hat'' symbol on top.} to differ from that of the SM sector $v$.  We can parametrize the amount of breaking by the dimensionless ratio $\hat v/v$.

Current searches for invisible Higgs decays require $\hat{v}/v\gtrsim 3$~\cite{ATLAS:2020kdi}, which corresponds to a mild tuning, which we define as $2v^2/(v^2 + \hat{v}^2) \approx 20\%$.  Measurements of Higgs couplings~\cite{CMS:2020gsy}, such as the $Z$-Higgs coupling, also require $\hat{v}/v \gtrsim 2$. This region is consistent with electroweak precision bounds~\cite{Contino:2017moj} and flavor bounds~\cite{Csaki:2015gfd}. The rough requirement of a natural theory, with tuning $\gtrsim 10\%$~(1\%), sets an approximate upper bound of $\hat{v}/v \lesssim 5$~($15$).

The specific components of the twin sector relevant for the study of cosmological observables are: the twin radiation, in the form of twin photons $\hat\gamma$ and twin neutrinos $\hat\nu$; and the twin matter, in the form of twin protons $\hat p$ and twin neutrons $\hat n$, which, after a period of twin BBN, form twin hydrogen ${\rm \hat{H}}^+$ and twin helium nuclei ${\rm \hat{H}e}^{++}$.  These nuclei bind with twin electrons $\hat{e}$ during twin recombination, which means that the universe today has a fraction of its energy budget in the form of twin atoms, ${\rm \hat{H}}$ and ${\rm \hat{H}e}$, that may or may not bind together gravitationally to form larger structures (like twin stars and twin galaxies)\footnote{See~\cite{Curtin:2019lhm,Curtin:2019ngc,Winch:2020cju,Hippert:2021fch} for more discussion on twin stars and Ref.~\cite{Chacko:2021vin} for a discussion of twin galaxy formation.} depending on the details of the nuclear physics in the twin sector.

While it would be efficient for the twin baryons to make up the entirety of the universe's dark matter, this need not be the case. In fact, given our lack of understanding of how the SM baryon density is set, it is best to leave the density of twin baryons as a free parameter. We thus define a new variable to describe the abundance of twin matter in the universe today, $\hat r$, the fraction of the universe's dark matter that is composed of twin baryons: 
\begin{equation*}
    \hat r\equiv \frac{\Omega_{\rm all\,\,twin\,\, baryons}}{\Omega_{\rm DM}}.
\end{equation*}
If the twin and SM sectors were exactly alike, our naive expectation would be that there would be equal amounts of twin baryons and SM baryons, in which case $\hat r \approx 0.2$; that is, the twin baryons would only contribute about $5\%$ of the universe's total energy budget, which would be roughly 20\% of the dark matter budget. However, we will allow $\hat r$ to vary over its entire permissible range of $0<\hat r\leq 1$.

The contribution of any new form of radiation to the energy budget of the universe is most often expressed as a change in the effective number of extra neutrinos $\Delta N_{\rm eff}$.  In the MTH model, the contribution of twin radiation to $\Delta N_{\rm eff}$ is a complicated function of redshift that depends on the relative relativistic degrees of freedom in the two sectors (SM and twin). However, for the processes of interest for this analysis, \ie{} twin neutrino decoupling (for the twin BBN calculation) and recombination, it can be inferred from the ratio of the temperatures of the two sectors. Using the definition of $\Delta N_{\rm eff}$,
\begin{eqnarray}
    \hat{g}_*\,\hat T^4 = 2\times
    \frac{7}{8}\left(\frac{4}{11}\right)^{4/3}  (\Delta \hat N)\, T^4 
	\quad\implies\frac{\hat{T}}{T} = 
	\left(\frac{\Delta \hat N}{7.4}\right)^{1 / 4},
	\label{eq.Ttwin}
\end{eqnarray}
% %
% \begin{equation}
% 	\frac{\hat{T}}{T} = 
% 	\left(\frac{\Delta \hat N}{7.4}\right)^{1 / 4},
% 	\label{eq.Ttwin}
% \end{equation}
%
where $\hat g_* = 3.36$ is the twin relativistic degrees of freedom, $\hat T~(T)$ is the temperature of the twin (SM) sector and $\Delta\hat N$ is the contribution of the twin sector to $\Delta N_{\rm eff}$. 

In this paper we will treat the twin neutrinos as  free-streaming radiation. We will not consider the possibility of twin neutrinos being warm dark matter \cite{Cheng:2018vaj}, which can further suppress the matter power spectrum. Just as in the SM, before twin electron-positron annihilation, the twin neutrino and twin photon temperature are equal: $\ie~T_{\hat \nu} = \hat T$. However, after the twin electrons and positrons annihilate, the twin photon temperature increases (relative to the twin neutrinos) due to the resulting entropy dump, leading to $T_{\hat\nu} = \left(4/11 \right)^{1/3} \hat T$. Thus, dividing the contributions to $\Delta\hat N$ into those due to twin photons and neutrinos,
\begin{equation}
	\Delta \hat N = \Delta N_{\hat \gamma} + \Delta N_{\hat \nu},
\end{equation}	
we find
\begin{align}
	\Delta N_{\hat \gamma} &= \frac{1}{(7/8)}\left(\frac{11}{4}\right)^{4/3} \left( \frac{\hat T}{T}\right)^4 
	=4.4\, \frac{\Delta \hat N}{7.4}, \\
	\Delta N_{\hat \nu} &= 3\left(\frac{11}{4}\right)^{4/3} \left( \frac{T_{\hat\nu}}{T_{~}}\right)^4 
 	= 3\, \frac{\Delta \hat N}{7.4}.
 	\label{eq.Nnu}
\end{align}
As discussed in Sec.~\ref{sec.intro}, there are several mechanisms for arranging the temperature differential between the two sectors.  We will proceed under the assumption that $\Delta\hat N$ is a free parameter.  Thus, the free parameters of the MTH model relevant for this cosmological study are: $\hat v/v$, $\hat r$, and $\Delta \hat N$.

%%%%%%%%%%%%%%
\begin{figure}
  \centering
  \includegraphics[width=0.7\linewidth]{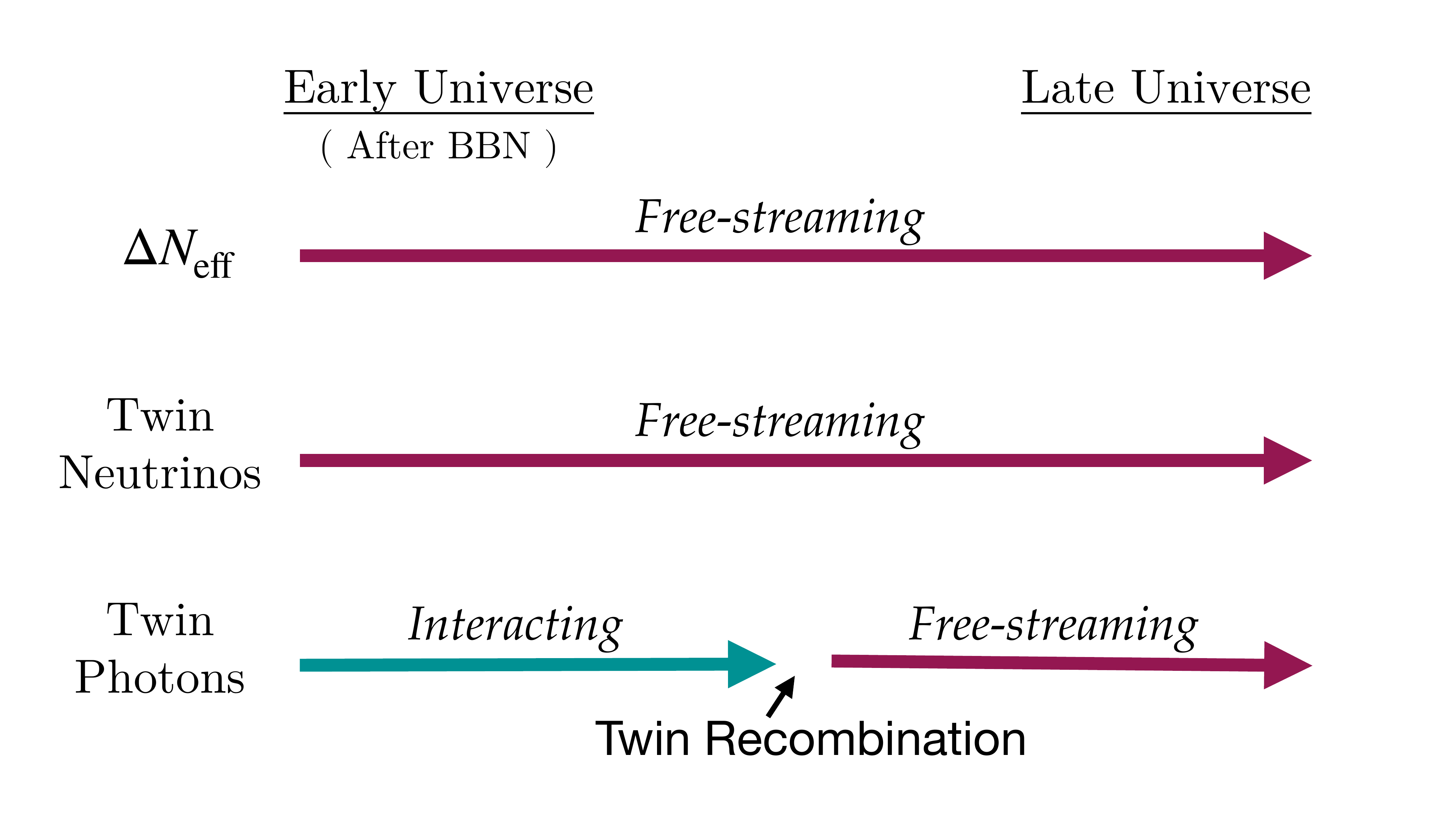}
  \caption{Pictorial representation of the phases of twin radiation.}
  \label{fig.radiation}
\end{figure}
%%%%%%%%%%%%%%

\paragraph{}
Throughout the rest of this work, we will be comparing the MTH model to two benchmark models.  The first is the canonical $\Lambda$CDM model, while the second model is a small variation on $\Lambda$CDM in which we add a new source of ultra-relativistic, free-streaming dark radiation with its own temperature set by \nur{}.  The reasons for comparing to this latter model are two-fold. Firstly, the MTH model contains both dark radiation (twin photons and twin neutrinos) and interacting dark matter (twin baryons) and by comparing to the \lnur{} model (with $\Delta N_{\rm eff} = \Delta\hat N$), we can better isolate the effects of the twin baryons, while keeping the radiation energy density, and thus the expansion rate of the universe, the same in both models.  Secondly, it has been shown~\cite{DiValentino:2021izs} that allowing $\Delta N_{\rm eff}>0$ helps to ease the Hubble tension. By comparing our model to the \lnur{} model, we can identify when and how the extra structure of the twin sector plays any role in fitting to the cosmological parameters.

One unique aspect of the MTH model is, in fact, the behavior of the dark radiation. While the twin neutrinos in the MTH model are always free-streaming, the twin photons behave like interacting radiation prior to twin recombination, but then stream freely afterwards. We indicate this schematically in Fig.~\ref{fig.radiation}.  In this way, the dark radiation of the MTH model differs from that in our benchmark \lnur{} model.

%%%%%%%%%%%%%%%%%%%%%%%%%%%%%%%
\section{Mirror Twin Cosmology}
\label{sec.MTHcosmology}
%%%%%%%%%%%%%%%%%%%%%%%%%%%%%%%

The matter and CMB power spectra are highly sensitive to thermal processes of the early universe, namely BBN and recombination.  As the twin sector contains a near-perfect copy of the SM, equivalent thermal processes, such as twin BBN and twin recombination, will also occur.  The details and signatures of these twin processes, however, can be quite different from those of the SM.  The difference between the two sectors predominantly stems from the facts that the twin particles are heavier and/or colder than their SM counterparts and that we directly observe the power spectra in SM matter and radiation instead of via the twin counterparts. 

These early universe twin processes leave imprints on cosmological observables.  Therefore, an accurate account of these processes is vital for studying the MTH model using cosmological datasets.  We first discuss the important physics of twin BBN and twin recombination that can affect their cosmological signatures.  Our estimates closely follow those of Ref.~\cite{Chacko:2018vss}.  We will then be prepared to discuss the evolution of perturbations in the MTH model.  In each part of the discussion, we provide a brief overview of our implementation of the MTH model into \CLASS{}.
 
%%%%%%%%%%%%%%%%%%%%%%%%%%%%%%%%%%
\subsection{Twin BBN}
\label{subsec.bbn}

Big Bang nucleosynthesis is the primordial process which leads to the formation of helium nuclei along with traces of other light elements.  For the purposes of this work, the only output from BBN that we need to consider is the primordial mass fraction of helium, $Y_{\rm p}(^4\rm He) \equiv \rho_{\mathrm{He}}/(\rho_{\mathrm{He}}+\rho_{\mathrm{H}} )$.  In order to compute $Y_{\rm p}(^4\rm He)$, we use the \CLASS{} implementation of the BBN code \texttt{PArthENoPE}~(v1.0)~\cite{Pisanti:2007hk}, assuming a neutron lifetime of 885.7~seconds.

Twin BBN is the parallel process that occurs for twin matter.  During twin BBN, twin helium nuclei (\ie~nuclei made up of two twin protons and two twin neutrons) are formed, possibly along with traces of heavier twin nuclei.  Just as in the SM, we expect the primordial mass fraction of the heavier twin elements to be negligible and the twin matter energy density to be dominated by twin hydrogen \twinH and twin helium \twinHe.  As such, we ignore all heavier twin elements in calculating twin BBN.  With that simplification, the primordial abundance of twin helium, $\hat Y_{\rm p}(^4$\twinHe)$\equiv \rho_{\mathrm{\hat He}}/(\rho_{\mathrm{\hat He}}+\rho_{\mathrm{\hat H}})$, can be computed in a straightforward manner.

The successful formation of twin helium also requires the twin neutron to live long enough to form twin deuterium.  In the SM sector, the deuterium bottleneck means that some neutrons have decayed before they can form deuterium.  As discussed in~\cite{Chacko:2018vss}, the twin neutron yield at the twin deuterium formation depends on the ratio of the twin neutron lifetime over the twin deuterium formation time, where the later number can be estimated using the lattice result for the deuterium binding energy (see {\it e.g.},~\cite{Savage:2015eya} for a review). From the estimate in~\cite{Chacko:2018vss}, both of these time scales are roughly proportional to ${\hat v}^{-1}$ when $(\hat v/v)\gtrsim 2$. The resulting neutron yield is therefore insensitive to the $(\hat v/v)$ value and is similar to or only slightly smaller than the ratio in the SM case. The decay of twin neutrons before twin deuterium formation can at most suppress $\hat Y_{\rm p}(^4$\twinHe) by an additional $\mathcal{O}(10\%)$ relative to the calculated ratio and will not significantly change the conclusions of our cosmological study.

The twin helium mass fraction, $\hat Y_{\rm P}$, primarily depends on the freeze-out number density of twin neutrons $\hat n$, as almost all the twin neutrons after freeze-out combine with twin protons $\hat p$ to form \twinHe~nuclei.  Taking this to be the case, the primordial abundance of twin helium can be computed using,
\begin{equation}
	\hat Y_{\rm p}({ }^{4} \mathrm{\hat He}) \equiv \frac{\rho_{\mathrm{\hat He}}}{\rho_{\mathrm{\hat He}}+\rho_{\mathrm{\hat H}}} \approx \frac{4  n_{\mathrm{\hat He}}}{ n_{\mathrm{\hat H}}+4  n_{\mathrm{\hat He}}}=2  X_{\hat n}^{\text{freeze-out}},
\label{eq.BBNYpHe}
\end{equation}
where $ X_{\hat n}^{\text{freeze-out}}$ is defined as the freeze-out value of the twin neutron fraction, $ X_{\hat n} \equiv  n_{\hat n}/ ( n_{\hat n} +  n_{\hat p})$.   The evolution of $ X_{\hat n}$, in turn, can be calculated using the following Boltzmann equation~\cite{Bernstein:1988ad,Sarkar:1995dd,Mukhanov:2003xs,Fradette:2017sdd},
\begin{equation}
\frac{d X_{\hat n}}{d T}
=\frac{ \Gamma_{\hat n \hat\nu_{ e} \rightarrow \hat p \hat e^{-}}+
\Gamma_{\hat n \hat e^{+} \rightarrow \hat p {\bar{\hat \nu}}_{e}}}{T H(T)}
\left( X_{\hat n}-\left(1- X_{\hat n}\right) e^{-\hat Q / \hat T}\right),
\label{eq.BBNXn}
\end{equation}
where $\Gamma_{\hat n \hat\nu_{ e} \rightarrow \hat p \hat e^{-}}$ and $ \Gamma_{ \hat n \hat e^{+} \rightarrow \hat p {\bar {\hat \nu}}_{e}}$ are the twin neutron depletion rates and $\hat Q \equiv  m_{\hat n} -  m_{\hat p}$.  In the above equation, as long as $\hat T \gg \hat Q$, the twin neutrons and twin protons remain in equilibrium, with $ X_{\hat n} = 1/2$.  The depletion rates mimic the SM calculation:
\begin{equation}
\Gamma_{\hat n \hat\nu_{e} \rightarrow \hat p \hat e^{-}}=\frac{1+3 g_{a}^{2}}{2 \pi^{3}} \hat G_{F}^{2} \hat Q^{5} J(1 ; \infty), 
\quad\quad\quad
\Gamma_{\hat n \hat e^{+} \rightarrow \hat p {\bar {\hat \nu}}_{e}}=\frac{1+3 g_{a}^{2}}{2 \pi^{3}} \hat G_{F}^{2}
\hat Q^{5} J\left(-\infty ;-\frac{ m_{\hat e}}{\hat Q}\right),
\label{eq.depletion1}
\end{equation}
where
\begin{equation}
J(a, b) \equiv \int_{a}^{b} \sqrt{1-\frac{( m_{\hat e} / \hat Q)^{2}}{q^{2}}} \frac{q^{2}(q-1)^{2} d q}{\left(1+e^{\frac{\hat Q
}{ T_{\hat\nu}}(q-1)}\right)\left(1+e^{-\frac{\hat Q}{\hat T} q}\right)}.
\label{eq.depletion2}
\end{equation}
Though it is not obvious from the above equations, $J(a,b)$ in the above equation has a strong temperature dependence; this ensures that as the temperature decreases, the coefficient on the right of Eq.~\eqref{eq.BBNXn} drops below 1, so that freeze-out can happen.

\begin{figure*} [!htb]
  \centering
  \includegraphics[width=0.7\textwidth]{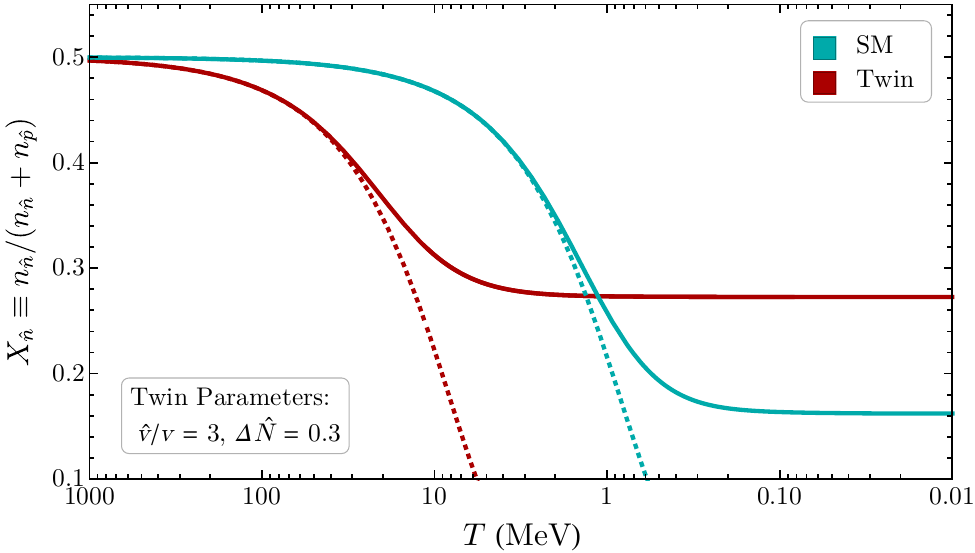}
  \caption{An example of the evolution of the twin neutron fraction $X_{\hat{n}}$ (solid red line) as a function of temperature showing twin neutron freeze-out. Also shown is the evolution of the SM neutron fraction $X_{n}$ (solid teal line). The dotted lines indicate the equilibrium values of the neutron fraction for the two sectors.}
  \label{fig.BBN_LCDMvsMTH}
\end{figure*}

Next, we briefly discuss our estimate of the mass difference, $\hat Q$, between the twin neutron and twin proton. Although the gauge couplings of the twin sector are equal to those of the SM at the cutoff scale of the theory (about $5-10$ TeV), the running of the couplings in the two models is different because of the higher twin VEV.  Due to this, the confinement scale for twin QCD ($\hat \Lambda_{\rm QCD}$) is higher than that of the SM ($ \Lambda_{\rm QCD}$).  Using the renormalization group running of the strong coupling at one-loop for the twin sector, one finds
\begin{equation}
\frac{\hat\Lambda_{\rm QCD}}{\Lambda_{\rm QCD}} \approx 0.68+0.41 \log \left(1.32+\hat v/v\right).
\end{equation}
Using this $\hat \Lambda_{\rm QCD}$ and the lattice QCD result in Fig.~3 of Ref.~\cite{Borsanyi:2014jba}, we can fit the mass splitting function between neutron and proton and estimate $\hat Q$ to be~\cite{Chacko:2018vss}
\begin{equation}
\frac{\hat Q}{Q} \approx 1.68 (\hat v/v) -0.68, 
\quad\quad\quad
Q \equiv m_n - m_p = 1.29~\mathrm{MeV}.
\end{equation}
Finally, the mass of the twin electron and the twin Fermi constant are related to their SM counterparts by,
\begin{equation}
m_{\hat e} = (\hat v /v) \; m_e,
\quad\quad\quad
\hat G_F = (\hat v/v)^{-2} \; G_F.
\label{eqn:twinmass}
\end{equation}
An example of the evolution of $X_{\hat n}$ for the twin sector with \vevratio~= 3 and \Ntwin~= 0.3 is shown in Fig.~\ref{fig.BBN_LCDMvsMTH}.  Just like the SM sector, the twin sector starts with an equal number density of twin neutrons and twin protons since their mass difference is small compared to their temperature and $\hat n \leftrightarrow \hat p$ rates are in balance.  However, as the universe cools, the rate of $\hat p \to \hat n$ process drops, suppressing the twin neutron fraction.  In time, the scattering rate drops below the Hubble rate; ultimately leading to the freeze-out of twin neutrons.  It is worth noting that the freeze-out of the twin neutrons happens much earlier than that of SM neutrons.  We find that for $1~{\rm MeV} \lesssim T \lesssim 10^4~{\rm MeV}$, the depletion rates $ \Gamma_{ \hat n \hat\nu_{ e} \rightarrow \hat p \hat e^{-}}$ and $\Gamma_{\hat n \hat e^{+} \rightarrow \hat p {\bar {\hat \nu}}_{e}}$ are smaller than the SM values by a factor of about $10^3$ -- $10^4$. Due to this, the coefficient on the right-hand side of Eq.~\eqref{eq.BBNXn} drops below unity at a higher temperature, leading to a earlier freeze-out.

%%%%%%%%%%%%%%%%%%
%%%%%%%%%%%%%%%

\begin{figure*} [!htb]	
  \centering
  \begin{subfigure}{0.45\textwidth}
    \includegraphics[width=\textwidth]{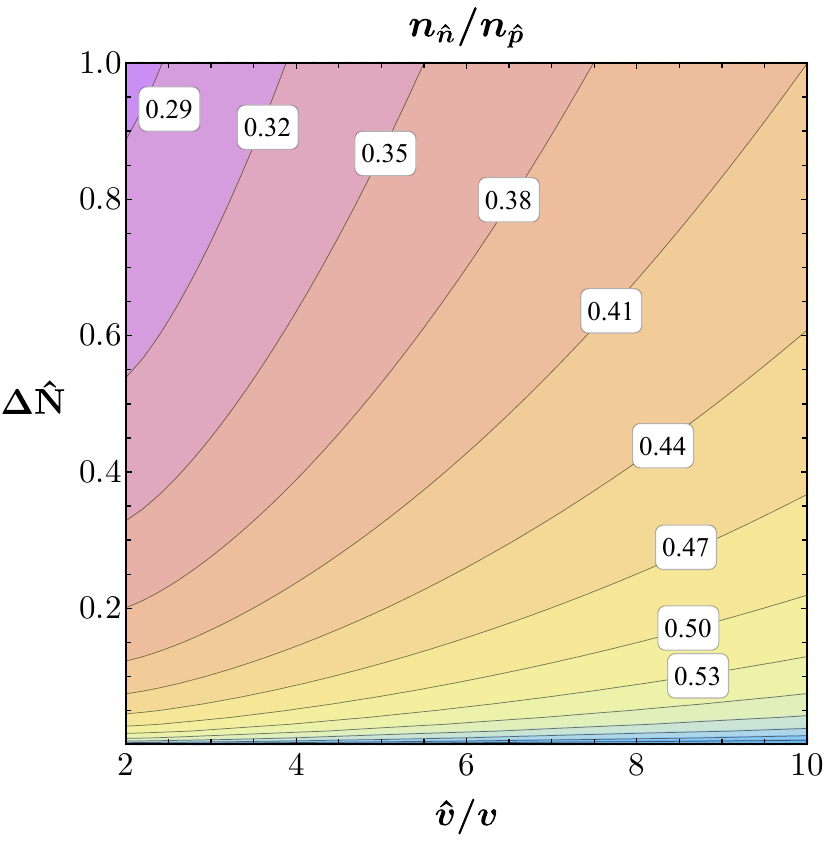} 			
  \end{subfigure}
  \quad\quad\quad  %%%%%%%%%%%%%%%%%%%%
  \begin{subfigure}{0.45\textwidth}
    \includegraphics[width=\textwidth]{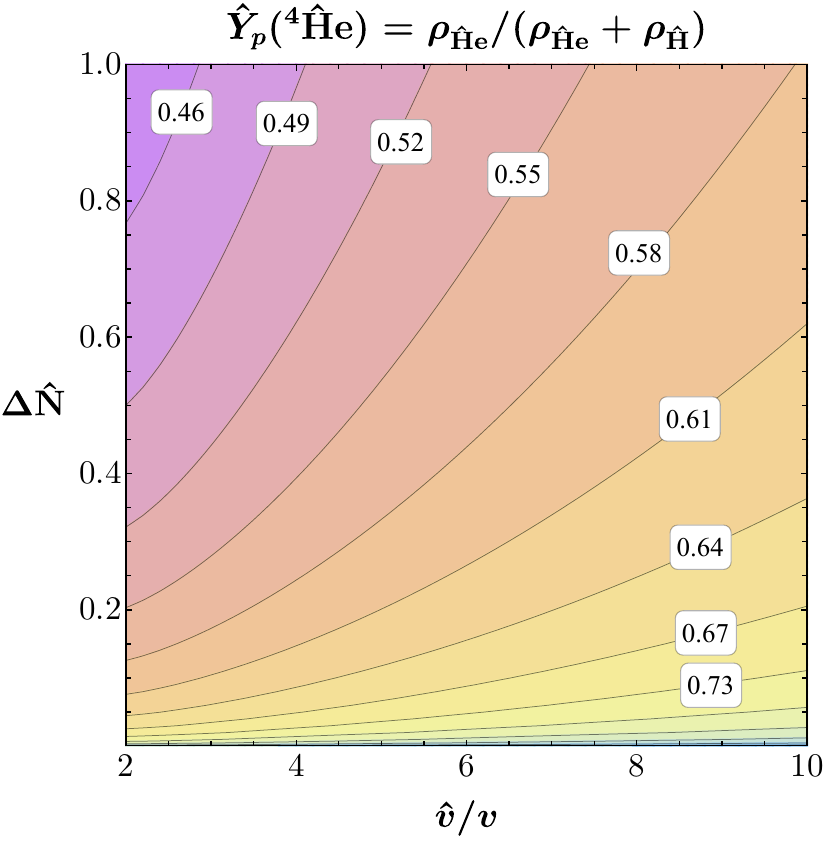}
  \end{subfigure}
  \caption{The twin neutron to twin proton fraction (left) and the primordial abundance of twin helium (right) obtained from twin BBN.}
  \label{fig.TBBN} 
\end{figure*}

Finally, using Eqs.~\eqref{eq.BBNYpHe} and~\eqref{eq.BBNXn}, we compute the primordial mass fraction of twin helium for the parameter space of \vevratio~and \Ntwin.  The results of this analysis are shown in Fig.~\ref{fig.TBBN}.\footnote{Note that the results shown in Fig.~\ref{fig.TBBN} are different from those in Ref.~\cite{Chacko:2018vss}, and the size of $\hat Y_{\rm p}(^4{\rm \hat{H}e})$ is lower than the estimate in~\cite{Chacko:2018vss} by $\approx 30\%$. This discrepancy is the result of an error in the calculations of Ref.~\cite{Chacko:2018vss}. Specifically, for the exponential in Eq.~\eqref{eq.BBNXn} we use twin photon temperature $\hat T$, whereas Ref.~\cite{Chacko:2018vss} incorrectly used the SM photon temperature $T$.}  A colder (\ie~lower \Ntwin{}) or heavier (\ie~higher \vevratio) twin sector has a larger primordial mass fraction of twin helium.  To incorporate twin BBN in \CLASS{}, we calculate $\hat Y_{\rm p}({ }^{4} \mathrm{\hat He})$ in \texttt{Mathematica} for a range of values of \vevratio~and \Ntwin{} using the procedure discussed above and store the output in a file.  This file is then read and interpolated in \CLASS{} to obtain $\hat Y_{\rm p}({ }^{4} \mathrm{\hat He})$ for the twin parameters of interest.

%%%%%%%%%%%%%%% 
%%%%%%%%%%%%%

%%%%%%%%%%%%%%%%%%%%%%%%%%%%%%%
\subsection{Twin Recombination}
\label{subsec.recombination}

Like in the SM sector, twin nuclei formed during twin BBN combine with twin electrons to produce twin neutral atoms.  We call this process twin recombination and it occurs in three distinctive steps.  First, doubly-ionized twin helium ($ {\rm \hat He~III}$) nuclei combine with a twin electron to form singly-ionized twin helium ($ {\rm \hat He~II}$).  Then, ${\rm \hat He~II}$ nuclei combine with a twin electron to form neutral twin helium ($ {\rm \hat He~I}$) atoms.  Finally, all of the remaining twin protons in the universe combine with free twin electrons to form twin hydrogen (${\rm \hat H~I}$) atoms. 

The dynamics of twin recombination can be studied by tracking the evolution of the ionization fraction of twin electrons, $x_{\hat e} \equiv n_{\hat e}/n_{\rm{\hat H}}$.   Here, $n_{\rm\hat H}$ is the combined number density of neutral and ionized twin hydrogen atoms.  The ionization fraction, in turn, can be calculated by tracking the capture of twin electrons into excited atomic levels, followed by their decay to the ground state.  A precise calculation of twin recombination would require a level-by-level treatment of both twin helium and hydrogen atoms.

The recombination of twin hydrogen can be well-described by a three-level system of the twin hydrogen atom, as was originally done by Peebles for the standard hydrogen recombination~\cite{Peebles:1968ja}.  However, a similar account of twin helium recombination is beyond the scope of this work as, even in the SM, helium recombination depends on a complex network of transitions between the various energy levels.  In our analysis, we approximate the twin helium recombination using the Saha equations, which are known to accurately predict the onset of recombination.

We do not expect this approximation to have a significant effect on our results.  This is because the effects of twin recombination on our observables are dominated by the physics of the last recombination (\ie~\twinH-recombination), the time at which twin photons decouple.  That is to say the current data is not sensitive to the details of \twinHe{} recombination.  To accurately compute the twin hydrogen recombination we use Peebles-like three-level twin hydrogen atom.  Note that for the recombination in the SM sector, we utilize the \CLASS{} implementation of \texttt{RECFAST}~\cite{Seager:1999bc} version 1.5.2 to compute the evolution of $x_e$ for the SM.  This calculation of $x_e$ is based on a level-by-level treatment of both hydrogen and helium atoms~\cite{Seager:1999km}.

The ionization fraction of twin electrons ($ x_{\hat e}$) resulting from twin helium recombination is computed using the following Saha equations~\cite{Seager:1999km}:
\begin{align}
    \frac{\left(x_{\hat e}-1- f_{\mathrm{\hat He}}\right)  x_{\hat e} n_{ \mathrm{\hat H}}}{1+2 f_{\mathrm{\hat He}} - x_{\hat e}}
    &=
	\left( \frac{ m_{ \hat{e}}  \hat T}{2 \pi } \right)^{3 / 2}
	e^{-\epsilon_{\mathrm{\hat HeII}} /  \hat T},
	\quad\quad\quad
	(\mathrm{\hat He~III \to \hat He~II}), \label{eq.SahaHeIII}\\
	\frac{\left( x_{\hat{e}}-1\right)  x_{\hat{e}} n_{ \mathrm{\hat H}}}{1+ f_{\mathrm{\hat He}}- x_{\hat{e}}}
	&=
	4\left( \frac{m_{ \hat{e}}  \hat T}{2 \pi } \right)^{3 / 2}
	e^{-\epsilon_{\mathrm{ \hat HeI}} /  \hat T},
	\quad\quad\quad
	(\mathrm{\hat He~II \to \hat He~I}), \label{eq.SahaHeII}
\end{align}
where, $ f_{\mathrm{\hat He}} \equiv n_{\mathrm{\hat He}}/\hat n_{\mathrm{\hat H}}$, $ m_{\hat e}=(\hat v/v) m_{e} $ is the mass of twin electrons, and $ \epsilon_{\mathrm{\hat HeII}}$ and $ \epsilon_{\mathrm{\hat HeI}}$ are the binding energies of the first and second twin electron, respectively, in the twin helium atom. 
Since the binding energies are directly proportional to the mass of the twin electrons, the latter two quantities can be written as,
\begin{align}
{\epsilon}_{\mathrm{\hat HeII}} &= \left(\frac{\hat v}{v} \right)\epsilon_{\mathrm{HeII}} 
\approx \left(\frac{\hat v}{v} \right) 54.4 \text{~eV}, \\
{\epsilon}_{\mathrm{\hat HeI}} &= \left(\frac{\hat v}{v} \right)\epsilon_{\mathrm{HeI}} 
\approx \left(\frac{\hat v}{v} \right) 24.6 \text{~eV}.
\end{align}
For twin hydrogen recombination, we follow the procedure outlined in Ref.~\cite{Chacko:2018vss}.  The evolution of $ x_{\hat e}$ during twin hydrogen recombination is computed using the following relation
\begin{equation}
\frac{d  x_{\hat e}}{d t}=-\frac{ \Lambda_{\hat\alpha}+ \Lambda_{2 \hat\gamma}}{ \Lambda_{\hat\alpha}+ \Lambda_{2 \hat \gamma}+4 \hat\beta} \hat\alpha^{(2)}
\left[n_{\mathrm{\hat H, tot}}  x_{\hat e}^{2}-\left(1- x_{\hat e}\right)\left(\frac{ m_{\hat e} \hat T}{2 \pi}\right)^{3 / 2} e^{- \epsilon_{\mathrm {\hat H}} / \hat T}\right],
\label{eq.recH}
\end{equation}
where $t$ is the time, $ \Lambda_{\hat\alpha}$ is the decay rate of the $\hat{\text{H}}(2p)$ state to the $\hat{\text{H}}(1s)$ state due to the emission of a Lyman-$\alpha$ twin photon, $ \Lambda_{2\hat\gamma}$ is the two twin photons decay rate of the $\hat{\text{H}}(2s)$ state to the $\hat{\text{H}}(1s)$ state, and $\hat\alpha^{(2)}$ and $\hat\beta$ are the recombination and photoionization rates of the $n = 2$ states of twin hydrogen. The last two quantities can be approximated as~\cite{Ma:1995ey}
\begin{align}
	\hat \alpha^{(2)} &=
	0.448 \frac{64 \pi}{\sqrt{27 \pi}} \frac{\alpha^{2}}{ m_{\hat e}^{2}}
	\left(\frac{\epsilon_{\rm\hat H}}{\hat T}\right)^{1 / 2} 
	\ln \left(\frac{\epsilon_{\rm\hat H}}{\hat T}\right), \\
	\hat \beta &= \frac{\hat \alpha^{(2)}}{4}
	\left(\frac{ m_{\hat e} \hat T}{2 \pi}\right)^{3 / 2} 
	e^{- \epsilon_{\rm \hat H} / 4 \hat T},
\end{align}
where, $\alpha$ is the fine structure constant and is equal to the SM value, and $ \epsilon_{\rm \hat H}$ is twin hydrogen atom binding energy, given by
\begin{equation}
	{\epsilon}_{\rm\hat H}=(\hat v/v) \epsilon_{\rm H}.
\end{equation}
The Lyman-$\alpha$ and two twin photon transition rates are given by~\cite{1951ApJ...114..407S,CyrRacine:2012fz,Chacko:2018vss}
\begin{align}
	 \Lambda_{\hat \alpha} &= \frac{H\left(3\,  \epsilon_{\rm \hat H}\right)^{3}}
	 {(8 \pi)^{2}  n_{\mathrm{\hat H}, \text { tot }} (1- x_{\hat e})},\\
	\Lambda_{2 {\hat \gamma}} &= \left(\frac{\hat{\alpha}_{\mathrm{em}}}{\alpha_{\mathrm{em}}}\right)^{6}\left(\frac{\hat{\epsilon}_{\rm H}}{ \epsilon_{\rm H}}\right) \Lambda_{2 \gamma}=\left(\frac{\hat v}{v}\right) \Lambda_{2 \gamma}~,
\end{align}
where $\Lambda_{2 \gamma} = 8.227~{\rm sec}^{-1}$.

To compute twin recombination in \CLASS{}, we have incorporated Eqs.~\eqref{eq.SahaHeIII}, \eqref{eq.SahaHeII}, and \eqref{eq.recH} in its \texttt{thermodynamics} module.\footnote{For the numerical computation of \twinH-recombination in \CLASS{}, we first start with the Saha equation instead of Eq.~\eqref{eq.recH} for $1\leq  x_{\hat e} \leq 0.99$ in order to avoid numerical fluctuations. 	This works because it is well-known that the Saha equation accurately predicts the onset of recombination.  In fact, a similar procedure is also used by \texttt{RECFAST} to compute SM H-recombination.  For completeness, we here provide the Saha equation for twin hydrogen,
\begin{equation} 
	\frac{  x_{\hat e}^2}{1-x_{\hat e}}=
	\frac{(2 \pi m_{ \hat{e}} \hat T)^{3 / 2}}{ n_{ \mathrm{\hat H}}} 
	e^{-\epsilon_{\mathrm{\hat H}} /\hat T},
    \quad\quad\quad
	(\mathrm{\hat H~II \to \hat H~I}). \nonumber \label{eq.SahaH}
\end{equation}}

%%%%%%%%%%%%%%%%%%%%%%
\begin{figure*} [!htb]	
	\centering
	\begin{subfigure}{0.8\textwidth}
		\includegraphics[width=\textwidth]{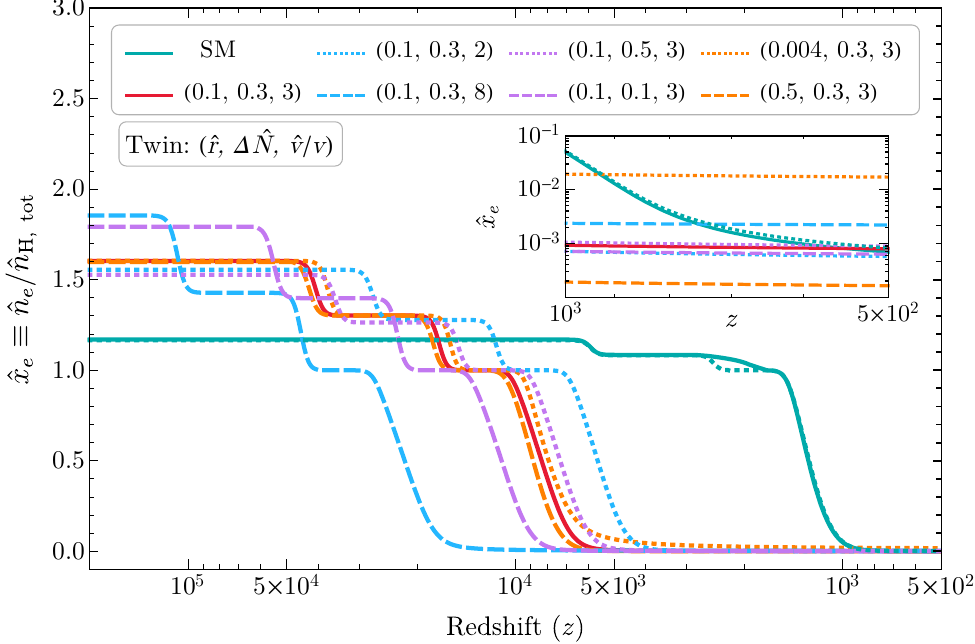}
		\caption{} \label{fig.TwinRecXe}
	\end{subfigure}
	\vspace{1em}
	\begin{subfigure}{0.8\textwidth}
		\includegraphics[width=\textwidth]{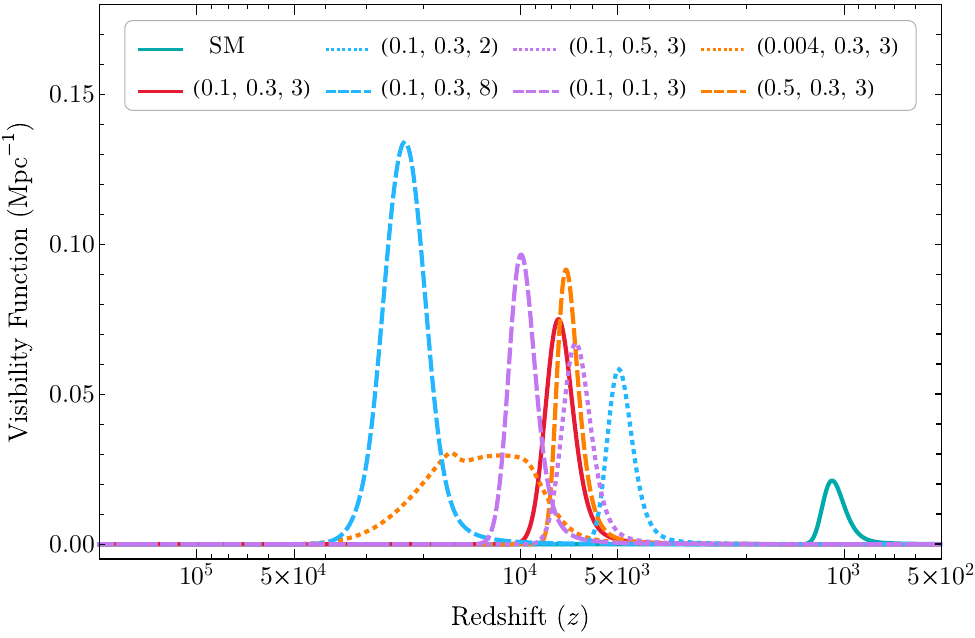} 
		\caption{} \label{fig.TwinRecVis}
	\end{subfigure}
	\caption{The ionization fraction (Fig.~\ref{fig.TwinRecXe}) and 
		visibility function (Fig.~\ref{fig.TwinRecVis}) for $\Lambda$CDM and the MTH model with various twin parameters labelled according to $( \hat r$, $\Delta \hat N$, $\hat v/v )$.  The $y-$axis corresponds to $x_e$ for the $\Lambda$CDM model.}
		\label{fig.rec}
\end{figure*}
%%%%%%%%%%%%%%%%%%%%%%

In Fig.~\ref{fig.rec}, we show a few examples of $ x_{\hat e}$ as computed by \CLASS{} for various combinations of twin parameters.  In the figure, the values of twin parameters are indicated in the following format: $(\hat r, \Delta \hat N, \hat v/v)$.  In the lower panel, we also show the visibility function, $g(z)$, as a function of redshift.\footnote{We define the visibility function $g(\eta)$ for the twin sector at a conformal time ($\eta$) as,
\begin{equation}
	g(z) = - \frac{d\tau}{d\eta}e^{-\tau}, 
	\quad\quad\quad
	\text{~where~} \tau(\eta) \equiv \int_{\eta}^{\eta_0} d\eta\, a n_{\hat e} \hat \sigma_T
\end{equation}
where $a$ is the scale factor.  The visibility function is normalized with respect to $\eta$. Therefore, in Fig.~\ref{fig.TwinRecVis}, where the visibility function is plotted as a function of redshift, the different curves do not appear to be normalized to 1.}  The visibility function is a probability distribution and is defined as the probability that a photon (or twin photon) in the CMB was last scattered at a particular redshift, $z$.  Although recombination happens over a finite range of time, the nominal redshift of recombination is defined as the point at which the visibility function reaches its maximum.  In the SM sector, this maximum happens towards the end of the recombination of hydrogen.  However, as discussed below, this is not necessarily the case for the twin sector.   In fact, as we will see, the visibility function will provide an unique insight in to the physics of twin recombination, particularly when only a tiny fraction of the dark matter is made up of twin baryons.

We next discuss a few salient features of Fig.~\ref{fig.rec}.  Here, we use the red curve corresponding to $(\hat r, \Delta\hat N, \hat v/v) = (0.1, 0.3, 3)$ as our benchmark twin model, while the dotted and dashed curves show the results obtained by changing just one twin parameter relative to the benchmark model.  The SM-labelled curve shows the evolution of $x_e$ (rather than $ x_{\hat e}$) due to recombination in the $\Lambda$CDM model.\footnote{Note that in the SM, $x_e$ starts to grow again at $z \lesssim 100$ due to reionization. Though a similar situation may happen in the twin sector, we do not consider that to be the case in this work.  We assume that the twin atoms do not reionize after twin recombination. However, even if they did, we do not expect this process to have much impact on our results.}  This curve is based on the built-in functionality in \CLASS{} to compute $x_e$ and utilizes level-by-level treatment of both the hydrogen and helium atoms.  In addition, we show a dotted line in the same color as the SM curve (most visible around $z=2\times 10^3$).  This curve shows the evolution of $x_{\hat{e}}$ for the twin sector model in the limit when it behaves like the SM, which happens for $(\hat r, \Delta \hat N, \hat v/v)$ = (0.188, 7.4, 1).  As we can see from the figure, the evolution of $x_{\hat e}$ in such a model closely follows the evolution of $x_e$ in the SM, except around HeII $\to$ HeI recombination, which is expected from the earlier discussion.

Here are a few additional important observations from Fig.~\ref{fig.rec}:
\begin{itemize}
	
	\item
	The first thing to notice in Fig.~\ref{fig.TwinRecXe} is that the starting value of $ x_{\hat e}$ (at the largest redshifts) is not the same for the different twin models shown in the figure.  This is because at those redshifts, all the twin electrons are ionized while the denominator, in the definition of $x_{\hat  e}$, does not include the twin protons bound into the \twinHe{} nuclei:
	\begin{equation}
	 x_{\hat e} \equiv \frac{ n_{\hat e}}{ n_{\rm{\hat H}}} = 
	\frac{ n_{\rm\hat H} + 2 n_{\rm\hat He}}{ n_{\rm{\hat H}}} = 
	1+2\frac{ \hat Y_p }{4(1- \hat Y_p)}, \quad\quad\quad
	\text{(at early times)}, \nonumber
	\end{equation}
	where $\hat Y_p \equiv \hat Y_p (^4{\rm \hat He})$. Since these models have different abundances of twin helium (see twin BBN in Sec.~\ref{subsec.bbn}), their starting $ x_{\hat e}$ are different. Heavier and/or colder twin sectors lead to larger starting $x_{\hat e}$, whereas lighter and/or hotter twin sectors result in smaller values. Note that the starting value of $x_{\hat e}$ (and $\hat Y_p$) is independent of the value of $\hat r$ (compare the orange and red lines in the figure).
	
	\item Twin recombination happens earlier on increasing $\hat v/v$ or lowering $\Delta \hat N$.  This is because a twin sector with a higher VEV has higher atomic binding energies causing twin recombination to take place at earlier times.  On the other hand, a smaller $\Delta\hat N$ leads to a colder twin sector.  The twin photons in such a sector do not have enough energy to reionize a twin electron that has attached to a nucleus. Thus decreasing $\Delta \hat N$ also leads to an earlier recombination.  Finally, note that the shape of $x_{\hat e}(z)$ is roughly independent of the value of $\hat r$ besides a small overall shift in $z$ (compare the orange and red lines in the figure).
	
	\item The visibility function in Fig.~\ref{fig.TwinRecVis} mostly behaves like a narrow Gaussian function and peaks towards the end of hydrogen recombination, as expected.  However, the $(\hat r, \Delta\hat N, \hat v/v) = (0.004, 0.3, 3)$ (orange-dotted) cases show some peculiar deviations from this expected behavior.  In this case, the visibility function shows a small peak around \twinHe-recombination.  It happens because the number density of twin particles in this case, due to small \rhat, is quite small.  Therefore, when there is a further drop in $ n_{\hat e}$ due to recombination, a fraction of the twin photons decouple.	This ultimately leads to a small peak in the visibility function during twin helium recombination.\footnote{Note that the opposite of this scenario happens for the $(\hat r, \Delta\hat N, \hat v/v) = (0.5, 0.3, 3)$ case (orange-dashed), where the visibility function peaks after those of $\hat r$ = 0.004 and $\hat r$ = 0.1 (dotted orange and red lines). This is surprising given that $ x_{\hat e}$ drops the earliest among the three (see upper panel). This happens because in the model with $\hat r = 0.5$, the number density of ionized twin electrons is so large that $ x_{\hat e}$ has to drop to a significantly smaller value before the twin photons can decouple.}	Physically, it means that a fraction of the twin photons actually last scatter the twin electrons that were going to form neutral \twinHe{} atoms.  Finally, though the evolution of $ x_{\hat e}$ in Fig.~\ref{fig.TwinRecXe} is roughly independent of $\hat r$, the visibility function has broken this degeneracy in $\hat r$.
	
\end{itemize}
Before concluding this subsection, we want to discuss the temperature of twin baryons, $T_{\hat b}$, which will be useful in computing the evolution of matter perturbations.  In \CLASS{}, we compute the evolution of $ T_{\hat b}$~\cite{Peebles:1968ja} using
\begin{equation}
	\frac{d T_{\hat b}}{d t}=	\frac{8}{3}
	\frac{ \hat \sigma_{\rm T} a \hat T^{4}}
	{  m_{\hat e} }  x_{\hat e}
	\left(\hat T- T_{\hat b}\right)
	-\frac{2  T_{\hat b}}
	{a} \frac{d a}{d t},
	\label{eq.recTb}
\end{equation}
where $t$ is the time, $a$ is the scale factor, and $\hat\sigma_{\rm T}$ is the twin Thomson scattering cross-section, related to its SM value by $\hat \sigma_{\rm T} = \sigma_{\rm T}(\hat v/v)^{-2}$.

The first term in Eq.~\eqref{eq.recTb} accounts for the heat transfer from the twin photons to twin baryons due to Thomson scattering, while the second term corresponds to the adiabatic cooling of matter due to the expansion of the universe.  At high temperatures, before twin recombination, the first term dominates and the temperature of twin baryons closely follows the temperature of twin photons: $ T_{\hat b} \simeq \hat T$.  As $ x_{\hat e}$ drops when twin electrons start to recombine, the first term becomes sub-dominant and the twin baryons decouple from the twin photons.  From here on, twin baryons cool adiabatically due to the expansion of the universe.

To compute $ T_{\hat b}$ in \CLASS{}, we start with $ T_{\hat b} = \hat T$ for $x_{\hat e} > 0.1$, and then evolve Eq.~\eqref{eq.recTb} for smaller $x_{\hat e}$ to find $T_{\hat b}$ as a function of redshift.

%%%%%%%%%%%%%%%%%%%%%%%%%%%%%%%%%%%%%%%%%%%%%%
\subsection{Perturbations in the MTH Universe}
\label{subsec.perturbations}

The twin sector of the MTH model leaves its imprints on cosmological observables like LSS and the CMB through its effect on metric (gravity) perturbations.  The energy density perturbations of the twin sector modify the metric perturbations, which in turn, affect the matter perturbations and subsequently, the CMB.  The energy density of the twin sector is carried by three species: twin baryons, twin photons, and twin neutrinos.  In this subsection, we will compute the evolution of perturbations for these species using the Boltzmann equations. 

We work in the conformal Newtonian gauge where $\psi$ and $\phi$ characterize the two scalar perturbations on the background metric\footnote{We follow the general formalism and notation of Ref.~\cite{Ma:1995ey}.}
\begin{equation}
	ds^2 = a^2(\tau) [ - ( 1 + 2 \psi ) d \tau^2 + ( 1 - 2 \phi ) \delta_{ij} dx^i dx^j ]  \; ,
\end{equation}
where $\tau$ denotes conformal time.

The Boltzmann equations describing the evolution of twin photon perturbations are:
\begin{align}
	\dot{\delta}_{\hat{\gamma}} + \frac{4}{3} \theta_{\hat{\gamma}} - 4 \dot{\phi} &= 0, \\
	\dot{\theta}_{\hat{\gamma}} + k^2 (\sigma_{\hat{\gamma}} - \frac{1}{4} \delta_{\hat{\gamma}} ) - k^2 \psi &= -a  n_{\hat e}\hat \sigma_T (\theta_{\hat{\gamma}} - \theta_{\hat{b}} ), \\
	\dot{F}_{\hat{\gamma},\ell} + \frac{k}{2 \ell + 1} ( ( \ell + 1 ) F_{\hat{\gamma},\ell+1} - \ell F_{\hat{\gamma},\ell-1} ) &= -a n_{\hat e}\hat \sigma_T \alpha_\ell F_{\hat{\gamma},\ell}, 
	\qquad (\text{for } \ell \geq 2)
\end{align}
where $\delta_{r} \equiv \delta \rho_r/\bar{\rho}_r\,$, $\theta_{r}\equiv  \partial_i v^i_r $, and $\sigma_{r}$ are the density perturbation, velocity divergence, and shear stress of species $r$. $F_{\hat{\gamma},\ell}$ is the $\ell^{\text{th}}$ moment of twin photon temperature perturbation.  $k$ is the comoving wave number of the perturbation, $ n_{\hat e} (\equiv x_{\hat e}  n_{\rm\hat H})$ is the number density of the ionized twin electrons, and $\hat \sigma_T$ is the Thomson scattering cross-section for the twin sector.  The overhead dot represents the derivative with respect to conformal time.  The coefficient $\alpha_{\ell}$ contains information about the angular dependence of the $\hat{\gamma}-\hat{b}$ scattering cross section. For the twin sector, we find $\alpha_{2} = 9/10$ and $\alpha_{\ell} = 1$ for all the higher $\ell$ modes. 

The Boltzmann equations governing the twin baryons perturbations are
\begin{align}
	\dot{\delta}_{\hat{b}} + \theta_{\hat{b}} - 3 \dot{\phi} &= 0, \\
	\dot{\theta}_{\hat{b}} + \frac{\dot{a}}{a}\theta_{\hat{b}} - c^2_{\hat{b}} k^2 \delta_{\hat{b}}  - k^2 \psi &= 
	- \frac{4 \rho_{\hat {\gamma}}}{3 \rho_{\hat b}} a  n_{\hat e}\hat \sigma_T (\theta_{\hat{b}} - \theta_{\hat{\gamma}} ),
\end{align}
where $c_{\hat{b}}$ is the adiabatic sound speed of $\hat{b}$.  From the above Boltzmann equations for the twin photons and twin baryons, it is clear that the twin sector communicates to the SM only through the metric perturbations, $\phi$ and $\psi$.

In this work, we compute the evolution of perturbations for the MTH universe in \CLASS{} by exploiting the \CLASS{} implementation of interacting dark matter-dark radiation (IDM-DR) model.  This implementation is based on an effective theory of structure formation, known as {\sc ETHOS}~\cite{Cyr-Racine:2015ihg}, which is a framework that encapsulates a broad category of dark matter models so that the cosmological signature of these models can be computed using only a handful of effective parameters.  Due to the inherent complexities in the twin sector related to twin BBN and twin recombination, we do not directly use the {\sc ETHOS} effective parameters to parametrize the twin sector.  Rather, we first compute twin BBN and twin recombination in \CLASS{} without invoking the IDM-DR model.  This is done using the procedure outlined earlier in this section.  Then, to compute the evolution of the perturbations in the twin sector, we map all perturbation-related aspects of the twin sector to the relevant IDM-DR parameters.

In Table~\ref{tab.ETHOS}, we provide the mapping used in order to utilize the ETHOS framework for studying the perturbations in the MTH model.  The twin photons of the MTH model behave like dark radiation in the {\sc ETHOS} framework and twin baryons behave like interacting dark matter.  The opacity parameters due to the interactions in the dark sector can be computed using the physics of twin BBN and twin recombination.  In principle one can also simultaneously study the self interactions of the DR using the {\sc ETHOS} framework.  Since twin photon self-interactions ($\hat{\gamma} \hat{\gamma} \leftrightarrow \hat{\gamma} \hat{\gamma}$) are negligible in such a model, we turn off the DR self interactions part of the code.

%%%%%%%%%%%%%%%%%%%%%
\begin{table*} [!htb]
	\renewcommand*{\arraystretch}{2}
	\centering
		\begin{tabular*}{\textwidth}{ @{\extracolsep{\fill}} |c||c|c|c|c|c|c|c|c|c|}
			\hline 
			ETHOS   & DR    & $\chi$
			& $\dot{\kappa}_{\mathrm{DR}-\mathrm{DM}}$
			& $\dot \kappa_\chi$
			& $c_\chi^2$
			& $\alpha_{\ell = 2}$
			& $\alpha_{\ell \geq 2}$
			& $\dot{\kappa}_{\rm{DR- DR}}$
			& $\beta_{\ell }$
			\\
			\hline
			MTH     & $\hat \gamma$ & $\hat b$
			& $-a \hat n_{e}\hat \sigma_T$
			& $- \displaystyle\frac{4 \rho_{\hat {\gamma}}}{3 \rho_{\hat b}} a \hat n_{e}\hat \sigma_T$
			& $c_{\hat b}^2$
			& 9/10
			& 1
			& 0
			& 0\\[7 pt]
			\hline
		\end{tabular*}
\caption{The mapping used between the {\sc ETHOS} framework~\cite{Cyr-Racine:2015ihg} and the MTH model to utilize the interacting dark matter-dark radiation (IDM-DR) Boltzmann equations for the twin sector. $\chi$ is the IDM in the {\sc ETHOS} framework.}
\label{tab.ETHOS}
\end{table*}
%%%%%%%%%%%%%
 
Finally, to compute the perturbations in twin neutrinos, we assume that the twin neutrinos are massless and behave like free-streaming radiation.  Therefore, they obey the same Boltzmann equations as the SM neutrinos~\cite{Ma:1995ey}.  To account for twin neutrinos in \CLASS{}, we incorporate their effect in the effective number of free-streaming ultra-relativistic degrees of freedom ($N_{\rm eff}$).

With this setup in place, the final $N_{\rm eff}$ in the MTH model is
\begin{equation}
	N_{\rm eff} = 3.046 + \Delta\hat N_{\nu},
\end{equation} 
where 3.046 is the contribution from the SM neutrinos\footnote{Recent studies~\cite{Gariazzo:2019gyi} have found the contribution of the SM neutrinos to the relativistic degrees of freedom to be $\Delta N_{\rm eff} = 3.044$. This will have no impact on our results.} and $\Delta\hat N_{\nu}$ is that from the twin neutrinos.  Note that $\Delta\hat N_{\nu}$ is not a free parameter is our setup, but is determined by \Ntwin, as shown in Eq.~\eqref{eq.Nnu}.

To summarize, we use the following mapping to compute the evolution of perturbations in the twin photons, twin baryons, and twin neutrinos:
\begin{equation}
	\hat\gamma \leftrightarrow \text{DR}, 
	\quad\quad\quad
	\hat b \leftrightarrow \chi,
	\quad\quad\quad
	N_{\rm eff} = 3.046 + \Delta \hat N_{\nu},
\end{equation}
where DR and $\chi$ are the dark radiation and interacting dark matter species of the {\sc ETHOS} framework.

%%%%%%%%%%%%%%%
\begin{figure*} [t]	
  \centering
  \includegraphics[width=0.7\textwidth]{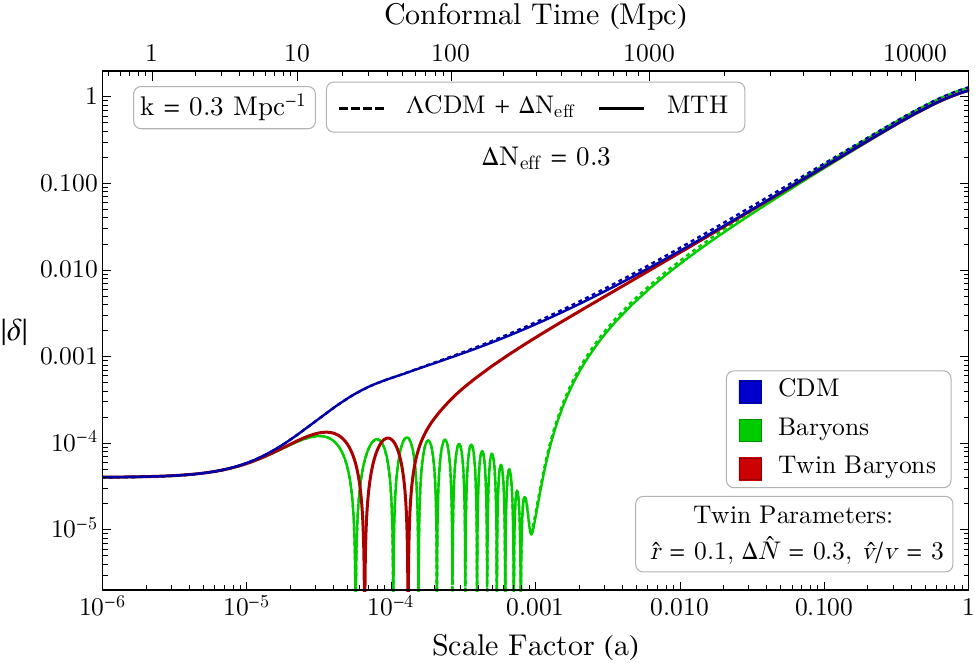}
  \caption{An example of the linear evolution of density perturbations of CDM, baryons, and twin baryons for \lnur{} (dashed) and the MTH model (solid).}
  \label{fig.evolutionofdelta}
\end{figure*}
%%%%%%%%%%%%%%%

An example of the evolution of matter density perturbations ($\delta$) at the linear order for a $k$-mode of 0.3 $\rm{Mpc}^{-1}$ obtained using this setup in \CLASS{} is shown in Fig.~\ref{fig.evolutionofdelta}.   The solid lines show the perturbations in each of the matter species present in the MTH model: CDM, baryons and twin baryons.  The dashed lines correspond to the two matter species of the $\Lambda$CDM + $\Delta N_{\rm eff}$ model: CDM and baryons.  Here $\Delta N_{\rm eff}~(= \Delta\hat N = 0.3$) is additional free-streaming ultra-relativistic radiation that we have added to the $\Lambda$CDM model so that it has the same background quantities as the MTH model.  To obtain this plot, the six $\Lambda$CDM input parameters $\{\Omega_{\rm dm}h^2,~\Omega_{\rm b}h^2,~\theta_s,~\ln(10^{10}A_s),~n_s,~\tau_{\rm reio}\}$, defined later in Sec.~\ref{sec.signals}, are the same for the two models, while the additional twin parameters for the MTH model are $(\hat r, \Delta\hat N,\hat v/v)$ = (0.1, 0.3, 3).  By using $\hat r = 0.1$ for the MTH model, we are assuming that 90\% of the dark sector is composed of CDM ($\Omega_{\rm cdm} = 0.9\,\Omega_{\rm dm}$), while the remaining 10\% is made up of twin baryons. Whereas for the $\Lambda$CDM+ $\Delta N_{\rm eff}$ model, 100\% of the dark sector is made up of CDM ($\Omega_{\rm cdm} = \Omega_{\rm dm}$).

Before the $k$-mode shown in Fig.~\ref{fig.evolutionofdelta} enters the horizon at conformal time $\eta$ $\sim3$ Mpc, all of the matter species in both the models have equal density perturbations.  However, once it enters the horizon, the different matter species evolve differently.  The density perturbations in the CDM of both models start to grow as the CDM starts to clump together due to gravitational self-attraction and thus starts to form over-dense regions.  Just like the SM baryons, density perturbations in the twin baryons start to undergo oscillations due to their interaction with the twin photons, a process we will call twin baryon acoustic oscillations (twin BAO).  While gravity tries to clump the twin baryons just as it does the CDM, the pressure from twin photons pushes them out.  This tug-of-war between gravity and photon pressure leads to twin BAO.  However, after twin recombination at $a\sim 10^{-4}$, the twin photon pressure drops due to the drop in the number density of ionized twin electrons and thus, twin BAO stops.  From this point on, twin baryons start to fall into the gravitational potential wells created by the CDM, while twin photons start to free stream.   In principle, these twin photons would lead to a twin CMB, sensitive to $z \sim 10^4$, though we lack the means to detect it directly.

Even without directly detecting the twin photons, we can study the twin sector indirectly by searching for its signatures in LSS and the CMB.  This will be the main topic of discussion of Sec.~\ref{subsec.CMB-LSS}.  However, we can already see at least two hints of these signatures in Fig.~\ref{fig.evolutionofdelta}.  First is an overall suppression in the matter density perturbations of the MTH model as compared to those of the $\Lambda$CDM+ $\Delta N_{\rm eff}$ model.  As can be seen in the figure, the $|\delta|$ for the MTH model (solid lines) is less than that for the $\Lambda$CDM + $\Delta N_{\rm eff}$ for the present day universe ($a=1$).  This is primarily because of the fact that not all the DM in the MTH models clumps, but a fraction of it oscillates.  Due to this, the potential wells created by the DM are not as deep in the MTH model and, thus, one finds that the density perturbations are suppressed.  Second is the phase at which TBAO stops due to recombination. Since the nearby $k$-modes would freeze out at a slightly different phases, we expect to see an oscillatory pattern on plotting density perturbations at a function of $k$ (such a plot will be discussed in Sec.~\ref{subsec.CMB-LSS}).  Note that these two features of the twin sector will only be observed for the $k$-modes that enter the horizon before the twin recombination.  For modes that enter the horizon after twin recombination, we do not expect them to behave any differently than the $\Lambda$CDM + $\Delta N_{\rm eff}$ model.

%%%%%%%%%%%%%%%%%%%%%%%%%%%%%%%
\section{Cosmological Signals}
\label{sec.signals}
%%%%%%%%%%%%%%%%%%%%%%%%%%%%%%%

In this section, we perform a Markov chain Monte Carlo (MCMC) scan of the MTH parameters and compare their predictions with observations.  We use the cosmological inference package \texttt{MontePy-} \texttt{thon}~(v3.4)~\cite{Brinckmann:2018cvx}, interfaced with our version of \CLASS{} that is capable of solving for the MTH universe. The MTH model cosmology is specified by nine parameters: the $\Lambda$CDM parameters $\{\Omega_{\rm cdm} h^2$, $\Omega_b h^2$, $\theta_s$, $\ln(10^{10}A_s)$, $n_s$, $\tau_{\rm reio}\}$~\cite{Aghanim:2018eyx} plus the twin sector parameters $(\hat r,~\Delta\hat N,~\hat v/v)$.  For the MCMC scans, we use flat priors on all the three MTH parameters with the following ranges unless mentioned otherwise: $\hat r \in [0.001, 1]$, $\hat N \in [0.001, 1]$, and $\hat v/v \in [2, 15]$.

%%%%%%%%%%%%%%%%%%%%%%%%%%%%
\subsection{Constraints from CMB Measurements}
\label{subsec.CMB-LSS}

We first discuss the features of the MTH model in the CMB and the matter power spectrum.  This will help us later in understanding the results of the MCMC scans.  As discussed in Sec.~\ref{subsec.perturbations}, the matter density perturbations ($\delta_m$) of the universe are influenced by the twin sector through gravity.  The basic idea is that twin BAO modifies the metric perturbations (gravity) which in turn affect the evolution of the perturbations in the CDM and SM baryons.  Furthermore, we predicted that these effects would show up in two forms: an overall suppression of the matter power spectrum, and an oscillatory pattern (in $k$-space) for the $k$-modes that enter the horizon before twin recombination.\footnote{For earlier work on LSS in the context of mirror models, see~\cite{Khlopov:1989fj,Ignatiev:2003js,Berezhiani:2003wj,Ciarcelluti:2004ik,Ciarcelluti:2004ip,Foot:2012ai,Foot:2014mia}. Detailed studies of LSS for the general case of atomic dark matter may be found in~\cite{Cyr-Racine:2012tfp,Cyr-Racine:2013fsa}.}

%%%%%%%%%%%%%%%%%%
\begin{figure*}[t]	
  \centering
  \includegraphics[width=0.7\textwidth]{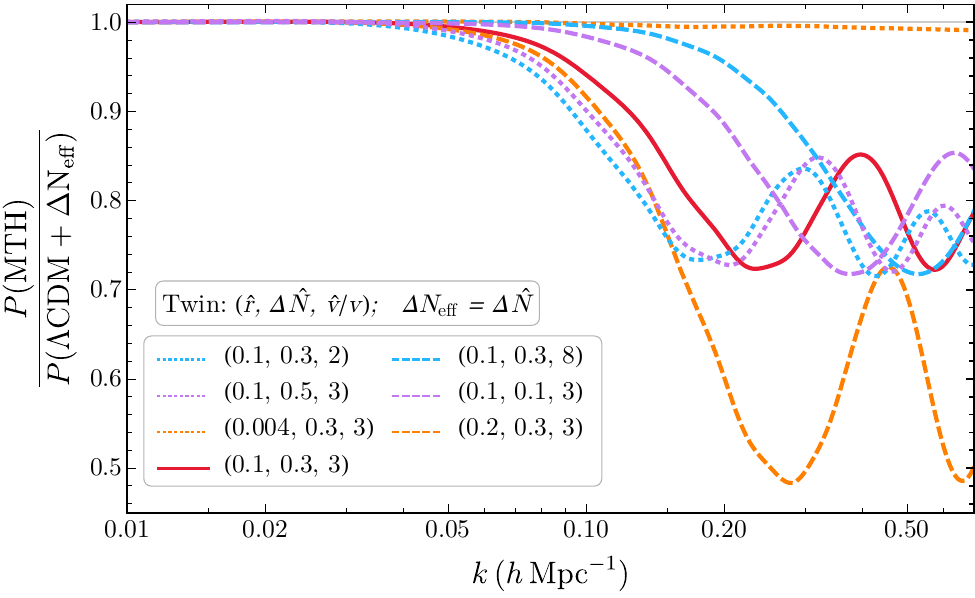}
  \caption{The linear matter power spectrum for the MTH model relative to that of the $\Lambda$CDM $+ \Delta N_{\rm eff}$ model for various twin parameters. }
  \label{fig.MPSratio}
\end{figure*}
%%%%%%%%%%%%%%%%%%

In Fig.~\ref{fig.MPSratio}, we show the linear matter power spectrum, $P(k) \equiv \langle|\delta_m(k)|^2\rangle$, in the MTH model relative to the $\Lambda$CDM + $\Delta N_{\rm eff}$ model for a variety of twin parameters, where $\Delta N_{\rm eff}$ is taken to be equal to the corresponding value of $\Delta\hat N$.  The values of the $\Lambda$CDM parameters for all the curves (including those for $\Lambda$CDM + $\Delta N_{\rm eff}$ model) are the same; these numbers are $\{\Omega_{\rm dm}h^2$, $\Omega_b h^2$, 100$\,\theta_s$, $\ln(10^{10}A_s)$, $n_s$, $\tau_{\rm reio}\}$ = \{0.119, 0.0224, 1.04, 3.05, 0.965, 0.0576\}.  Note that the first parameter used here is $\Omega_{\rm dm}$, instead of $\Omega_{\rm cdm}$. Though $\Omega_{\rm cdm} = \Omega_{\rm dm}$ for the \lcdm{} model, for the MTH model, $\Omega_{\rm cdm} = (1-\hat r) \Omega_{\rm dm}$ with the twin baryons making up the rest of the DM in the MTH model. Finally, the values of the three twin parameters used for each curve are shown in the figure itself.
There are a number of salient features of the matter power spectrum demonstrated in Fig.~\ref{fig.MPSratio}.
\begin{itemize}
	
\item The MTH models behave like the \lnur{} model at small $k$, but then start to deviate at larger $k$. Large $k$-modes that enter the horizon before twin recombination, undergo twin BAO and thus, lead to a suppressed matter power spectrum. The MTH models with smaller \vevratio{} and/or higher \Ntwin{} start to deviate from the \lnur{} model for smaller $k$, since decreasing \vevratio{} and/or increasing \Ntwin{} lead to a later twin recombination.
	
\item If we ignore the oscillations for a moment, the models with same \rhat{} have a similar overall suppression.  As shown analytically in Ref.~\cite{Chacko:2018vss}, the suppressed matter power spectrum is $\propto (1-\hat{r})^2$. For instance, the matter power spectrum for all models with $\hat r =0.1$ is suppressed to $ (1-0.1)^2 \approx 80\%$.  The $\hat r = 0.004 $ and $\hat r = 0.2$ models also show the expected behavior with the ratio suppressed to $\sim (1-0.004)^2\approx 99\%$ and $\sim (1-0.2)^2 \approx60\%$, respectively.

\item In addition to an overall suppression, we also see oscillations in the matter power spectrum. These oscillations are caused by the fact that different $k$-modes stop twin BAO (due to twin recombination) simultaneously but at different phases.  Like standard BAO, the oscillation period of twin BAO is $\Delta k\approx 2\pi/\hat{r}_s$, where $\hat{r}_s$ is the sound horizon of twin particles at the time of last scattering.
Therefore, the period of oscillations in Fig.~\ref{fig.MPSratio} should also be $\Delta k\approx 2\pi/\hat{r}_s$~\cite{Chacko:2018vss}.
We verify this prediction by comparing the period of oscillation of (\rhat, \Ntwin, \vevratio) = (0.1, 0.3, 3) case from the figure, shown in red curve, with the calculated value of the sound horizon. 
Using our modified version of \CLASS{}, we find $\hat r_s = 17.4\, h^{-1}\,$Mpc for this case. 
This implies, the period of oscillations, $\Delta k = 2\pi/\hat r_s = 0.36\, h {\rm\,Mpc}^{-1},$ which approximately agrees with the period of oscillation seen in the red curve.
Finally, note that there is a subdominant oscillations for each curve; this is caused by the interference of twin BAO with the standard BAO.

\end{itemize}
Though there are appreciable alterations in the matter power spectrum due to the dynamics of the twin sector, the current measurements of LSS still allow fractional suppression of the matter power spectrum at the $\mathcal{O}(1-10)\%$ level. Moreover, at $k \gtrsim 0.1 h {\,\rm Mpc}^{-1}$, non-linear corrections start to dominate, as described in Sec.~\ref{subsec.nonlinear}. A careful treatment of the non-linear correction to $P(k)$ in the MTH model is beyond the scope of this paper, but we will attempt to estimate the size of the non-linear corrections and examine the robustness of our results to those estimates. 

%%%%%%%%%%%%%%%%%
\begin{figure*}[t]	
  \centering
  \begin{subfigure}{0.45\textwidth}
      \includegraphics[width=\textwidth]{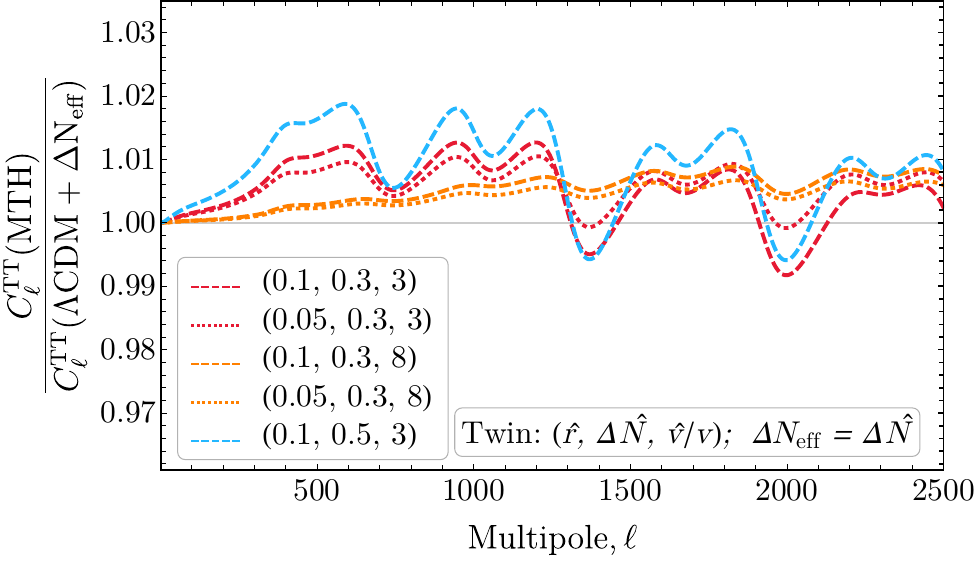}
      \caption{}
      \label{fig.CMBratioTT}
  \end{subfigure}
  \quad\quad\quad
  \begin{subfigure}{0.45\textwidth}
      \includegraphics[width=\textwidth]{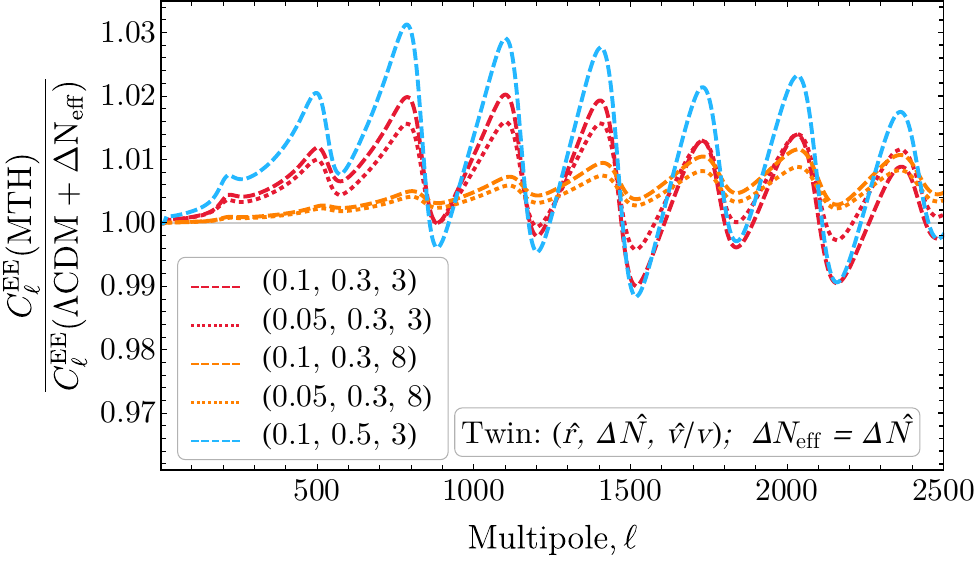} 
      \caption{}
      \label{fig.CMBratioEE}
  \end{subfigure}
  \caption{The ratio between the MTH model and the $\Lambda$CDM+$\Delta N_{\rm eff}$ model of the CMB temperature (Fig.~\ref{fig.CMBratioTT}) and polarization (Fig.~\ref{fig.CMBratioEE}) power spectra.}
  \label{fig.CMBratio}
\end{figure*}
%%%%%%%%%%%%%%%%%

On the other hand, the Planck collaboration~\cite{Aghanim:2018eyx} has measured the temperature and polarization maps of the CMB very precisely.  Using these maps, the collaboration has already placed stringent limits~\cite{Aghanim:2018eyx} on the parameter space of the $\Lambda$CDM model.  This has motivated us to study the signatures of the twin sector in the CMB maps. In Figs.~\ref{fig.CMBratioTT} and~\ref{fig.CMBratioEE}, we show the temperature and polarization anisotropies, respectively, in the MTH model relative to the $\Lambda$CDM + $\Delta N_{\rm eff}$ model. As before, for each curve $\Delta N_{\rm eff} $ $=$\,\Ntwin, and $\Lambda$CDM parameters are the same as in the above discussion.

%%%%%%%%%%%%%%%%%
\begin{figure*}[t]	
  \centering
  \begin{subfigure}{0.45\textwidth}
      \includegraphics[width=\textwidth]{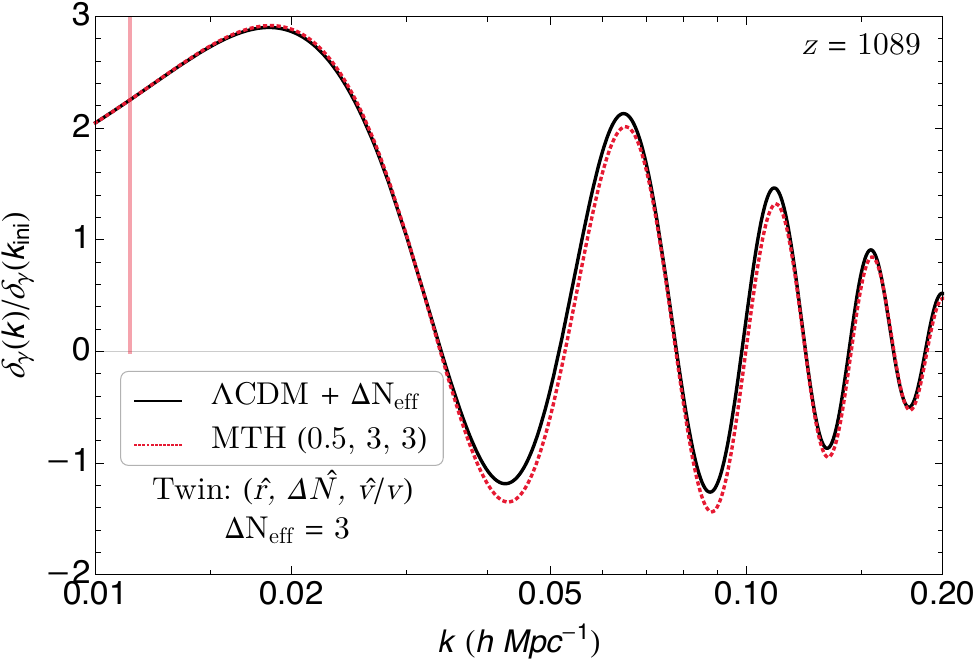}
      \caption{}
      \label{fig.photonPert}
  \end{subfigure}
  \quad\quad\quad
  \begin{subfigure}{0.47\textwidth}
      \includegraphics[width=\textwidth]{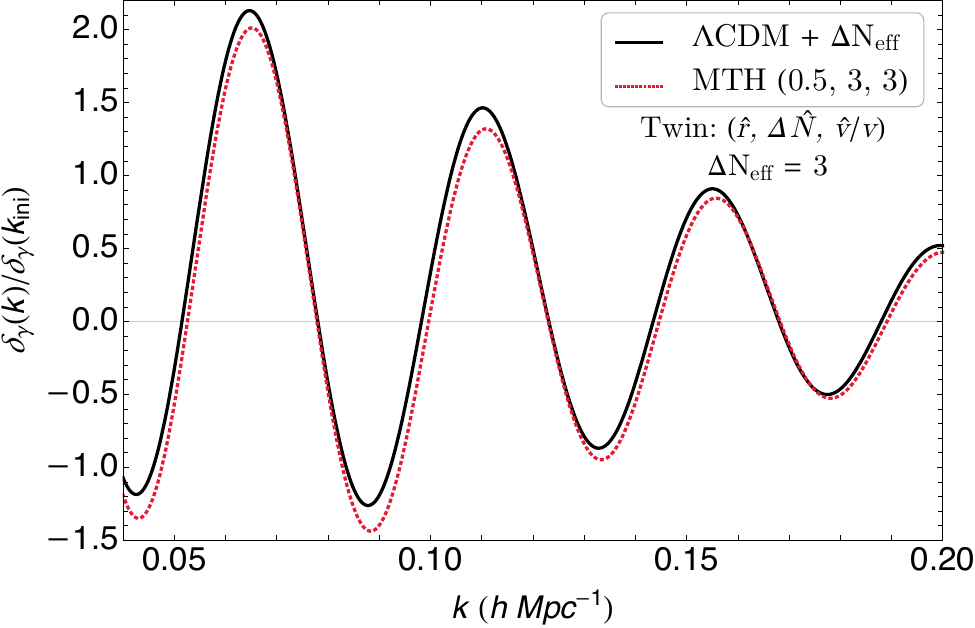} 
      \caption{}
      \label{fig.photonPertZoom}
  \end{subfigure}
  \caption{Photon perturbations right before SM recombination, calculated in the Newtonian gauge and normalized by the mode $k_{\rm ini} = 10^{-5}~h{\rm Mpc}^{-1}$ outside the horizon.  The curves represent the $\Lambda$CDM $+\Delta N_{\rm eff}$ model using $\Delta N_{\rm eff}=3$ (black solid) and the MTH model with $(\hat r,\Delta\hat{N},\hat{v}/v)=(0.5,3,3)$ (red dotted).  The $\Lambda$CDM parameters are set to the same values as in previous plots. 
  The left plot ranges logarithmically from $k=0.01~h {\rm Mpc}^{-1}$ to $k=0.20~h {\rm Mpc}^{-1}$ and the right plot ranges linearly from $k=0.04~h {\rm Mpc}^{-1}$ to $k=0.20~h {\rm Mpc}^{-1}$.  The vertical red line corresponds to the mode $k=0.011~h{\rm Mpc}^{-1}$% (corresponding to $\ell\approx k\chi_*=195$) 
  which enters the horizon when the twin visibility function reaches its peak value.  Compared to the free-streaming scenario, the MTH curve has a lower equilibrium point of oscillation due to the suppressed gravity perturbation caused by twin BAO. The MTH curve also has a relative phase shift to larger $k$-modes due to the smaller fraction of free-streaming radiation.}
  \label{fig.dTk}
\end{figure*}

\begin{figure*}[t]	
  \centering
  \begin{subfigure}{0.45\textwidth}\includegraphics[width=\textwidth]{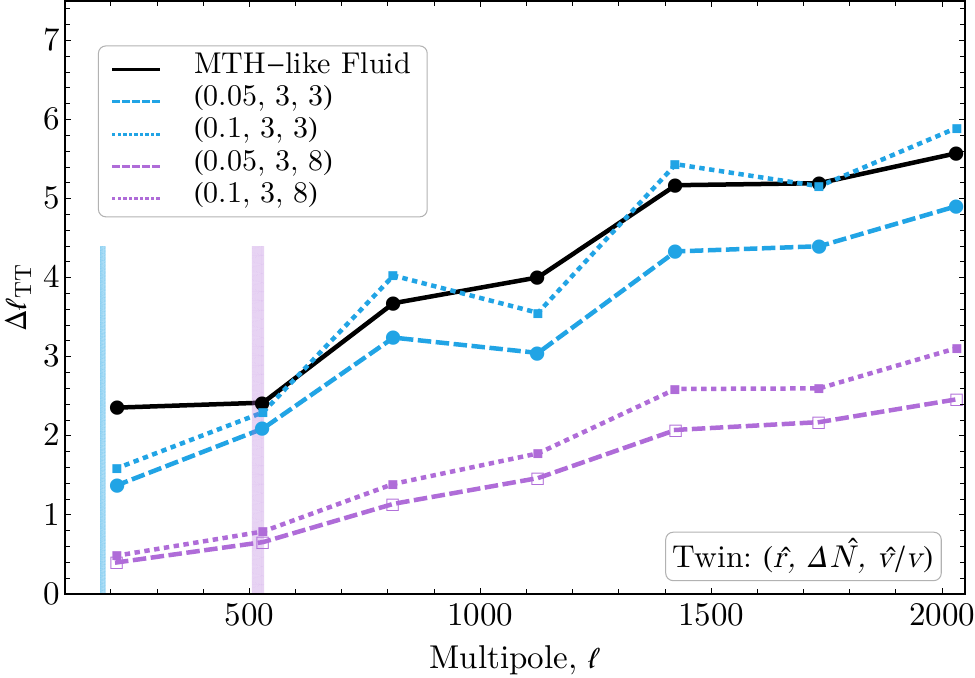}
      \caption{}
      \label{fig.phaseDiffTT}
  \end{subfigure}
  \quad\quad\quad
  \begin{subfigure}{0.47\textwidth}
      \includegraphics[width=\textwidth]{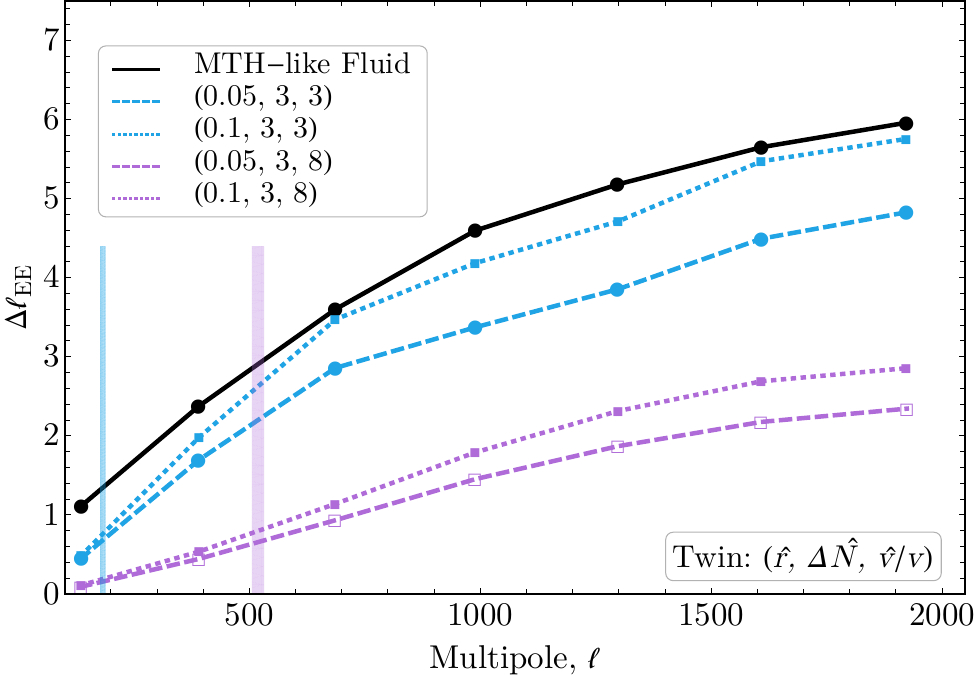} 
      \caption{}
      \label{fig.phaseDiffEE}
  \end{subfigure}
  \caption{Shift of the $\ell$-modes of the peaks relative to the free-streaming radiation scenario in the $C_\ell^{\rm TT}$ (left) and $C_\ell^{\rm EE}$ (right) spectra, assuming $\Delta N_{\rm eff}=3$. We connect the result at each peak by  straight lines for visualization purposes. The blue and purple curves are from the MTH models with $(\Delta\hat{N},\hat{v}/v)=(3,3)$ and $(3,8)$, respectively. Both the dotted blue and purple curves have $\hat r=0.1$ and the dashed curves have $\hat r=0.05$. The thick vertical lines indicate the $\ell$-modes that mainly receive contributions from perturbations of the mode $k=\tau_{\rm twin\,rec}^{-1}$. The left (right) boundary of the vertical lines indicates $\ell_{\rm rec}\equiv\tau_0/\tau_{\rm twin\,rec}$, which is defined to indicate the modes that came into horizon around the twin recombination time, for the dotted (dashed) MTH curves. The $\delta\ell$ values asymptote to zero when $\ell$ approaches $\ell_{\rm rec}$ and below. The black curve shows the phase shift of a dark-radiation model that contains the same proportion of free-streaming vs. fluid energy density as the $\hat\nu$ and $\hat\gamma$. The MTH curves get closer to the black curve as twin recombination is pushed to later times (smaller $\ell_{\rm rec}$).}
  \label{fig.dTll}
\end{figure*}

As we can see in Fig.~\ref{fig.CMBratio}, the MTH model leads to the deviations of $\mathcal{O}(1\%-3\%)$ from the $\Lambda$CDM + $\Delta N_{\rm eff}$ model, even though both models share the same \nur{}$=\Delta\hat N$. Thus these deviations are intrinsic to the MTH model and not simply an artifact of adding additional radiation. The deviations seen in Fig.~\ref{fig.CMBratio} mainly come from the combination of two effects that modify the phase and amplitude of the oscillating $C_{\ell}^{\rm TT,EE}$ spectra. First, unlike the case in which the extra radiation from $\Delta N_{\rm eff}$ is free streaming, the twin photons remain as scattering radiation until twin recombination. Compared to the free-streaming $\Delta N_{\rm eff}$ scenario, the presence of scattering radiation in $\Delta\hat N$ shifts the $\delta_{\gamma}$ oscillation to higher $k$-modes~\cite{Bashinsky:2003tk,2013PhRvD..87h3008H,Baumann:2015rya,Chacko:2015noa}. Secondly, the suppression of the gravity perturbation, $\psi$, due to twin BAO also shifts the equilibrium point of the $\delta_{\gamma}$ oscillation, as it is driven by radiation pressure and gravity. 

We can understand these differences in more detail.  In Fig.~\ref{fig.dTk}, we show the photon transfer function in the $\Lambda$CDM$+\Delta N_{\rm eff}$ model (black solid line), taking $\Delta N_{\rm eff}=3$, and in the MTH model, with $(\hat r,\Delta\hat{N},\hat{v}/v)=(0.5,3,3)$ (red dashed line), at $z=1089$ (right before SM recombination).  For large $k$-modes, where perturbations enter the horizon before twin recombination (denoted in the plot as a vertical red line), the equilibrium point of the $\delta_\gamma$ oscillations in the MTH model shifts to a lower $\delta_\gamma$ value compared to the \lnur{} model. The suppression of gravity perturbations from twin BAO increases (decreases) the amplitude of perturbations for the expansion (contraction) modes compared to the \lnur{} model. In Fig.~\ref{fig.CMBratioTT}, the dips at $\ell\approx800,\,1400,\,2000$ correspond to the location of compression peaks ({\it i.e.}, the odd peaks) of the $C_\ell^{\rm TT}$ spectrum.  Since the high $\ell$-modes mainly come from the $k$-modes that enter the horizon during radiation domination, their metric perturbations damp quickly and the temperature fluctuations mainly come from $\delta_\gamma$.  The decrease of the compression amplitudes of $\delta_\gamma$ in the MTH case, therefore, generate dips in Fig.~\ref{fig.CMBratioTT} near the odd $C_\ell^{\rm TT}$ peaks. There is also an enhancement near the even peaks of $C_\ell^{\rm TT}$ ($\ell\approx 500,\,1100,\,1700,\,2400$) in Fig.~\ref{fig.CMBratioTT} due to the enhanced expansion modes in $\delta_\gamma$, though the peak structure in Fig.~\ref{fig.CMBratioTT} is complicated by the phase shift effect.

Besides the change of the amplitude, the $\delta_\gamma$ oscillation in the MTH model also shifts to higher $k$-modes due to the slower propagation of the twin photon compared to free-streaming radiation. Since the phase shift comes from the change of metric perturbations due to radiation propagation, the effect is more prominent for large $k$-modes and becomes less significant for modes that enter the horizon close to the matter dominated era~\cite{Baumann:2015rya}. Moreover, since twin photons behave as free-streaming radiation after twin recombination, the phase shift between the MTH and free-streaming radiation models is further decreased for $k$-modes entering after the twin recombination.

The effect of the phase shift can also be observed in $\ell$-space as $\delta\ell$.  We show the $\ell$-dependence of the TT spectra in Fig.~\ref{fig.phaseDiffTT} and of the EE spectra and in Fig.~\ref{fig.phaseDiffEE}.  For the models we consider in Fig.~7, the size of the phase shift is only as large as $\delta\ell\approx 1$. To better explain the $\ell$-dependence of the effect, we consider models in Fig.~\ref{fig.phaseDiffTT} and~\ref{fig.phaseDiffEE} with a large  $\Delta N_{\rm eff}=3$. We also fix $\theta_s$ when making the plots\footnote{As we verified numerically, Fig.~\ref{fig.dTll} is almost identical even if we choose to fix $h$ instead of $\theta_s$. Since the sound horizon $r_s$ mainly grows near SM recombination~\cite{Knox:2019rjx},  which takes place in the matter-dominated era, $r_s$ is quite insensitive to the difference between free-streaming and scattering radiation propagation discussed here. Given that all the models considered in Fig.~\ref{fig.dTll} have the same $\Delta N_{\rm eff}$, both the sound horizon and the angular diameter distance to the last scattering surface are almost identical among these models even if we fix the Hubble expansion rate today ({\it i.e.} the value of $h$) instead of $\theta_s$.}.  Here, $\delta\ell$ is defined as the relative location of the acoustic peaks $\delta\ell=\ell_{\rm peak,MTH}-\ell_{\rm peak, FS}$ with respect to the free-streaming radiation model. Each $\ell$-mode of the TT/EE spectra receives contributions mainly from perturbations with $k\approx \ell/\chi_*$, where $\chi_*=\tau_0-\tau_{\rm rec}\approx14$~Gpc is the conformal distance to the last scattering surface in flat space.  Like the photon transfer function, the phase shift in the TT/EE spectra between free-streaming and scattering radiation also increases from the low to the high $\ell$-modes and asymptotes to a constant value, which is determined by the composition of the two types of radiation.

The $\delta\ell$ for free-streaming radiation also drops significantly when $\ell\lesssim\ell_{\rm rec}$, where $\ell_{\rm rec}$ corresponds to the twin recombination scale (vertical blue and purple lines). Since the phase shift at each $\ell$-mode depends on the time integral of the metric perturbations, the MTH models that have an earlier twin recombination time receive a smaller phase shift.

In Fig.~\ref{fig.dTll} we also show curves (black) from a dark radiation model that has the same fraction of scattering vs. free-streaming radiation as $\hat\nu$ and $\hat\gamma$ before the twin recombination. Without having a significant change of metric perturbations due to the twin acoustic oscillations, the MTH model with $(\hat r,\hat v/v)=(0.1,3)$ (dotted blue), that has a late twin recombination, generates a similar $\delta\ell$ as the black curves, as expected. The resulting phase shift in $C_{\ell}^{\rm TT,EE}$ within the MTH scenario generates the ratios in Fig.~\ref{fig.CMBratio} that oscillate in $\ell$. The effect of the phase shift can be seen more cleanly in the EE plot, Fig.~\ref{fig.CMBratioEE}, where the peaks at $\ell\approx 800,1100,1400,1700,2000$ all correspond to the ``troughs" in the $C_\ell^{\rm EE}$ spectrum. Shifting the $C_\ell^{\rm EE}$ spectrum to higher $\ell$-modes causes a relatively large increase of the spectrum value at the troughs, therefore generates the peaks in the ratio plot. %\textcolor{violet}{I only have one question. Shouldn't oscillations in the polarization spectrum be out of phase (but having the same period) with those in the temperature spectrum? ($cf.$ Dodelson Figure 10. 18). It's just that I can't see this easily in Figure 7. Thanks!}

These deviations in the CMB spectra are much smaller than those in the matter power spectrum of Fig.~\ref{fig.MPSratio}.
Despite that, the CMB maps measured by the Planck collaboration are precise enough to detect these percent-level deviations.

To compare the MTH model with the CMB and BAO data, we use the following datasets:
\begin{itemize}

	\item Planck: high-$\ell$ TTTEEE, low-$\ell$ EE, low-$\ell$ TT, and lensing datasets of the Planck 2018 Legacy release~\cite{Aghanim:2019ame}.	
	
	\item BAO: Measurements of $D_V/r_s$ by 6dFGS at $z = 0.106$~\cite{Beutler:2011hx}, by SDSS at $z = 0.15$~\cite{Ross:2014qpa}, and by BOSS at $z = 0.2-0.75$~\cite{Alam:2016hwk}.
	
\end{itemize}
On comparing the predictions of the MTH model with these datasets, we find that the data does not prefer the MTH model relative to the $\Lambda$CDM model. The minimum $\chi^2$ value for the MTH model is similar to that of the $\Lambda$CDM model even though the MTH model has three additional free parameters.  This indicates that the MTH model is a slightly worse fit to these datasets.\footnote{Note that in the limit $\Delta \hat N$ goes to zero, the MTH model behaves like the $\Lambda$CDM model. This is because without twin photons, twin baryons do not thermalize and thus, behave like CDM.  Therefore, the minimum $\chi^2$ for the MTH model must be smaller or equal to that of the $\Lambda$CDM model.  An additional point to note here is that, due to computational constraints, we use a lower limits of $\Delta \hat N \geq 0.001$ and $\hat r \geq 0.001$ for all our MCMC scans. For the \lnur{} model, we take $\Delta N_{\rm eff} > 0$.}  Therefore, we use the Planck data to constrain the MTH parameter space.

%%%%%%%%%%%%%%%%%%%
\begin{figure*} [t]	
  \centering
   \includegraphics[width=0.5\textwidth]{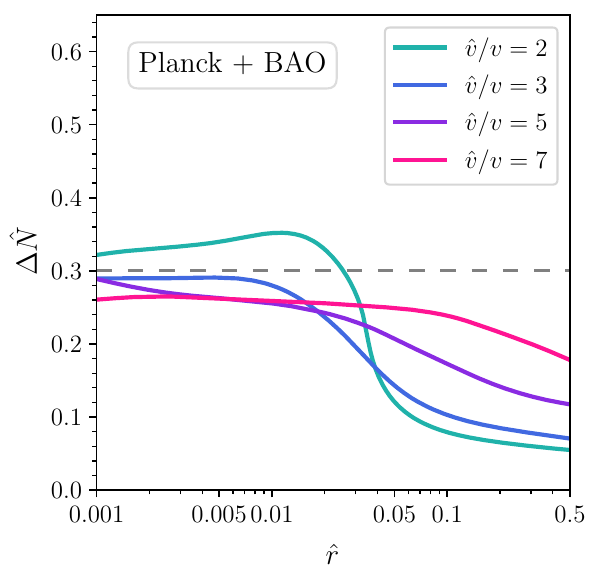}
	\caption{\label{fig.PlanckBAOConstraints}The constraints on the twin parameters obtained using the Planck and BAO datasets, for fixed values of $\hat v /v$ and using a log prior on $\hat{r}$. All of the parameter space above the lines is excluded at $95\%$~C.L. The dashed gray line indicates the $95\%$~C.L. bound on $\Delta N_{\rm eff} \,(\leq 0.30)$ using the same datasets, \ie{} Planck~2018 TTTEEE+lowE+lensing+BAO~\cite{Aghanim:2018eyx}. The simultaneous constraints on all three of the MTH parameters for this choice of datasets are shown in the Appendix.} 
\end{figure*}
%%%%%%%%%%%%%

In Fig.~\ref{fig.PlanckBAOConstraints}, we show the constraints on the twin parameters obtained using the Planck and BAO datasets.  The parameter space above the curves are excluded at 95\%~confidence level (C.L.).  To get these bounds we use $\hat v/v$ as a fixed parameter, and $\Delta \hat N$ (flat prior) and $\hat r$ (log prior, range: [0.001, 0.5]) as free parameters along with the six $\Lambda$CDM parameters.  Completing the MCMC scans with this setup, we plot our results as $95\%$~C.L. contours in the \Ntwin{} versus \rhat{} plane for various values of \vevratio{} to obtain Fig.~\ref{fig.PlanckBAOConstraints}.  In the figure, we also show with a dashed gray horizontal line the $95\%$~C.L. bound on additional free-streaming ultra-relativistic radiation from the Planck + BAO data of $\Delta N_{\rm eff} \leq 0.30$~\cite{Aghanim:2018eyx}.

As the figure shows, the Planck and BAO data already impose strong constraints on the twin parameter space, especially for $\hat r \gtrsim 0.1$.  For instance, in the MTH model with \vevratio{} = 3 and more than 10\% of the dark sector made up of twin baryons ($\hat r \gtrsim 0.1$), the current data forces $\Delta \hat N \leq 0.1$. The constraints on \Ntwin{} are, for the most part, stronger that those \nur{} (shown with a dashed gray line).  The only difference between the MTH and $\Lambda$CDM + $\Delta N_{\rm eff}$ model originates from the fact that the twin photons couple with twin baryons, while the radiation in the $\Lambda$CDM + $\Delta N_{\rm eff}$ model is free-streaming.  Therefore, the stronger constraints on \Ntwin{} tells us that the CMB data as measured by Planck is already sensitive to twin BAO.  This also explains the behavior of the bounds of \Ntwin{} at very small $\hat r$ and large \vevratio.  As \rhat{} approaches zero, there is not enough matter for the twin photons to collide with, and thus around $\hat r \sim 0.001$, the constraints on \Ntwin{} are similar to those on $\Delta N_{\rm eff}$.  Likewise, even when having a sizable $\hat{r}$, twin BAO stops earlier as \vevratio{} increase, and the bound on \Ntwin{} starts to approach those on the $\Delta N_{\rm eff}$.

In the figure, we also see a peculiar peak-like shape near $\hat r \sim 0.01$ where the constraints on \Ntwin{} are weaker than than those on \nur.  This happens because for this range of \rhat, twin photons behave like coupled-radiation through scattering with $\hat e$.  That is, for these \rhat{}, there is not enough energy density stored in the twin baryons to have a substantial impact on the metric perturbations.  However, the number density of these twin baryons is sufficiently high to keep the twin photons interacting and behaving like a fluid.  As shown in Ref.~\cite{Baumann:2015rya,Brust:2017nmv, Blinov:2020hmc,Ghosh:2021axu}, the constraints on the fluid-radiation are slightly weaker than those on free-streaming radiation, with the Planck and BAO data allowing $\Delta N_{\rm fluid} \lesssim 0.5$ within $95\%$~C.L.~\cite{Blinov:2020hmc}.  This is why the constraints on \Ntwin{} are weaker than the $\Delta N_{\rm eff}$ bound in parts of our parameter space.\footnote{The reason \Ntwin{} does not get as large as $\Delta N_{\rm fluid}$ is because \Ntwin{} gets contribution from the twin neutrinos as well, which are always free streaming. In addition, twin photons behave fluid-like only until recombination, after which they start to free-stream. The analysis of Ref.~\cite{Blinov:2020hmc}, in contrast, assumes that the radiation is always fluid-like.}

%%%%%%%%%%%%%%%%%%%%%%%%%%%%%%%%%%%%%%%%%%%%%%%%%%%%%%%%%%%%%%%%
\subsection{LSS Constraints and Application to the \texorpdfstring{${H}_{0}$}{H0} and \texorpdfstring{$S_8$}{S8} Tensions}
\label{subsec.tensions}

We have seen in Sec.~\ref{subsec.CMB-LSS} that the MTH model leads to a suppression in the matter power spectrum relative to both the \lcdm{} and the \lnur{} models.  It is important to check whether current LSS data further strengthens the bounds in Fig.~\ref{fig.PlanckBAOConstraints}. To that end, we will consider two very different sets of LSS data in order to observe their effect on our fits.  We begin by including:
\begin{itemize}

\item LSS: KiDS + Viking 450 (KV450)~\cite{Hildebrandt:2018yau} matter power spectrum shape data up to\\ $k_{\rm max}=0.3h\,\rm{Mpc}^{-1}$.

\end{itemize}
This combined analysis of data from the KiloDegree Survey (KiDS) and the VISTA Kilo-Degree Infrared Galaxy Survey (VIKING) incorporates detailed photometric redshift measurements with cosmic shear/weak-lensing observations to measure the matter power spectrum over a wide range of scales.  The 12 million galaxies in the survey are found at redshifts between $0.1$ and $1.2$, and probe a range of $k$ values including but not limited to those regions that contribute to $S_8$.  We use the full spectral information in our fits, up to a maximum value of $k=0.3~h\rm{~Mpc}^{-1}$, for reasons described below.  However, due to its utility in parametrizing the matter power spectrum, we will also show in our plots the value of $S_8$ obtained by the KV450 collaboration, namely $S_8 = 0.737^{+0.040}_{-0.036}$~\cite{Hildebrandt:2018yau}.  The other LSS measurement that we will consider is the Planck SZ data, which we will return to later in the section.

One important complication of including the LSS data has already been mentioned briefly: the expected presence of  non-linear corrections to the matter power spectrum that begin to become significant for $k \gtrsim 0.1~h\,\rm{Mpc}^{-1}$. The canonical \lcdm{} and \lnur{} models have been studied extensively over the previous decades and a number of estimates and parametrizations of the non-linear effects within these paradigms have been made. No such work has been done within the MTH model, and so we will handle this issue in two ways. First, we will only use data from KV450 probing scales $k<k_{\rm max} = 0.3~h\,\rm{Mpc}^{-1}$, limiting our exposure to that regime in which non-linear effects are most important. Second, we will complete all of our fits twice, once using only the linear evolution equations for the density perturbations, and once using HMcode~\cite{Mead:2016zqy}, which is a parametrization that provides a good fit to the non-linear effects in the canonical models.  This is also the code used by the KV450 collaboration in their analysis. 

While HMcode is not designed to provide the correct non-linear effects within the MTH model, we will use it as an estimate of the size of those corrections.  The Planck+BAO data will continue to play the dominant role in constraining the MTH model, so our main results will be nearly identical with or without the inclusion of the estimated non-linear corrections provided by HMcode. However, for the purposes of our discussion in this section, we will only use the linear evolution of the matter perturbations, for $k<k_{\rm max}$, returning to a more complete discussion of the non-linear corrections in the next section.

\begin{figure*}[t]	
  \centering
  \includegraphics[width=0.5 \textwidth]{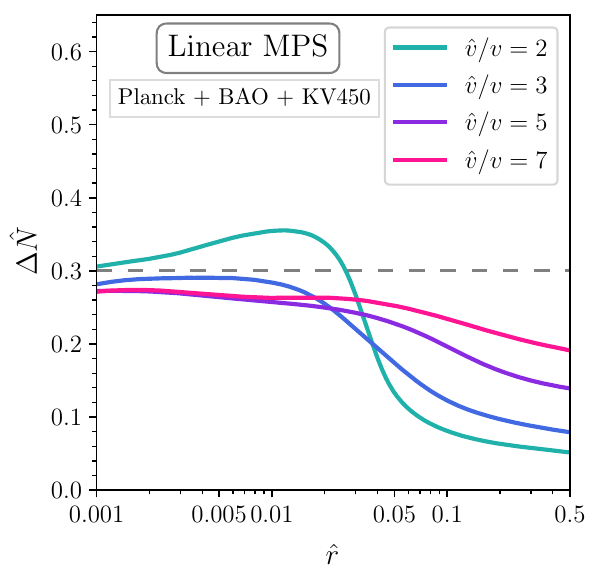} 
  \caption{Constraints on the twin parameters obtained using the Planck, BAO, and KV450 datasets, for fixed values of $\hat v /v$ and using a log prior on $\hat{r}$.  The linear matter power spectrum has been used in the fit.  All of the parameter space above the lines is excluded at $95\%$~C.L. The simultaneous constraints on all three of the MTH parameters for this choice of datasets are shown in the Appendix.}
  \label{fig.KV450linearNhat}
\end{figure*}

In Fig.~\ref{fig.KV450linearNhat}, we show bounds on the twin parameters after including the KV450 data with a linear matter power spectrum. The bounds on $\Delta\hat{N}$ as a function of $\hat{r}$ are nearly identical to the Planck+BAO constraints shown in Fig.~\ref{fig.PlanckBAOConstraints}.  Thus, the KV450 data does little to improve the bounds in this calculation. 

%%%%%%%%%%%%%%%%%
%%%%%%%%%%%%%%%%%

\begin{figure*} [t]	
  \centering
  \includegraphics[width=0.6\textwidth]{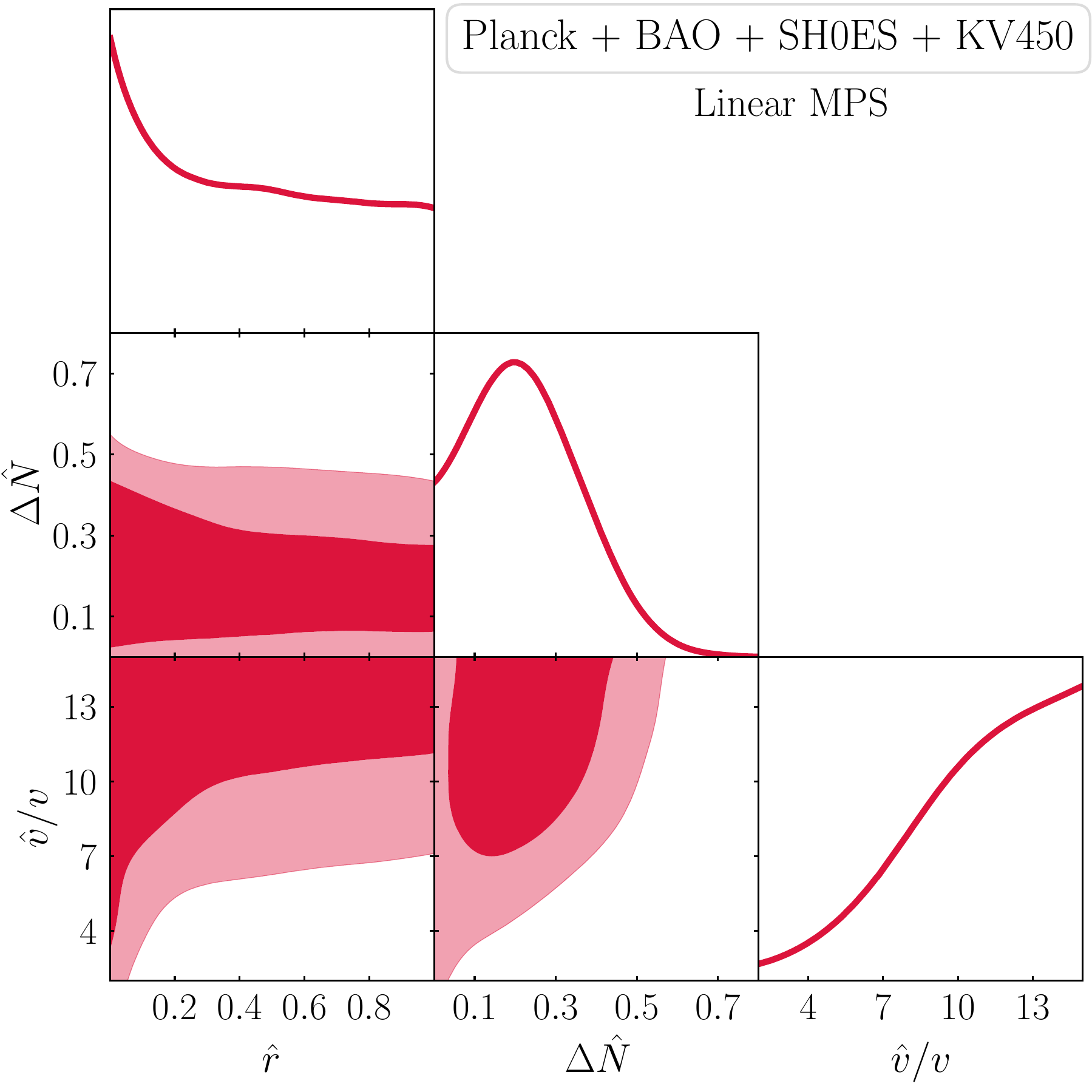}    
  \caption{Constraints on all three MTH parameters using the Planck, BAO, KV450, and SH0ES datasets.  The linear matter power spectrum has been used in the fit. The corresponding $H_0$ versus $S_8$ plot is shown in Fig.~\ref{fig.S8H0withKV}.}
  \label{fig.KV450linear}
\end{figure*}

Now we turn to the fits of the canonical cosmological parameters.  At SM recombination, the presence of $\Delta N_{\rm eff} > 0$ reduces the sound horizon compared to the \lcdm{} case, which results in a larger $H_0$ from the fit of Planck data, thus agreeing better with the SH0ES result. However, increasing $\Delta N_{\rm eff}$ also enhances the matter perturbations when fitting the Planck data and worsens the $S_8$ tension.  This simple observation has led to an enormous effort in the field to find cosmological models that can reduce $S_8$ while simultaneously increasing $H_0$ (see {\it e.g.},~\cite{Chacko:2016kgg,Raveri:2017jto,Lesgourgues:2015wza,Dessert:2018khu,Pandey:2019plg,Allali:2021azp,Clark:2021hlo}). One novel feature of the MTH model is that the presence of twin radiation leads to twin BAO, suppressing the matter power spectrum. We therefore want to check whether such behavior can help to relax both the $H_0$ and $S_8$ tensions.  For the purposes of these fits, we will include the SH0ES collaboration data as an additional constraint on the MCMC fit to the MTH model parameters:
\begin{itemize}

\item $H_0$ : $H_0$ by the SH0ES collaboration ($H_0 = 74.03 \pm 1.42$ $\mathrm{~km} \mathrm{~s}^{-1} \mathrm{Mpc}^{-1}$)~\cite{Riess:2019cxk}.

\end{itemize}
In Fig.~\ref{fig.KV450linear}, we show the relations among the bounds on the three MTH parameters once one includes both KV450 and SH0ES.  We see that the twin matter can make up a larger fraction of the dark matter (large $\hat r$) when either the twin electrons are heavier (larger $\hat{v}$) or the twin sector is colder (smaller $\Delta\hat{N}$); both of these lead to an earlier twin recombination. Although not statistically significant, we also see a small preference for $\Delta \hat N > 0$ peaking around $\Delta \hat N = 0.2$.
This is not surprising since extra radiation leads to a higher $H_0$, as required by the SH0ES dataset. 
In addition, we see that these datasets prefer higher values of $\hat{v}/v$, and therefore an earlier twin recombination. 
This indicates that, except for the twin radiation, there is not a strong preference for additional twin sector features. 
This observation is also reflected in the best-fit $\chi^2$ values, we find that the minimum $\chi^2$ of the MTH model is equal to that of the \lnur~model. Both the MTH and \lnur~model offer an improvement of $\sim 4$ units of $\chi^2$ over the \lcdm{} model.
We also see evidence (which will become clearer later) of a correlation between $\Delta\hat N$ and $\hat v/v$ in which both can grow larger together without worsening the fit.

%%%%%%%%%%%%%%%
%%%%%%%%%%%%%

\begin{figure*} [tb]	
    \centering
	\begin{subfigure}{0.48\textwidth}
		\includegraphics[width=\textwidth]{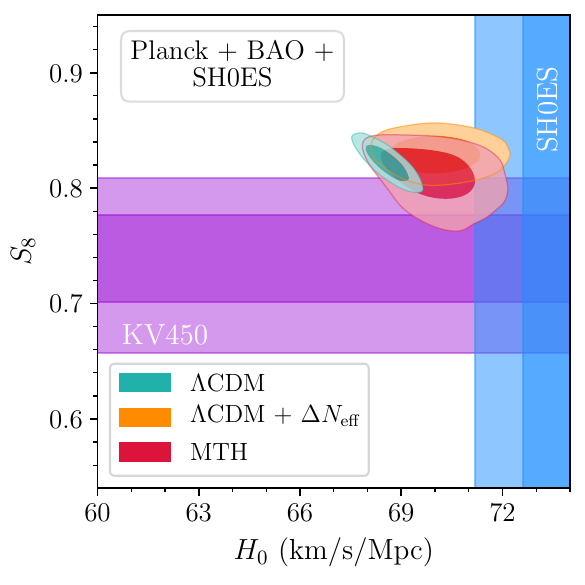}
		\caption{}\label{fig.S8H0noSZ}
	\end{subfigure}
	\begin{subfigure}{0.48\textwidth}
		\includegraphics[width=\textwidth]{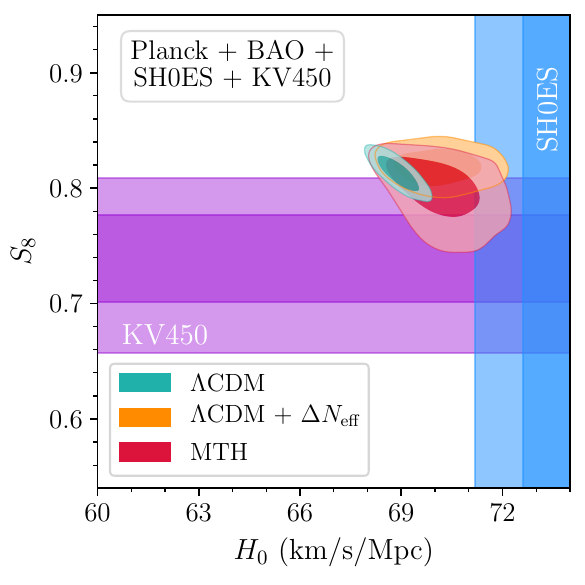}
		\caption{}\label{fig.S8H0withKV}
	\end{subfigure}
	\caption{The preferred ranges of the parameters $S_8\equiv\sigma_8(\Omega_m/0.3)^{0.5}$~and~$H_0$ for the three models: $\Lambda$CDM, \lnur$(>0)$, and the MTH model using the Planck, BAO, and SH0ES datasets (Fig.~\ref{fig.S8H0noSZ}), and with the addition of the KV450 data (Fig.~\ref{fig.S8H0withKV}). For reference, the measurements by $H_0$ by SH0ES and $S_8$ by KV450 are also shown in blue and violet, respectively. The MTH fit with KV450 data is done using the linear matter power spectrum, while the $\Lambda$CDM and \lnur$(>0)$ analyses include the non-linear corrections computed by Halofit.  We include non-linear corrections to the MTH fit from HMcode and Halofit in Fig.~\ref{fig.PlanckBAOH0_KV_SZ}.}
	\label{fig.PlanckBAOH0_KV_SZ_Linear}
\end{figure*}

Of particular interest are the fits to the cosmological parameters. In Fig.~\ref{fig.PlanckBAOH0_KV_SZ_Linear}, we show the $(H_0, S_8)$ results, where we follow the definition $S_8\equiv\sigma_8(\Omega_m/0.3)^{0.5}$ used in the KV450 literature~\cite{Hildebrandt:2018yau}. For these analyses, all three twin parameters $(\hat r, \Delta\hat N, \hat v/v)$ are treated as free parameters with flat priors.  In the figure, we also show the fits to the \lcdm{} and \lcdm{} + \nur{} models, and the measurements of $H_0$ (SH0ES, blue) and $S_8$ (KV450, purple). In the left figure, we include no information on the matter power spectrum in the fits.  It is clear from Fig.~\ref{fig.S8H0noSZ} that the \lcdm{} model is in sharp tension with the SH0ES result and in mild tension with the KV450 measurement of $S_8$. For the \lcdm{} + \nur{} model, $H_0$ can indeed be increased, as expected, but it comes at the price of raising $S_8$ from the \lcdm{} result, worsening that problem. The MTH contour, on the other hand, allows a smaller $S_8$ for a larger $H_0$, even without including $S_8$ (or any equivalent LSS information) in the fit. It is important to note here that, with the Planck+BAO+SH0ES datasets, the minimum $\chi^2$ for the MTH model is equal to that of the \lnur{} model. In addition, these data prefer the higher $\hat{v}/v$ or the lower \rhat{} parts of the twin parameter space. This indicates that the only feature of the twin sector preferred by these data is the extra radiation in the form of twin radiation.

In the right figure, we now add the KV450 data to our fits. Unsurprisingly, all three best-fit regions are now pulled to smaller $S_8$, but only the MTH contour obtains good agreement with the KV450 measurement, while it does no worse than the \lnur{} contour in obtaining agreement with the SH0ES data. Thus we see the behavior we hoped for: the MTH model is able to increase the prediction of $H_0$ from the CMB data while simultaneously decreasing the prediction of $S_8$ or $\sigma_8$. As noted above, the best-fit $\chi^2$ of the MTH model is equal to that of the \lnur{} model in this case as well.

%%%%%%%%%%%%%%%
%%%%%%%%%%%%%

As interesting and hopeful as this result seems, it is important to note that, while the KV450 matter power spectrum does show a mild disagreement with the \lcdm{} model, it is only at $2.3\sigma$~\cite{Hildebrandt:2018yau}.  There exists a different LSS dataset, known as the Planck SZ survey, which paints a similar picture but potentially exhibits stronger tension with the \lcdm{} value. The Sunyaev-Zeldovich (SZ) effect~\cite{1969Ap&SS...4..301Z,1970Ap&SS...7....3S} is the inverse Compton scattering of CMB photons by hot gas and is most significant when the photons pass through a galaxy cluster.  The measurements of the distribution of galaxies from the SZ effect relies on linking the observed SZ flux to the mass of a galaxy cluster, which introduces a mass bias factor $(1-b)$ that relates the observed signal to the true cluster mass.  The Planck~2013 report gives a measurement of $S_8$:
\begin{itemize}
\item Planck~2013 SZ : $S^{\rm SZ}_8 \equiv \sigma_{8}\left(\Omega_{\mathrm{m}} / 0.27\right)^{0.3}=0.782 \pm 0.010$~\cite{Ade:2013lmv},
\end{itemize}
by fixing the mass bias to its central value from a numerical simulation, $(1-b)=0.8$.  When presenting results in the Planck~2015 report~\cite{Planck:2015lwi}, the collaboration allowed $(1-b)$ to vary with a Gaussian prior centered at $0.78$.  The central value of the resulting $S^{\rm SZ}_8$ is actually somewhat smaller, but has a much larger uncertainty:
\begin{itemize}
\item Planck~2015 SZ : $S^{\rm SZ}_8 \equiv \sigma_{8}\left(\Omega_{\mathrm{m}} / 0.31\right)^{0.3}=0.774 \pm 0.034$~\cite{Planck:2015lwi}.
\end{itemize}
In order to demonstrate the MTH model's ability to fit a small $S_8$ with a large $H_0$, we use the Planck~2013 $S^{\rm SZ}_8$ value and errors in the following discussion in order to maximize the $S_8$ tension.  Furthermore, if uncertainty on the mass bias shrinks in the future while retaining the same central value, the value of $S^{\rm SZ}_8$ would be closer to the 2013 result.  As a comparison, we also conducted an MCMC study using the Planck~2015 $S^{\rm SZ}_8$ value, which gives a very similar result to Fig.~\ref{fig.KV450linear} since the 2015 value of $S^{\rm SZ}_8$ does not show a strong tension with the Planck+BAO results.

The Planck SZ data set differs from that of KV450 in several important ways. First, and most trivially, Planck SZ uses a somewhat different definition of $S_8$, as given above. One can easily relate this to the previously defined $S_8$, finding the Planck~2013 SZ measurement of:
\begin{equation}
S_8 = (0.782\pm 0.010) \left(\frac{\Omega_m}{0.35}\right)^{0.2}.
\end{equation}
If we use a typical value of $\Omega_m=0.3$, this corresponds to $S_8=0.758$, which is perfectly consistent with the KV450 value of $0.737$ given the errors on the latter value. However, the quoted errors on the Planck SZ result are roughly four times smaller than those from KV450, which will have a profound effect on our analysis. 

The second difference is that the Planck SZ dataset, as encoded in our fits using \texttt{MontePython}, is only a single number: $S^{\rm SZ}_8$ (or $\sigma_8$). This is in contrast to the KV450 data, which contains spectral information on the matter power spectrum. The single value $\sigma_8$ is mostly sensitive to the linear matter power spectrum over just a narrow band of $k$: $0.05\lesssim{}$  $k\, ({\rm in~} h \rm{~Mpc}^{-1})$ $\lesssim 0.3$.  

\begin{figure*}[t]	
  \centering
		\includegraphics[width=0.48\textwidth]{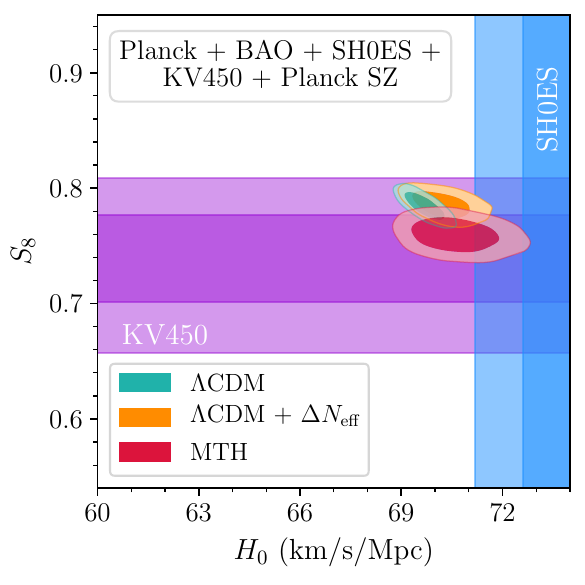}
  \caption{
  Same as Fig.~\ref{fig.S8H0withKV} but now including the Planck~SZ data.  We include non-linear corrections to the MTH fit from HMcode and Halofit in Fig.~\ref{fig.PlanckBAOH0_KV_SZ}.  }
  \label{fig.S8H0withSZ}
\end{figure*}

\begin{figure*}[bth]	
  \centering
  \begin{subfigure}{\textwidth}
		\includegraphics[width=0.99\textwidth]{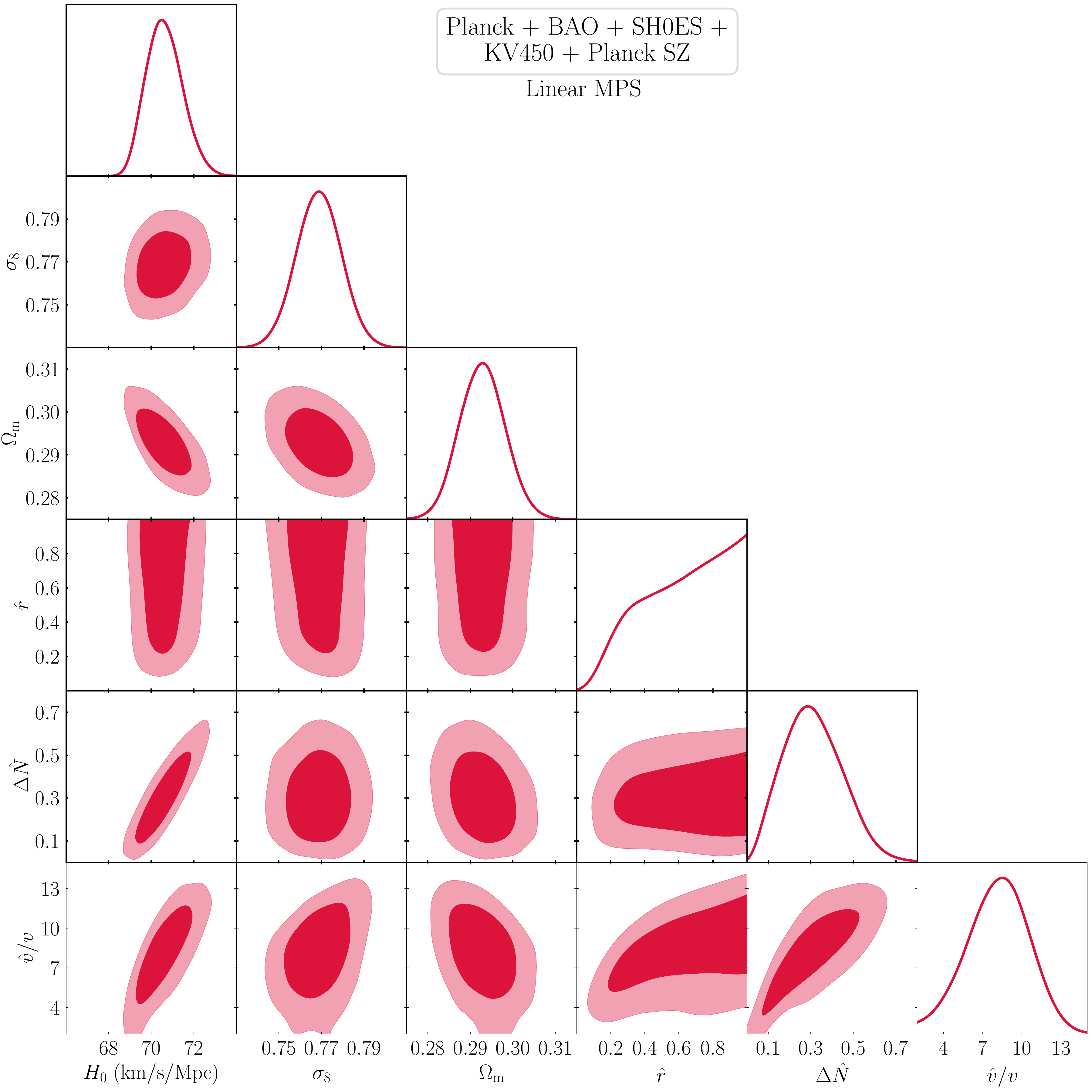}
  \end{subfigure}
  \caption{The preferred ranges of the twin parameters $H_0$, $\sigma_8$, $\hat r$, $\hat v/v$, and $\Delta \hat N$ obtained using the Planck, BAO, SH0ES, KV450, and Planck SZ datasets and the linear matter power spectrum.}
  \label{fig.AllparsPlanckBAOH0_KV_SZ_Linear}
\end{figure*}

Third, given the much smaller error bars in the Planck SZ measurement of $S_8$, the tension between the value inferred from the CMB measurements and that obtained from the low redshift measurements has increased substantially, to roughly $2-4\sigma$~\cite{Douspis:2018xlj, Blanchard:2021dwr}.\footnote{It has been argued \cite{Ade:2013lmv, vonderLinden:2014haa, Umetsu:2020wlf, Blanchard:2021dwr,Nunes:2021ipq} that this tension could be the result of systematics uncertainties, for instance, in mass bias. However, in this work, we assume that the tension is an indication of new physics.}  Given the key role the Planck SZ data plays in generating the $S_8$ tension, we also perform a set of fits including the Planck SZ value of $S_8$, in addition to the spectral measurements of KV450. In fitting to the MTH model parameters, we allow $\hat r$ all the way up to one, and also allow $\Delta\hat N$ up to one, both with flat priors. These fits are shown in Figs.~\ref{fig.S8H0withSZ} and \ref{fig.AllparsPlanckBAOH0_KV_SZ_Linear}. Due to the greater pull generated by the Planck SZ result, one sees in Fig.~\ref{fig.S8H0withSZ} that the best-fit regions for the MTH and for the canonical models are all shifted to lower $S_8$. For the \lnur{} model, a price is exacted in pulling down $S_8$, namely that the upper bound on $H_0$ is suppressed. On the other hand, the MTH model still shows agreement with the SH0ES data at the $1\sigma$ level. This is an important test: the MTH model is able to weaken significantly both the $H_0$ and $S_8$ tensions, even when we push the $S_8$ tension to its extreme by including the Planck SZ data.  

%%%%%%%%%%%%%%%
%%%%%%%%%%%%%%%

%%%%%%%%%%%%%%%
%%%%%%%%%%%%%%%

Fig.~\ref{fig.AllparsPlanckBAOH0_KV_SZ_Linear} shows the regions of MTH parameter space preferred by this full complement of datasets, including Planck SZ.  Several important features are notable.  As before, there is a preference for a heavier twin sector (larger $\hat v/v$) in order to obtain larger values of $\hat r$, though the preference is mild.  The prior relationship between larger $\hat r$ and smaller $\Delta\hat N$ is no longer evident. It appears that the bound on $\Delta\hat N$ is mostly independent of $\hat r$, which is evidence that it is dominated by fitting the SH0ES value of $H_0$ and not the matter power spectrum. 

Two other results jump out of these plots. First is the closure of the contours at small $\hat r$, and small $\hat v/v$. The figure indicates a larger than $2\sigma$ preference for non-zero $\hat r$ -- that is, the data clearly prefers a universe with twin matter over one without. Since the $\hat r,\Delta\hat N\to 0$ limit corresponds to the \lcdm{} model, we see the MTH scenario clearly differentiating itself from the standard cosmological scenario and even, within the ansatz of MTH, ruling out the standard cosmological scenario at greater than 95\% C.L. The range of values for $\hat r$ is particularly interesting. At the low end, we see a clear need for some fraction of the dark matter (at least about 20\%) to be in the form of twin matter. On the other hand, the data also allows $\hat r$ as large at 1, meaning that all of the universe's dark matter could be twin matter. Though we have limited our LSS datasets to $k\leq 0.3\,h\,$Mpc$^{-1}$, there is reason to believe that including data probing larger $k$, such as the Lyman-$\alpha$ bound~\cite{Garny:2018byk} or measurements of the sub-halo mass function (SHMF)~\cite{DES:2020fxi}, would cut out models with very large $\hat r$.  Such models show a pronounced suppression in the matter power spectrum at large $k$, caused by the delay in structure formation in the twin sector, which only starts after twin recombination.  If the DM halo is mainly composed of twin atoms, it will also violate bounds on self-interacting DM (SIDM) such as the Bullet Cluster constraint (for a review of the SIDM bounds, see~\cite{Tulin:2017ara}).

The $k$ values associated to these small scale structure constraints, however, would also be in the range in which non-linear corrections to the power spectrum would play a particularly important role and we are not in a good position to estimate those corrections reliably as $\hat r\to 1$.  We may obtain some indication about the smaller scale structure bounds by comparing the suppression of the linear power spectrum at large $k$-modes between the MTH and \lnur{} predictions to the warm DM (WDM) result. For example, in the study of the SHMF~\cite{DES:2020fxi}, the smallest satellite galaxies have $k\approx 50\,{\rm Mpc}^{-1}$ and the WDM suppression to the linear power spectrum should be less than $\approx50\%$ around that $k$-mode~\cite{Carena:2021bqm}. The Lyman-$\alpha$ constraint also sets a comparable bound on the 1D power spectrum that is mainly affected by the suppression of the 3D power spectrum at $k\sim 10~{\rm Mpc}^{-1}$~\cite{Murgia:2017lwo}. Since the MTH scenarios we consider do suppress the matter power spectrum for the better fit to the Planck SZ data at $k\sim 0.1\,{\rm Mpc}^{-1}$, there is a larger suppression at higher $k$-modes that asymptotes to $(1-\hat r)^2$ (see Fig.~\ref{fig.MPSratio}), and we can consider $\hat r\lesssim 25\%$ as a very rough estimate of the smaller structure constraints. When presenting $\chi^2$ results in Table~\ref{tab.PlanckBAOH0_KV450_SZ}, we show the MTH result both with the $\hat r$ prior up to $1$ and $0.25$.  As we will see, the resulting $\chi^2$ of the best-fit points in the two cases are similar to each other. We also numerically verify that the $68\%$~C.L. lower bounds on the $\hat r$ are very similar in the two cases, which is expected from the similar $\chi^2$ of the two best-fit values. A more dedicated study of MTH's Lyman-$\alpha$ and SHMF signals may further strengthen the $\hat r$ bound.

Besides the non-zero $\hat r$, the second feature of Fig.~\ref{fig.AllparsPlanckBAOH0_KV_SZ_Linear} is the now-tight correlation between $\Delta\hat N$ and $\hat v/v$, in which we can increase $\hat v/v$ at the price of simultaneously increasing $\Delta\hat N$, and vice-versa. Though this correlation will have profound implications, it is simple to understand.  The calculation of twin recombination in Sec.~\ref{sec.MTHcosmology} is highly sensitive to the ratio $\hat v/\hat T$. The narrow contour in the plot shows that the data prefers a small range of redshifts for twin recombination. Taking the peak location of the twin visibility function as one of the derived parameters of the MCMC scans, we find the preferred redshift to be $1\times10^4 \lesssim z_{\text{rec, twin}} \lesssim 3 \times 10^4$ at $95\%$~C.L. for the datasets used in this fit. This may be interpreted as a prediction of the redshift at which the twin photons change from fluid-like behavior to freely streaming.

The reason this correlation matters is that the solution to the Hubble tension requires that $\Delta\hat N$ be large.  Within the MTH model, this will then require that $\hat v/v$ also be comparatively large and therefore, higher amounts of fine-tuning {\it vis \`a vis}\/ the hierarchy problem.  However, this fine tuning can be avoided by using different electroweak symmetry breaking mechanisms~(for example, see Ref.~\cite{Goh:2007dh, Beauchesne:2015lva, Harnik:2016koz}) that allow the MTH model with a large twin VEV to be a more natural model.  That is to say, such a solution to the Hubble tension could provide the motivation for further investigation into such mechanisms.

In Table~\ref{tab.PlanckBAOH0_KV450_SZ}, we show the mean and best-fit values of the model parameters along with $H_0$, $S_8$ (and $\sigma_8$), and the matter density for the three cosmological models we have studied here and plotted in Fig.~\ref{fig.S8H0withSZ}.  Note that while the \lnur{} model allows for a larger $H_0$ as compared to the \lcdm{} model, it does little to help the $S_8$ tension (the best-fit value of $S_8$ is slightly better than that of the $\Lambda$CDM model, but the mean is slightly worse). In fact, regions of the parameter space of the \lnur{} model where the Hubble constant is increased more substantially are also regions in which $S_8$ also tends to increase somewhat, making the $S_8$ tension worse.  Clearly, the ability of the \lnur{} model to solve the $H_0$ and $S_8$ tensions simultaneously is quite limited, especially on including the Planck SZ data which greatly prefers a smaller $S_8$.

The MTH model, on the other hand, allows for a substantially larger $H_0$ while simultaneously producing a much smaller $S_8$.  Therefore, the MTH model can accommodate a larger $H_0$ as compared to the \lnur{}, as shown in Table~\ref{tab.PlanckBAOH0_KV450_SZ}.  In the lower part of the table, we show the total $\chi^2$ of the best-fit points of the three models, along with the breakdown of contributions from the different datasets. As mentioned above, we show two best-fit points in the MTH model with $\hat r\leq1$ and $\hat r\leq0.25$ that are motivated by the small scale structure constraints. Although the best-fit point in the $\hat r\leq0.25$ scan is slightly larger ($\Delta\chi^2=2.1$) than the other case, it still provides a much better fit to the data with only three extra parameters and $\Delta\chi^2 \approx-20$ compared to \lcdm{} + \nur{}. Since the \lcdm{} model lives inside the parameter space of the MTH model, one can quote a significance at which the best-fit regions are preferred over the \lcdm{} fit. With 3 extra parameters and $\Delta\chi^2 \approx-20$, we find a $\sim 4\sigma$ significance at which the best-fit MTH regions are preferred over the \lcdm{} fit. This again shows that there is a strong statistical preference for a fraction of the dark sector to be made up of twin particles. The breakdown of the contributions of the different datasets to the total $\chi^2$ shows that the fit to both the SH0ES and Planck SZ data is improved in the MTH model, and it is these two data points that primarily drive the strong preference for the MTH model over the other two.  Importantly, since so little of the improved $\chi^2$ is coming from the KV450 data, we can anticipate that there will be very little sensitivity in this fit to the non-linear corrections to the matter power spectrum.  We will see this to be the case in the next section.

\begin{table}[!htbp]
	\renewcommand*{\arraystretch}{1.4}
	\centering
 	\setlength\tabcolsep{4pt}
	\resizebox{\textwidth}{!}{%
	\begin{tabular}{|c|c|c|c|c|c|c|c|} 
		\hline
		& \multicolumn{2}{c|}{$\Lambda$CDM}& \multicolumn{2}{c|}{$\Lambda$CDM + \nur} & \multicolumn{3}{c|}{MTH}\\ 
		\hline
		Param        & best-fit & mean$\pm\sigma$ & best-fit & mean$\pm\sigma$ &\begin{tabular}{@{}c@{}}best-fit \\ $(\hat r\leq1)$\end{tabular} 
		& 
		\begin{tabular}{@{}c@{}}mean$\pm\sigma$ \\ $(\hat r\leq1)$\end{tabular}  
		& 
		\begin{tabular}{@{}c@{}}best-fit \\ $(\hat r\leq0.25)$\end{tabular} \\
		          %   &  & &  &  & $(\hat r\leq 1)$ & $(\hat r\leq 1)$ & $(\hat r\leq0.25)$\\
		\hline 
		$100~\Omega_b h^2$ & $2.262$ & $2.269_{-0.013}^{+0.014}$  & $2.278$  & $2.275_{-0.014}^{+0.014}$   & $2.275$ & $2.275_{-0.016}^{+0.015}$  & $2.283$\\ 
		$\Omega_{dm } h^2$ & $0.1167$ & $0.1162_{-0.00083}^{+0.00081}$ & $0.1175$ & $0.1175_{-0.0015}^{+0.0011}$  & $0.1245$ & $0.1233_{-0.0029}^{+0.0024}$ & $0.1245$\\ 
		$100~\theta{}_{s }$ & $1.042$  & $1.042_{-0.00029}^{+0.00029}$ & $1.042$  & $1.042_{-0.00032}^{+0.00034}$  & $1.041$ & $1.041_{-0.0004}^{+0.00042}$ & $1.042$\\ 
		$\ln(10^{10}A_{s })$  & $3.015$ & $3.022_{-0.014}^{+0.015}$   & $3.013$  & $3.024_{-0.014}^{+0.015}$     & $3.043$ & $3.051_{-0.015}^{+0.015}$  & $3.051$\\ 
		$n_{s }$      & $0.9759$ & $0.9737_{-0.0038}^{+0.0036}$ & $0.9748$ & $0.9765_{-0.0046}^{+0.0042}$  & $0.9717$ & $0.974_{-0.0049}^{+0.0045}$ & $0.9727$\\ 
		$\tau{}_{reio }$  & $0.04599$ & $0.04783_{-0.007}^{+0.0078}$ & $0.04016$ & $0.04728_{-0.007}^{+0.0081}$  & $0.05222$ & $0.05562_{-0.0076}^{+0.0073}$ & $0.05402$\\ 
		$H_0$        & $69.41$ & $69.68_{-0.39}^{+0.39}$    & $69.94$  & $70.23_{-0.63}^{+0.51}$   & $70.53$ & $70.63_{-0.96}^{+0.77}$ & $70.67$\\ 
		$\sigma_8$     & $0.8035$ & $0.8035_{-0.0051}^{+0.0054}$  & $0.8025$ & $0.8068_{-0.0058}^{+0.0058}$ & $0.7547$ & $0.7687_{-0.011}^{+0.011}$  & $0.7605$\\ 
		$S_8$     & $0.7889$ & $0.7845_{-0.0078}^{+0.0081}$  & $0.7846$ & $0.7856_{-0.0082}^{+0.008}$  & $0.7496$ & $0.7594_{-0.01}^{+0.01}$ & $0.7543$\\ 
		\hline
		$\hat r$      & $-$    & $-$              & $-$    & $-$               & $0.9403$ & \{0.20, 1\} & $0.2488$\\ 
		$ \hat v /v$    & $-$    & $-$              & $-$    & $-$               & $8.768$ & \{3.46, 12.7\}  & $6.314$\\ 
		$\Delta{N}$     & $-$    & $-$              & $0.05941$ & $0.08787_{-0.088}^{+0.023}$  & $0.3409$ & \{0.052, 0.57\}  & $0.3461$\\ 
		\hline 
		$\bm{\chi^2_{total}}$  & \multicolumn{2}{c|}{3094.31} & \multicolumn{2}{c|}{3092.47}  & \multicolumn{2}{c|}{3072.06} & 3074.18\\
		\hline \hline
		Planck         & \multicolumn{2}{c|}{2779.12} & \multicolumn{2}{c|}{2781.91}  & \multicolumn{2}{c|}{2780.45} & 2781.47\\
		BAO           & \multicolumn{2}{c|}{8.27}   & \multicolumn{2}{c|}{9.24}   & \multicolumn{2}{c|}{5.94}  & 6.11\\
		SH0ES          & \multicolumn{2}{c|}{10.60}  & \multicolumn{2}{c|}{8.28}    & \multicolumn{2}{c|}{6.06} & 5.60\\		
		Lensing         & \multicolumn{2}{c|}{13.31}  & \multicolumn{2}{c|}{13.88}   & \multicolumn{2}{c|}{9.85} & 9.58 \\		
		KV450          & \multicolumn{2}{c|}{268.36}  & \multicolumn{2}{c|}{266.82}  & \multicolumn{2}{c|}{269.38} & 271.40 \\		
		Planck SZ        & \multicolumn{2}{c|}{14.65}  & \multicolumn{2}{c|}{12.34}   &  \multicolumn{2}{c|}{0.38} & 0.01\\		
		\hline 
	\end{tabular} }
	\caption{The mean and best-fit values for the \lcdm{}, \lcdm{} + \nur{} and MTH models obtained using the Planck, BAO, SH0ES, KV450, and Planck SZ datasets. $\Delta N$ corresponds to \nur{} for $\Lambda$CDM + \nur{} model and $\Delta \hat N$ for the MTH model. For the MTH parameters, the range allowed at $95\%$~C.L. is shown in brackets. In the lower part of the table, total $\chi^2$ of the best-fit points of the three models, along with the breakdown of contributions from the different datasets, is shown.  $\chi^2$ for the Planck Lensing data is shown separately with the label ``Lensing,'' and is not included in the combined $\chi^2$ for Planck.}
	\label{tab.PlanckBAOH0_KV450_SZ}
\end{table}
%%%%%%%%%%%%%%%%%%%%%%

It is clear from both the figures and from Table~\ref{tab.PlanckBAOH0_KV450_SZ} that a wide range of MTH parameters provide a strong fit to the cosmological data. The VEV ratio, $\hat v/v$, can fall as low as $3$ (within the region preferred by fine-tuning arguments while still consistent with Higgs coupling measurements) but can become as large as $11$, where we obtain a significant weakening of the Hubble tension. The values of $\Delta\hat N$ are tightly correlated with the VEV ratio and can range almost up to $0.6$ while remaining consistent with data.
Note that this is a weaker bound on $\Delta\hat N$ than found in Figs.~\ref{fig.PlanckBAOConstraints} and \ref{fig.KV450linearNhat} due to the inclusion of SH0ES in the current fit.

% The range of values for $\hat r$ is particularly interesting. At the low end, we see a clear need for some fraction of the dark matter (at least about 20\%) to be in the form of twin matter. On the other hand, the data also allows $\hat r$ as large at 1, meaning that all of the universe's dark matter could be twin matter. Though we have limited our LSS datasets to $k\leq 0.3\,h\,$Mpc$^{-1}$, there is reason to believe that including data probing larger $k$, such as the Lyman-$\alpha$ bound~\cite{Garny:2018byk} or measurements of the sub-halo mass function~\cite{DES:2020fxi}, would cut out models with very large $\hat r$.  Such models show a pronounced suppression in the matter power spectrum at large $k$, caused by the delay in structure formation in the twin sector, which only starts after twin recombination.  If the DM halo is mainly composed of twin atoms, it will also violate bounds on self-interacting DM (SIDM) such as the Bullet Cluster constraint (for a review of the SIDM bounds, see~\cite{Tulin:2017ara}). However, the $k$ values associated to these small scale structure constraints would also be in the range in which non-linear corrections to the power spectrum would play a particularly important role and we are not in a good position to estimate those corrections reliably as $\hat r\to 1$. Thus we urge caution in assuming from these results that the entirety of the universe's dark matter could be in the form of twin matter.

%%%%%%%%%%%%%%%%%%%%%%%%%%%%%%%%%%%%
\subsection{Non-Linear Corrections}
\label{subsec.nonlinear}

%%%%%%%%%%%%%%%%%
\begin{figure*}[t]	
  \centering
  \begin{subfigure}{0.48\textwidth}
      \includegraphics[width=\textwidth]{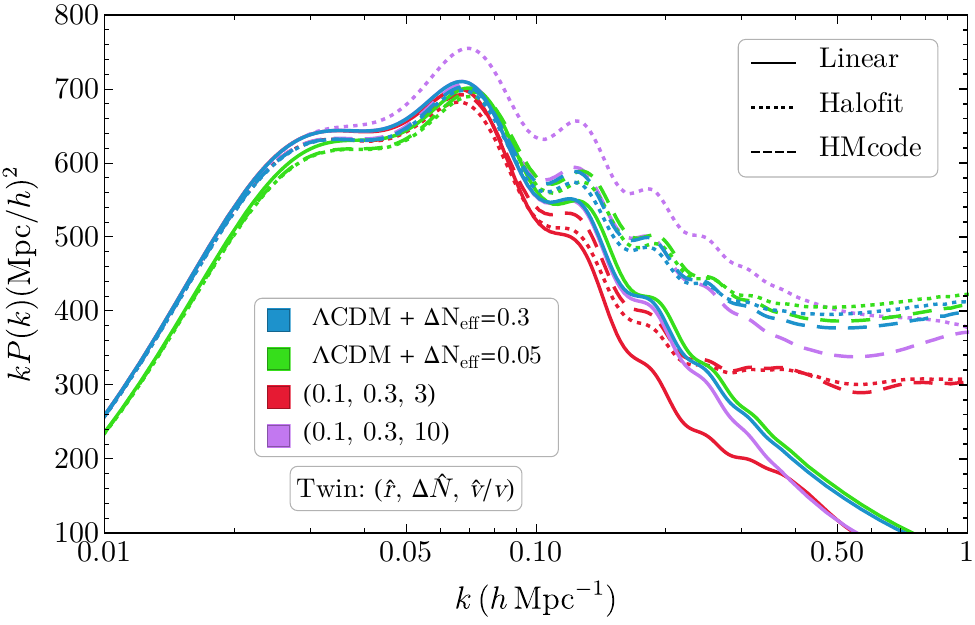}
      \caption{}
      \label{fig.AbsMPS_HaloFit_HMcode}
  \end{subfigure}
  \quad
  \begin{subfigure}{0.48\textwidth}
      \includegraphics[width=\textwidth]{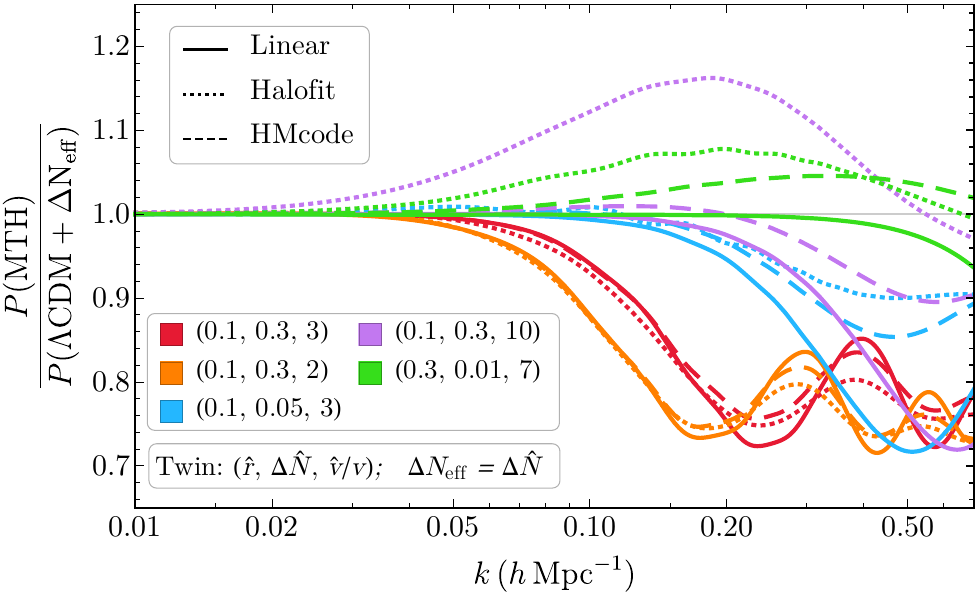} 
      \caption{}
      \label{fig.MPS_HaloFit_HMcode}
  \end{subfigure}
  \caption{Non-linear corrections to the matter power spectrum computed using Halofit and HMcode.}
  \label{fig.MPSnonlinear}
\end{figure*}
%%%%%%%%%%%%%%%%%

Although the growth of the matter density contrast can be well described by linear perturbation theory at large scales, its evolution becomes highly non-linear at smaller scales: $k \gtrsim 0.1~h\,\mbox{Mpc}^{-1}$. The exact computation of this non-linear evolution requires large $N$-body simulations, whose results can be matched to simple empirical parametrizations. This process has been done for the canonical \lcdm{} and \lnur{} models, but not for the MTH model. Nonetheless, it would be helpful to make some rough estimates of these effects and for that we turn to the parametrizations obtained for the \lcdm{} model. 

There are two built-in modules in \CLASS{} that can be utilized to compute the non-linear corrections, namely HMcode~\cite{Mead:2016zqy} and Halofit~\cite{Takahashi:2012em}. In this subsection we will discuss how our results change when including non-linear corrections from these modules. One should understand these results as an illustration of how non-linear corrections could effect our study instead of taking the numbers as precise results. Note that in our analysis, the non-linear correction is really only important when fitting the KV450 data, where we include the matter power spectrum up to $k \leq 0.3~h\,\mbox{Mpc}^{-1}$.

In Fig.~\ref{fig.MPSnonlinear}, we show a few examples of matter power spectra with non-linear corrections obtained using Halofit and HMcode.  From the left plot, it is clear that the non-linear corrections from Halofit and HMcode are quite similar for the \lnur{} model. However, for the MTH model, particularly the $\hat v /v=10$ case, we see that these corrections can be quite different.  The difference between the non-linear corrections from the two modules can be seen more clearly in the right plot, where the ratio of the matter power spectrum for the MTH model to that of the \lnur{} model is plotted.  To obtain this ratio plot, the matter power spectrum in both the numerator and denominator is computed using the same method (linear, Halofit, or HMcode).  By taking the ratio, we are extracting the features that are specific to the MTH model.

The results of these studies are somewhat surprising. Both HMcode and Halofit produce very similar non-linear corrections for cases in which the linear matter power spectrum of the MTH model begins to deviate significantly from the \lnur{} spectrum at $k \lesssim 0.1~h\rm{Mpc}^{-1}$ (orange and red lines in the figures).  Though not shown in the plots here, we find that the same also holds for the modes that start to show deviations only for $k \gtrsim 0.7~h\rm{Mpc}^{-1}$.  However, in the intermediate region (for instance, purple lines), the Halofit parametrization produces results that are both quite dissimilar to those of HMcode and to expectations. In particular, the Halofit calculation generates a rather incomprehensible enhancement of the power spectrum all the way down to $k\approx 0.03\,h\,$Mpc$^{-1}$, well outside the non-linear regime.  While we are not sure of the underlying cause of this behavior, neither of these codes was designed for the model studied here.  However, the odd and seemingly unphysical behavior of Halofit is the reason that we opted to use HMcode for the estimation of non-linear effects in most of our fits.

%%%%%%%%%%%%%%%%%%%%%%%%%%%%%%
\begin{figure*}[t]	
  \centering
    \includegraphics[width=0.48\textwidth]{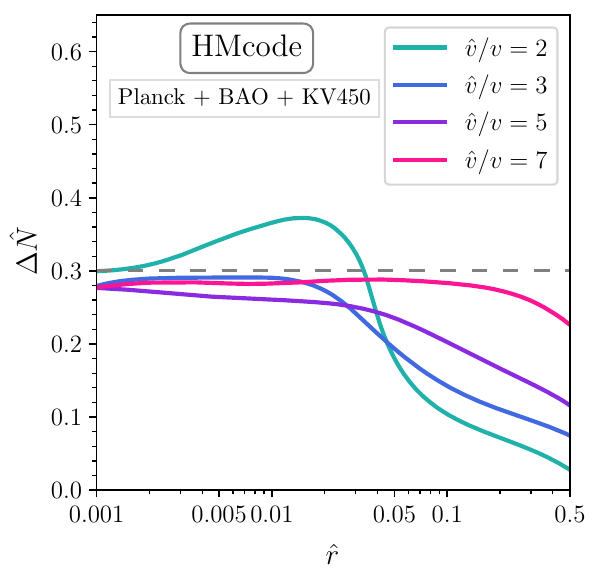}
  \caption{Constraints from Planck+BAO+KV450 datasets with non-linear power spectrum corrections from HMcode. }
  \label{fig.KV_Constraints_NonLinear}
\end{figure*}
%%%%%%%%%%%%%%%%%%%%%%%%%%%%%%

In Fig.~\ref{fig.KV_Constraints_NonLinear}, we show the constraints on the MTH model from the Planck+BAO+KV450 datasets, with the matter power spectrum calculated using the HMcode non-linear corrections. The bounds with HMcode only differ from the linear result shown in Fig.~\ref{fig.KV450linearNhat} by less than 10\%, at least until $\hat r\gtrsim 0.4$. In fact, that region where the mirror baryons represent a large fraction of the dark matter is also the regime in which we have the least trust in either HMcode or Halofit, since our model looks less and less like the \lcdm{} or \lnur{} models.

\begin{figure*}[t]	
    \centering
	\begin{subfigure}{0.45\textwidth}
		\includegraphics[width=\textwidth]{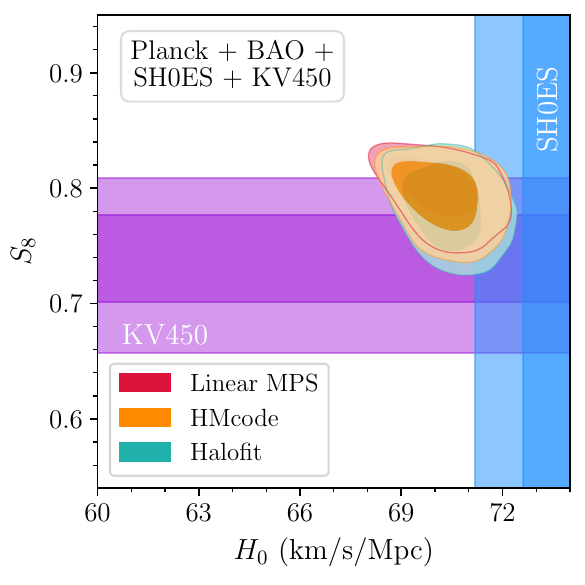}
		\caption{}\label{fig.PlanckBAOH0_KV_NL}
	\end{subfigure}
	\quad\quad\quad
	\begin{subfigure}{0.45\textwidth}
		\includegraphics[width=\textwidth]{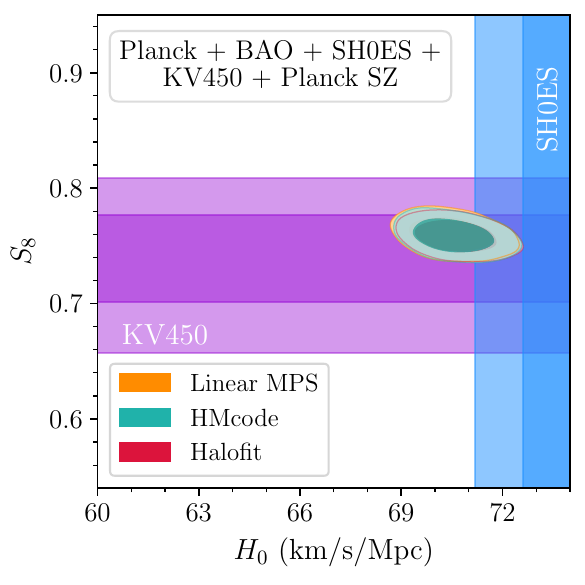}
		\caption{}\label{fig.PlanckBAOH0_KV_SZ_NL}
	\end{subfigure}
	\caption{Same plots as in Fig.~\ref{fig.S8H0withKV} (left) and Fig.~\ref{fig.S8H0withSZ} (right) but with two types of non-linear corrections to the matter power spectrum.}\label{fig.PlanckBAOH0_KV_SZ}
\end{figure*}

\begin{figure*}[t]	
  \centering
  \includegraphics[width=\textwidth]{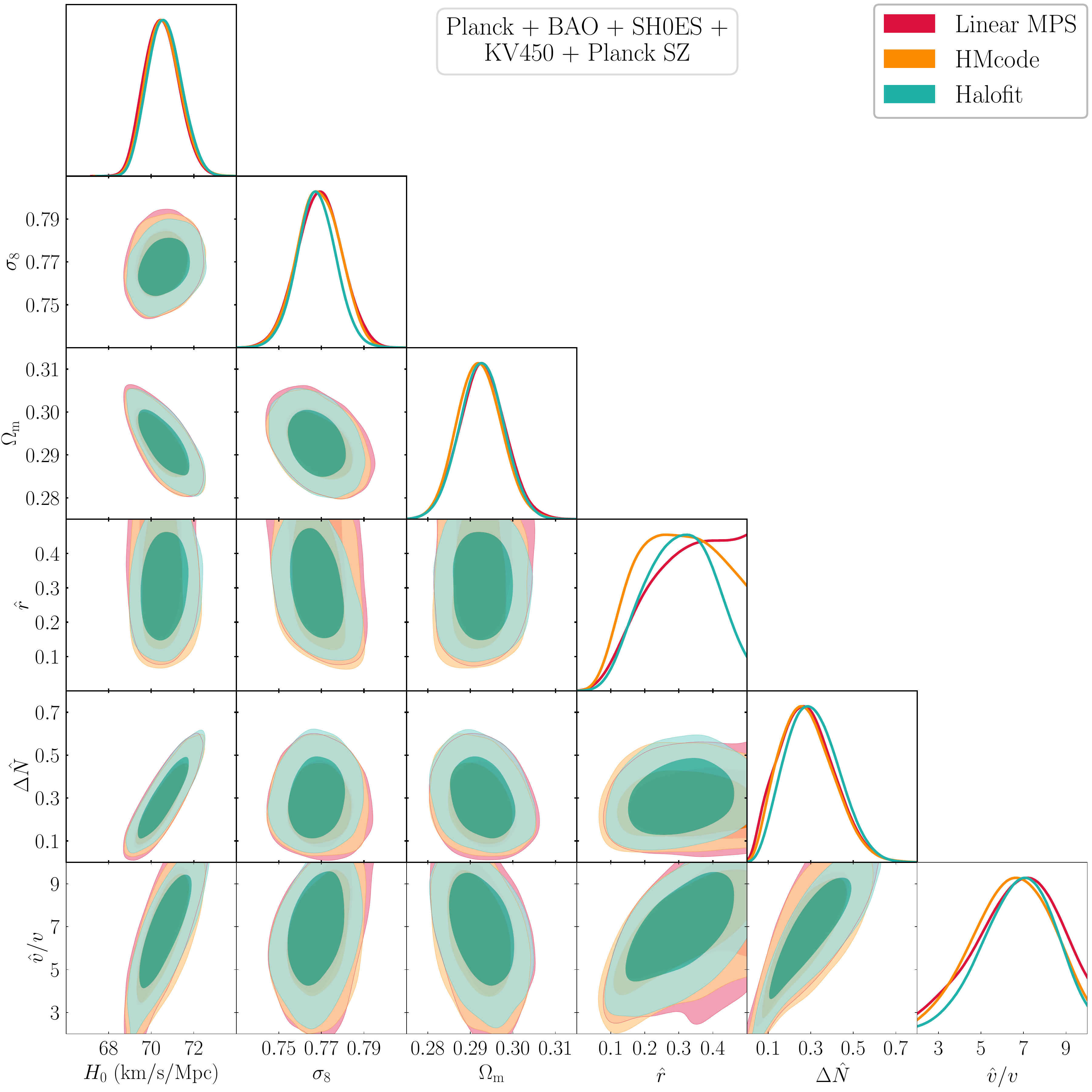}
  \caption{Same plots as in Fig.~\ref{fig.AllparsPlanckBAOH0_KV_SZ_Linear} but with two different non-linear corrections to the matter power spectrum. Since non-linear corrections for the MTH model using Halofit or HMcode becomes less reliable for large \rhat, we use an upper limit of 0.5 on \rhat{} to obtain this plot. For the VEV, we use a flat prior of $\hat v/v \in [2, 10]$.} 
  \label{fig.MTHparsPlanckBAOH0_KV_SZ_NonLinear}
\end{figure*}

As another check, in Fig.~\ref{fig.PlanckBAOH0_KV_SZ} we compare the regions of the $(H_0,S_8)$ plane preferred by the MTH model, with and without the non-linear corrections of Halofit and HMcode. We claimed earlier that because the bounds are driven primarily by the Planck+BAO and the Planck SZ data, which are both independent of the non-linear corrections, the preferred regions in $H_0$ versus $S_8$ should show little dependence on whether we do or do not include non-linear corrections. This is exactly what we see. In both the left figure (without Planck SZ) and the right figure (with Planck SZ), the contours for the linear and for the two trial non-linear fits are nearly the same. In particular, the ability of the model to achieve a larger $H_0$ and smaller $S_8$ simultaneously is completely independent of the form of the non-linear corrections or whether we include them at all, as expected. Similarly, we also compare the bounds on the MTH parameters in Fig.~\ref{fig.AllparsPlanckBAOH0_KV_SZ_Linear} to the results that include the non-linear corrections. As we can see in Fig.~\ref{fig.MTHparsPlanckBAOH0_KV_SZ_NonLinear}, both HMcode and Halofit produce similar bounds for $\hat{r}\lesssim 0.3$ and require the same minimum $\hat{r}$ in order to relax the $S_8$ tension.

%%%%%%%%%%%%%%%%%%%%%%%%%%%%%%
%%%%%%%%%%%%%%%%%%%%%%%%%%%%%%

%%%%%%%%%%%%%%%%%%%%%%%%%%%%%%%
%%%%%%%%%%%%%%%%%%%%%%%%%%%%%%%

In all, we have been able to show that using well-known calculations of the non-linear effects in the matter power spectrum, our results change little from those obtained with a purely linear spectrum. And while these non-linear calculations (Halofit and HMcode) are not done with the MTH model in mind, they do provide us a rough sense of how much we might expect our results to change should a detailed calculation of non-linear effects in the MTH model be completed.

%%%%%%%%%%%%%%%%%%%%%%%%%%%%%%
\subsection{Future Prospects}
\label{subsec.future}

%%%%%%%%%%%%%%%%%%%%%%
\begin{figure*} [!htb]	
  \centering
  \includegraphics[width=0.6\textwidth]{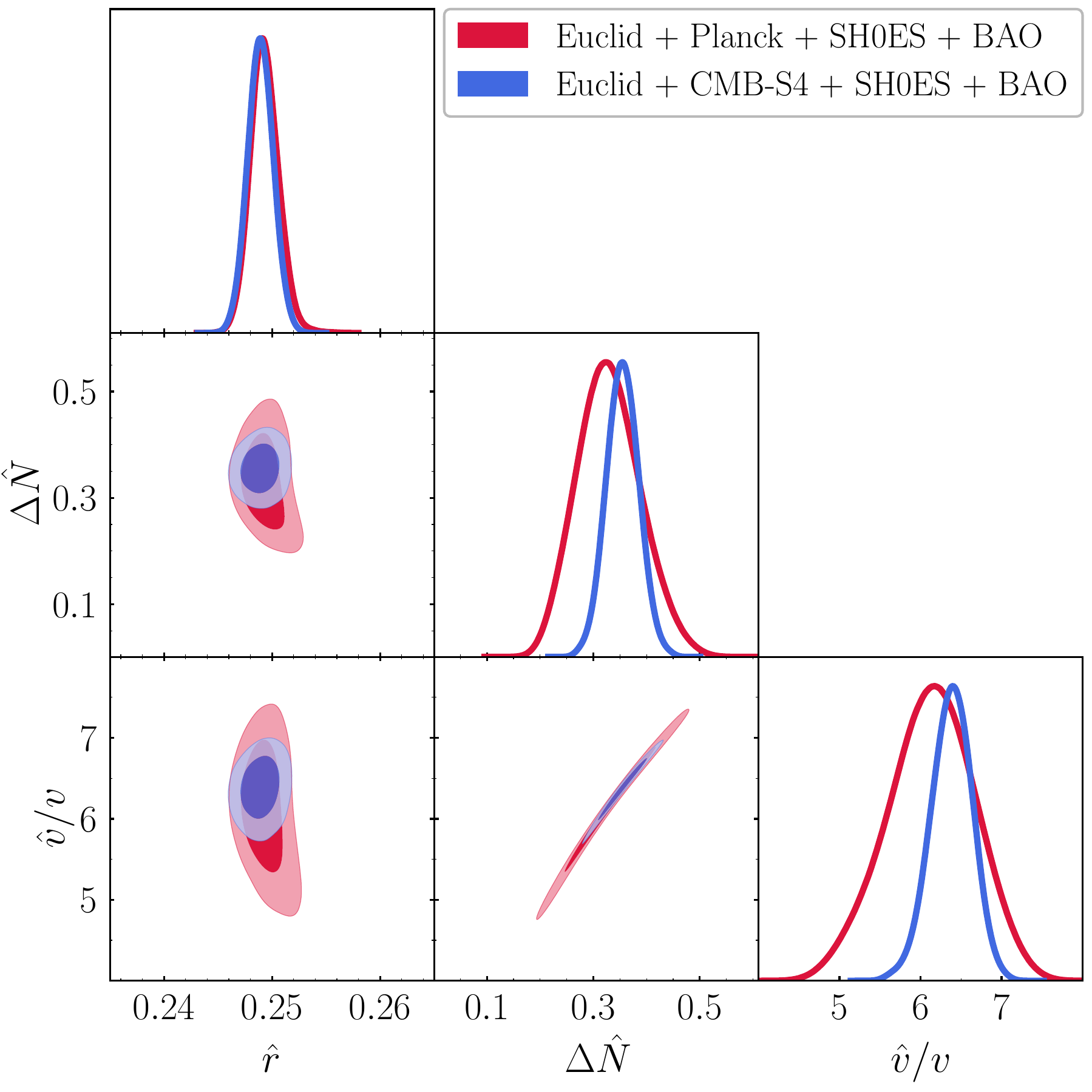}
  \caption{The predicted ranges for the twin parameters $\hat r$, $\Delta \hat N$, and $\hat v/v$ obtained using the Euclid and CMB-S4 mock data along with the observed Planck, BAO, and SH0ES data.  For both Euclid and CMB-S4, the mock data is generated using the best-fit values for the MTH model in Table~\ref{tab.PlanckBAOH0_KV450_SZ}.}
  \label{fig.MTHpars_forecast}
\end{figure*}
%%%%%%%%%%%%%

If the MTH model is the correct description of the universe, how precisely can we measure the twin parameters with future experiments and surveys? Will future surveys, such as Euclid or CMB-stage 4, be able to pin down the parameters of the MTH model, and if so, to what degree?

Working from the assumption that the $H_0$ and $S_8$ tensions from the Planck SZ and SH0ES data do indicate the existence of MTH particles, we generate mock data using the the best-fit values for $\hat r \leq 0.25$ shown in Table~\ref{tab.PlanckBAOH0_KV450_SZ} for the nine free parameters describing the cosmology of the MTH model.  Those best-fit values are $\{ \Omega_{\rm dm}h^2$, $\Omega_b h^2$, 100$\,\theta_s$, $\ln(10^{10}A_s)$, $n_s$, $\tau_{\rm reio}$, \rhat, \Ntwin, \vevratio \} = $\{$0.1245, 0.02283, 1.042, 3.051, 0.9727, 0.05402, 0.2488, 0.3461, 6.314$\}$.  After generating mock data to simulate such a universe, we then perform an MCMC study to constrain the parameters using the expected sensitivities to the matter power spectrum from the Euclid satellite~\cite{Amendola:2012ys}, and to CMB anisotropies from CMB-S4~\cite{Abazajian:2016yjj}.  For the purposes of this analysis, we only use the linear matter power spectra in the scans.  In Fig.~\ref{fig.MTHpars_forecast}, we show the preferred regions of the twin parameters when including the Euclid LSS mock data in addition to the Planck+BAO+SH0ES data in red (larger ellipses), and when including the Euclid and CMB-S4 data in addition to SH0ES+BAO in blue (smaller ellipses). Both regions should be compared to Fig.~\ref{fig.AllparsPlanckBAOH0_KV_SZ_Linear} which includes the Planck+BAO+SH0ES+KV450 datasets. 

What is immediately clear is that even with only the Euclid data, one can greatly improve the fits, closing the contours on the MTH parameters and thereby ruling out the \lcdm{} and \lnur{} models with high confidence. In particular, Euclid data should be able to restrict $\hat r$ to a narrow band with errors of order $\pm2$\%. However, the uncertainty in $\Delta\hat N$ will remain sizeable, of order $\pm 30$\%, while the uncertainty in \vevratio{} is roughly $\pm 20\%$. (All errors are quoted at $68\%$~C.L. unless otherwise stated.)

Including the CMB-S4 data will help only marginally in shrinking the errors on $\hat r$ but will greatly shrink the errors in $\Delta\hat N$ and \vevratio{}, halving the errors as compared to the case above, leaving an uncertainty in $\Delta\hat N$ of order $\pm15$\% and in \vevratio{} of order $\pm 10\%$. Moreover, with the CMB-S4 data, the redshift of twin recombination can be inferred with a relative precision of about 10\% at 95\% C.L.

In both cases, the limiting factor in pushing down the errors in both \Ntwin{} and \vevratio{} is the degeneracy in these two measurements which we have encountered before. Each dataset reduces the allowed range of \vevratio{} and \Ntwin{}, but the errors on \Ntwin{} and \vevratio{} still remain highly correlated due to their underlying connection to the redshift at which twin recombination occurs.

In order to break this degeneracy, one turns to collider physics. The high-luminosity run of the LHC (HL-LHC) should be able to probe the invisible Higgs branching ratio down to approximately the 4\% level at $95\%$~C.L.~\cite{CMS:2018tip}, and measure the deviation of the $Z$-Higgs coupling to the $2\%$ level at $68\%$~C.L.~\cite{CMS:2018qgz,ATLAS:2018jlh}. These two observables allow independent measurements of the $\hat v$ value if $\hat v/v\lesssim 5$.  Though the best-fit point itself (with \vevratio{} = 6.3) may evade the measurements at the HL-LHC, we know from Fig.~\ref{fig.AllparsPlanckBAOH0_KV_SZ_Linear} that \vevratio{} as low as 3.5 still fits the combined cosmological data much better than the \lcdm{} model.  For instance, if $3\lesssim\hat v/v\lesssim 5$, the HL-LHC can observe the Higgs decay into invisible twin particles and measure \vevratio{} at the $O(10\%)$ level.  Future lepton colliders (see Ref.~\cite{deBlas:2019rxi} for a review of expected precision at these colliders), in their role as Higgs factories, should be able to measure \vevratio{} to much higher precision.  Such measurements would help to improve the cosmological measurement of the twin parameters and strengthen the connection between the model's cosmological signatures and its role in stabilizing the electroweak hierarchy.

%%%%%%%%%%%%%%%%%%%%
\section{Conclusion}
\label{sec.conclusions}
%%%%%%%%%%%%%%%%%%%%

In this paper, we have presented the most thorough study to date of the cosmology of the Mirror Twin Higgs model by incorporating the model into the Boltzmann code \CLASS{} and performing a MCMC study with current CMB, BAO, and LSS datasets.  In particular, we consider a scenario in which the twin baryons constitute a subcomponent of the universe's DM. The contribution of the twin photons and neutrinos to dark radiation is suppressed due to asymmetric reheating, but remains large enough to be detected in future CMB experiments. 

This framework leads to distinctive signals in both the CMB and LSS. Before twin recombination, baryon acoustic oscillations in the twin sector lead to a suppression of structure formation at large scales and leave a residual oscillatory pattern in the matter power spectrum.  Twin BAO, the presence of the free-streaming twin neutrinos, and twin photons that propagate as a fluid before twin recombination all work together to modify the CMB power spectrum in observable ways. On performing the MCMC scan of the MTH parameters, we calculated the 
current Planck+BAO bounds on the twin parameters $(\hat r,\Delta\hat N,\hat v/v)$, showing our result in Fig.~\ref{fig.PlanckBAOConstraints}. The data roughly require the twin radiation abundance $\Delta\hat{N} \lesssim 0.3$ when $\hat r\lesssim 5\%$.  The bounds are a bit weaker comparing to the free-streaming radiation bound when $\hat v/v\lesssim 3$, in which case the twin photons remain as interacting radiation until a later time, weakening the bound.  The $\Delta\hat{N}$ bounds become stronger for larger $\hat r$ due to the modification of gravity perturbations from twin BAO. For instance, if $\Delta\hat N=0.2$, the energy density fraction of the twin baryons needs to be less than about $3\%$ ($10\%$) for $\hat v/v=3\,\,(5)$ (at $95\%$~C.L.). We find these bounds remain roughly the same after including the power spectrum constraints from KV450, as shown in  Figs.~\ref{fig.KV450linearNhat} and \ref{fig.KV_Constraints_NonLinear}, although the precise bounds require careful study of the non-linear corrections to the matter power spectrum at $k\gtrsim 0.1\,h \,{\rm Mpc}^{-1}$.

Earlier literature discussed the possibility of relaxing both the $H_0$ and $S_8$ tensions through the existence of the twin radiation and the twin BAO process~\cite{Chacko:2018vss}. Using our modified \CLASS{} code for the MTH cosmology, we can finally examine that claim through an MCMC study with the real cosmological data. What we find is that the presence of the MTH sector does allow an enhancement of the $H_0$ value while providing a smaller $S_8$ than the fits of the \lcdm{} and \lnur{} models. When fitting Planck+BAO+SH0ES, with or without including the KV450 data that only indicates a mild $S_8$ tension in the \lcdm{} fit of the CMB data, the contours of the MTH model in the $(S_8,H_0)$ plane simultaneously allow for a higher $H_0$ and a lower $S_8$ as compared to the \lcdm{} model (Figs.~\ref{fig.PlanckBAOH0_KV_SZ_Linear} and \ref{fig.PlanckBAOH0_KV_NL}). Moreover, when including the Planck~2013 SZ data with a fixed  mass bias, for which the data prefers a significant suppression in the matter power spectrum as compared to the \lcdm{} fit of the CMB data, the MTH model still allows a large $H_0$ value that better agrees with the SH0ES result (Figs.~\ref{fig.S8H0withSZ} and \ref{fig.PlanckBAOH0_KV_SZ_NL}). In this case, the best-fit parameters of the MTH model, shown in Table~\ref{tab.PlanckBAOH0_KV450_SZ}, improve the fits over both the \lcdm{} and \lnur{} models by $\Delta\chi^2 \approx -20$. The best-fit model (for $\hat r \leq 0.25$) has $\hat v/v=6.3$ , but $\chi^2$ changes slowly across a wide range of $\hat v/v$ values. For example, models with $\hat v/v=4$, a VEV ratio which induces relatively mild tuning when used to solve the Higgs hierarchy problem, still produce fits with $\Delta\chi^2\simeq-20$ compared to the canonical models.

Taken together, our study indicates a preference for the MTH model over \lcdm{} by $\sim 4\sigma$ when including the Planck SZ data. The fit can also be used to bound the redshift for twin recombination to $1\times10^4 \lesssim z_{\text{rec, twin}} \lesssim 3 \times 10^4$ at $95\%$~C.L.

In addition to calculating MTH bounds from existing data, we have also used the modified \CLASS{} code to estimate the precision to be expected of future cosmological measurements on the MTH parameter space. Taking the best-fit value of MTH parameters in Table~\ref{tab.PlanckBAOH0_KV450_SZ} as the true cosmological parameters, the Euclid mission can measure $\hat r$ ($\hat v/v$) at the $\sim 1\%$ ($10\%$) level precision and determine the peak of the visibility function of the twin recombination with $\sim 10\%$ level precision. The CMB-S4 data can further improve the $\Delta\hat N$ measurement and tighten the $\hat v/v$ uncertainty to be $\lesssim 10\%$. Higgs-related measurements at the HL-LHC can further improve the constraints, and tighten the relation between the twin sector as observed in cosmology and as a solution to the Higgs hierarchy problem.

Our code can be easily extended to study the cosmological signals for models of hidden naturalness which, like the mirror twin Higgs model studied here, also contain atomic dark matter and its accompanying dark radiation. There are already other solutions to the Higgs hierarchy problem, such as the $N$-naturalness scenario, which predict the existence of a SM-like dark sector that can generate cosmological signatures similar to those studied here, and to which this code can be extended.  In addition, it is a simple matter to imagine, and to implement in the code, multiple atomic DM species, each with their own binding energies and relative abundances.  It remains an open question, however, as to how well future cosmological data will be able to differentiate among various hidden naturalness scenarios and then measure their parameters. It is therefore urgent to collect more theoretical templates for the study of hidden naturalness cosmology before that future data arrives. The present work takes us one step towards the tantalizing possibility that the first hints of naturalness in the Higgs sector may come from precision cosmological data.

%%%%%%%%%%%%%%%%%%%%%%%%%%%%%%%%%%%%%%%
\begin{acknowledgments}
We thank Michael Geller, Subhajit Ghosh, Vivian Poulin, and Sandip Roy for the helpful discussions and Zackaria Chacko and David Curtin for useful comments on the draft.  ML acknowledges support from the Fermi Research Alliance, LLC, under Contract No.\ DEAC02-07CH11359 with the U.S. Department of Energy, Office of Science, Office of High Energy Physics. The research of YT is supported by the NSF grant PHY-2014165 and PHY-2112540. JK was supported by the National Research Foundation of Korea (NRF) grant funded by the Korea government (MSIT) (No.\ 2021R1C1C1005076), by National Science Foundation under grant PHY-1820860, and PHY-1230860. SB would like to thank the Universities Research Association (URA) for the Visiting Scholars Program award 
(No.\ 20-F-02), which was used to fund this research.  SB also acknowledges support in part by the DOE grant DE-SC0011784.
 
\end{acknowledgments}

%%%%%%%%%%%%%%%%%%%%%%%%%%%%%%%%%%%%%%%
\appendix
%\section{MCMC results with different datasets}
\section{Additional MCMC results}
\label{app.otherdata}
%%%%%%%%%%%%%%%%%%%%

%%%%%%%%%%%%%%%%%%%%%%%%%%%%%%
\begin{figure*}[t]	
    \centering
	\begin{subfigure}{0.45\textwidth}
		\includegraphics[width=\textwidth]{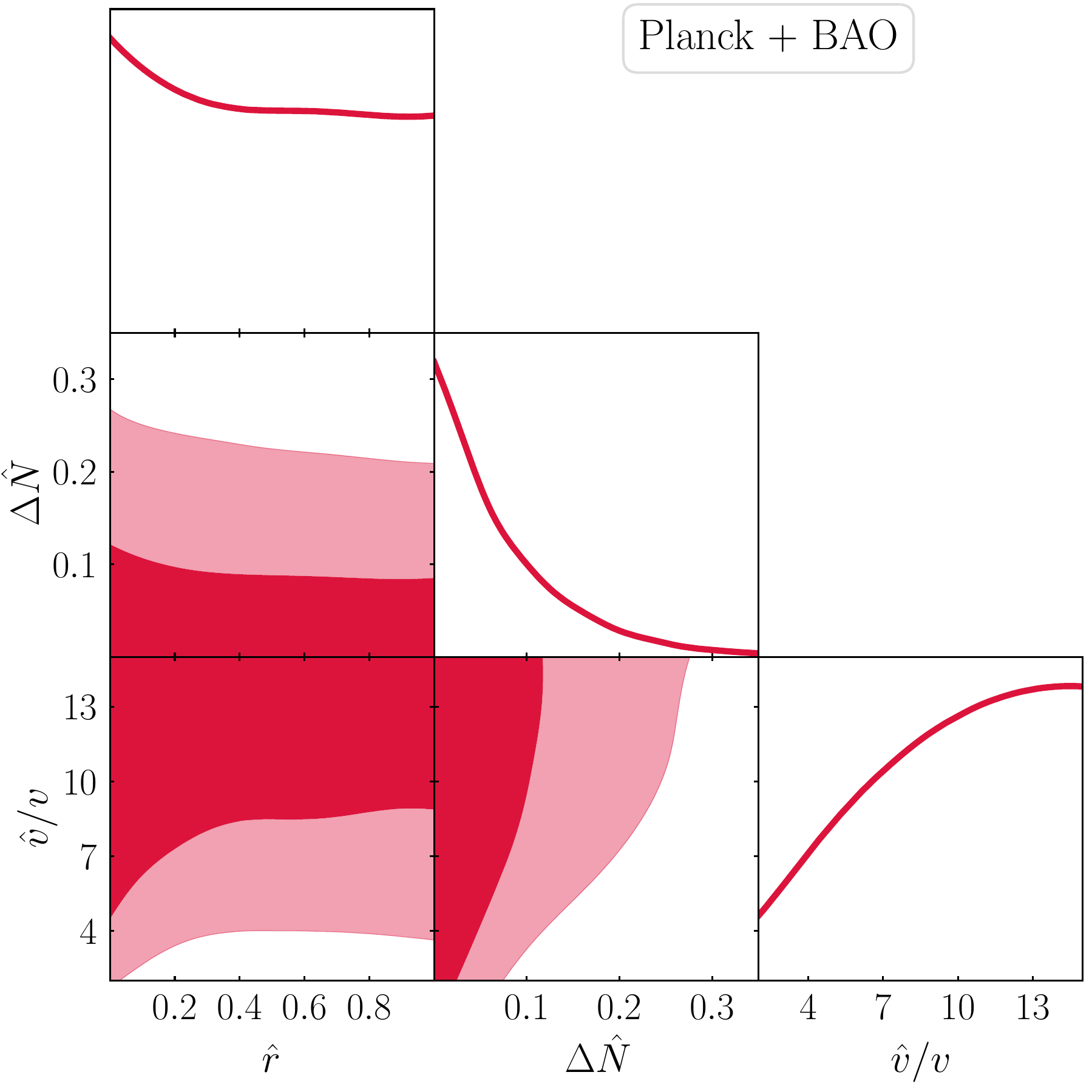}
		\caption{}\label{fig.MTHpars_Planck}
	\end{subfigure}
	\quad
	\begin{subfigure}{0.45\textwidth}
		\includegraphics[width=\textwidth]{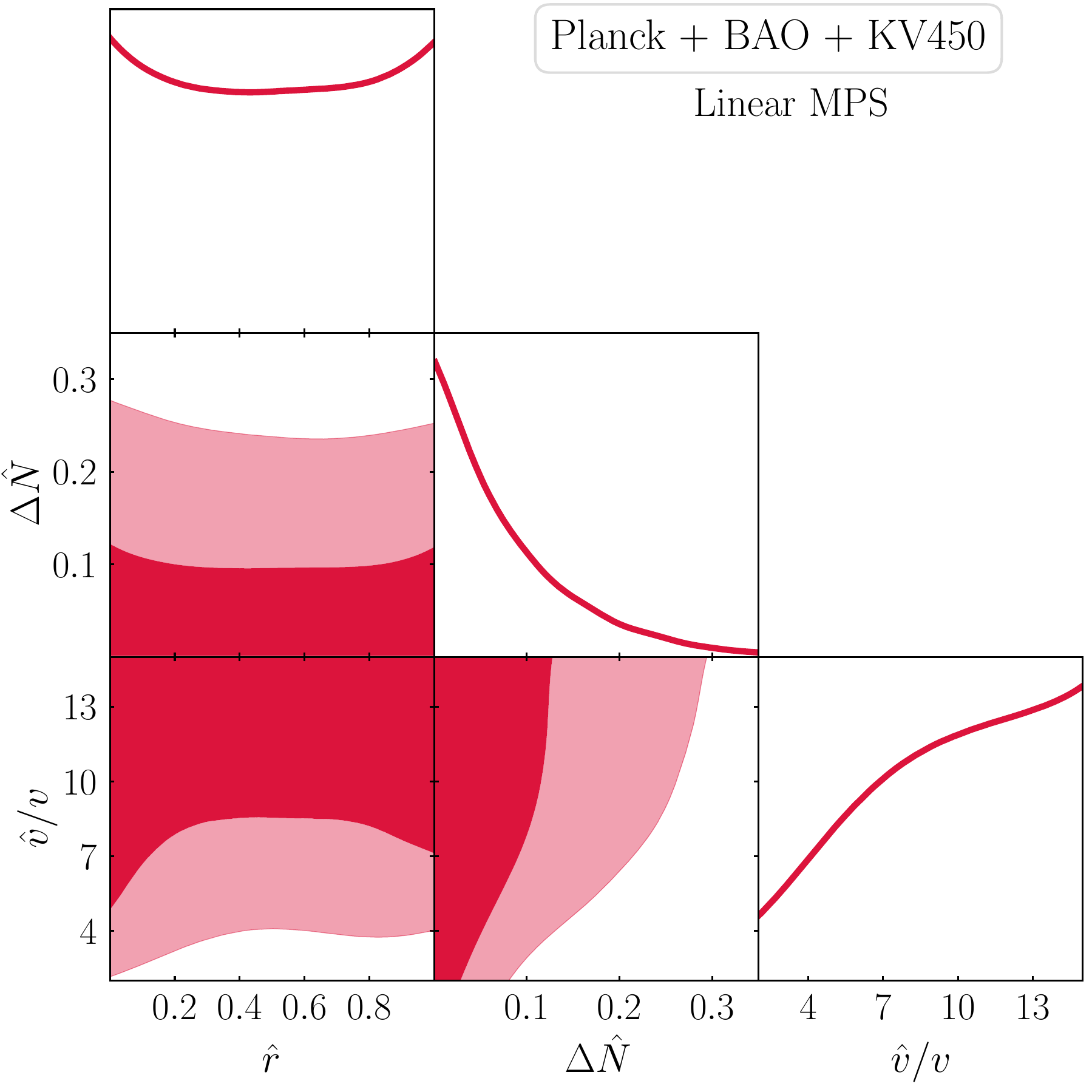}
		\caption{}\label{fig.MTHpars_Planck_KV}
	\end{subfigure}
	\caption{Constraints on all three MTH parameters using the Planck+BAO (left) and Planck+BAO+KV450 (right) datasets.}\label{fig.Appendix_MTHpars}
\end{figure*}
%%%%%%%%%%%%%%%%%%%%%%%%%%%%%%
In this appendix, we show the constraints on the twin parameters using the Planck+BAO (Fig.~\ref{fig.MTHpars_Planck}) and Planck+BAO+KV450 datasets (Fig.~\ref{fig.MTHpars_Planck_KV}), with all the three twin parameters taken to be free parameters. Results in which the vev ratio is taken to be fixed are given in Figs.~\ref{fig.PlanckBAOConstraints} and \ref{fig.KV450linearNhat}. 
\bibliography{MTH_cosmo}
\bibliographystyle{JHEP}
\end{document}